\documentclass[journal,10pt]{IEEEtran}
\usepackage{cite}
\usepackage{bm}
\usepackage{amsmath}
\usepackage{mathrsfs}

\usepackage{algorithmic}

\usepackage{array}

\ifCLASSOPTIONcompsoc
\usepackage[caption=false,font=normalsize,labelfont=sf,textfont=sf]{\part{title}subfig}
\else
\usepackage{subfigure}
\fi

\usepackage{url}
\usepackage{graphicx,amsmath,amssymb,amsfonts}
\usepackage{algorithmic,algorithm}
\usepackage{stfloats}
\usepackage{amsmath}
\allowdisplaybreaks[3]
\usepackage{xcolor}
\usepackage{multirow}
\usepackage{threeparttable}

\newtheorem{theorem}{\textbf{Theorem}}

\newtheorem{lemma}{\textbf{Lemma}}

\newtheorem{corollary}{\textbf{Corollary}}

\newtheorem{proposition}{\textbf{Proposition}}

\makeatletter

\newcommand{\Rmnum}[1]{\expandafter\@slowromancap\romannumeral #1@}
\newcommand{\tabincell}[2]{\begin{tabular}{@{}#1@{}}#2\end{tabular}}
\makeatother

\begin{document}
	
	\title{
		{\fontsize{23.5 pt}{\baselineskip}\selectfont 
			{Safeguarding NOMA Networks via Reconfigurable Dual-Functional Surface under Imperfect CSI}} 
	}
	\author{Wen~Wang,~Wanli~Ni,~Hui~Tian,~Zhaohui~Yang,~Chongwen~Huang,~and~Kai-Kit~Wong
		\thanks{W. Wang, W. Ni, and H. Tian are with the State Key Laboratory of Networking and Switching Technology, Beijing University of Posts and Telecommunications, Beijing, China (e-mail: \{wen.wang, charleswall, tianhui\}@bupt.edu.cn) {\emph{Corresponding author: Hui Tian.}}}
		
		\thanks{Z. Yang and K.-K. Wong are with the Department of Electronic and Electrical Engineering, University College London, London, United Kingdom (e-mail: zhaohui.yang@ucl.ac.uk; kai-kit.wong@ucl.ac.uk).}
		\thanks{C. Huang is with Zhejiang Provincial Key Lab of information processing communication and networking, Zhejiang University, Hangzhou, China (e-mail: chongwenhuang@zju.edu.cn).}
	}
	\maketitle
	\begin{abstract}
		This paper investigates the use of the reconfigurable dual-functional surface to guarantee the full-space secure transmission in non-orthogonal multiple access (NOMA) networks.
		In the presence of eavesdroppers, the downlink communication from the base station to the legitimate users is safeguarded by the simultaneously transmitting and reflecting reconfigurable intelligent surface (STAR-RIS), where three practical operating protocols, namely energy splitting (ES), mode selection (MS), and time splitting (TS), are studied.
		The joint optimization of power allocation, active and passive beamforming is investigated to maximize the secrecy energy efficiency (SEE), taking into account the imperfect channel state information (CSI) of all channels.
		For ES, by approximating the semi-infinite constraints with the $\mathcal{S}$-procedure and general sign-definiteness, the problem is solved by an alternating optimization framework.
		Besides, the proposed algorithm is extended to the MS protocol by solving a mixed-integer non-convex problem.
		While for TS, a two-layer iterative method is proposed.
		Simulation results show that:
		1) The proposed STAR-RIS assisted NOMA networks are able to provide up to 33.6\% higher SEE than conventional RIS counterparts;
		2) TS and ES protocols are generally preferable for low and high power domain, respectively;
		3) The accuracy of CSI estimation and the bit resolution power consumption are crucial to reap the SEE benefits offered by STAR-RIS.
	\end{abstract}
	\vspace{-1mm}
	
	\vspace{-2mm}
	\section{Introduction}
	\IEEEPARstart{I}{n} the forthcoming sixth-generation (6G) era, it is forecast that the connected devices will be extremely dense and the resulting energy demand will become untenable\cite{2022-6G}.
	Besides, communication security is a critical issue in 6G networks due to the broadcast nature of wireless channels\cite{6G_vision}.
	Owing to their prominent capability of achieving green communications and enhancing physical layer security (PLS), reconfigurable intelligent surfaces (RISs) are particularly appealing in industry and academia\cite{RIS-review-1,RIS-review-2,RIS-review-3}.
	Specifically, the RIS consisting of low-cost passive elements can work stably without dedicated energy supply. 
	By intelligently tuning the amplitudes and phase shifts of these elements, the RIS is able to modify the wireless channel and customize a favorable smart radio environment (SRE)\cite{JSAC-SRE}.
	With the aid of RISs, the signals from different links can be coherently combined at the legitimate users (Bobs) to constructively enhance the desired signal, or destructively at the eavesdroppers (Eves) to suppress information leakage\cite{Wen-WCL_SOP}.
	These characteristics position RISs as a key enabler for facilitating PLS in an economical and energy-efficient manner.
	
	Most existing researches focus on single-functional RISs (SF-RISs) whose only function is to reflect the incident signal\cite{RIS-review-3,JSAC-SRE,Wen-WCL_SOP}.
	To receive the signals from SF-RISs, receivers have to be located at the same side as the transmitters.
	Obviously, this geographical constraint, i.e., half-space SRE coverage, gravely restricts the flexibility and effectiveness of RISs, and even incurs serious performance loss\cite{STAR-RIS-TWC,STAR-CL,STAR-Liu,STAR-IOS}.
	For example, we consider a challenging scenario with multiple Bobs and Eves distributed at both sides of the SF-RIS.
	In this case, there are always some Bobs and Eves in the communication dead zone that are not supported by the SF-RIS.
	This leads to undesirable security threat and energy loss. 
	To address these issues, we pin our hopes on finding a dual-functional RIS (DF-RIS) to achieve full-space secure communications.
	In a nutshell, DF-RISs refer to the reconfigurable dual-functional surfaces that can conduct both reflection and transmission manipulation for incident signals.
	For example, an upgraded version of SF-RISs, termed simultaneously transmitting and reflecting RISs (STAR-RISs) has been proposed in \cite{STAR-RIS-TWC,STAR-CL,STAR-Liu,STAR-IOS}.
	In addition to inheriting the beneficial features of SF-RISs, DF-RISs have the following unique advantages:
	1) Given their simultaneous control of transmitted and reflected signals, DF-RISs are able to provide a full-space SRE\cite{STAR-CL};
	2) By offering more degrees-of-freedoms (DoFs) for signal manipulation, DF-RISs allow higher design flexibility\cite{STAR-Liu}.
	Moreover, the authors of \cite{IOS-TVT} proposed another kind of DF-RISs, called intelligent omni-surfaces (IOSs), which resemble STAR-RISs but employ the identical phase shifts for transmission and reflection.
	Intuitively, a question is raised: Can DF-RISs achieve more secure data transmission and higher energy efficiency than conventional SF-RISs?
	
	In the application of RISs assisted communications, base stations (BSs) and RISs usually need to simultaneously provide services for a massive number of users with stringent communication requirements, especially for future 6G networks\cite{NOMA-Mag}.
	With its advantages in enhancing spectrum efficiency and supporting massive connectivity, non-orthogonal multiple access (NOMA) has been regarded as a promising technology to address these challenges\cite{NOMA-WCNC,Ding-NOMA}.
	Distinctively different from conventional orthogonal multiple access (OMA) schemes, NOMA can serve multiple users within the same resource block\cite{NOMA_RIS_OMA}.
	Specifically, NOMA exploits the differences in the channel gain among the users for multiplexing and relies on successive interference cancellation (SIC) to decode the superimposed data stream.
	Inspired by these advantages, the employment of NOMA in RIS-enhanced networks is highly attractive and conceived to be a win-win strategy\cite{NOMA_RIS_Liu}.
	On the one hand, by allowing multiple users to share the same resources, NOMA constitutes an efficient access strategy for multi-user RIS aided networks.
	On the other hand, as a channel changing technique, RISs facilitate the implementation of NOMA by tuning the direction of users' channel vectors.
	This enables a smart NOMA operation to be carried out\cite{NOMA_RIS_UAV}.
	At this point, another question arises: Can the intrinsic integration of NOMA and DF-RISs stimulate their individual merits to achieve more reliable and energy-saving communications?
	
	
	\vspace{-2mm}
	\subsection{Related Works}
	\emph{1) Studies on SF-RIS Assisted NOMA Networks:}
	Enlightened by the benefits of integrating RISs into NOMA networks, their combination has been intensively investigated\cite{NOMA-Mag,NOMA-WCNC,Ding-NOMA,NOMA_RIS_OMA,Ni_NOMA_RIS_TWC,Mu_NOMA_RIS_TWC,SEE_TVT,NOMA-EE-deployment}.
	Initial investigations on SF-RIS assisted NOMA transmission were provided in \cite{NOMA-Mag,NOMA-WCNC,Ding-NOMA,NOMA_RIS_OMA}.
	Specifically, the authors of \cite{NOMA-Mag} performed a comprehensive discussion of futuristic use case scenarios for SF-RIS assisted NOMA networks and identified the main challenges.
	Moreover, the technical works in \cite{NOMA-WCNC,Ding-NOMA,NOMA_RIS_OMA} demonstrated that SF-RISs can achieve more flexible performance trade-off among the
	users by permuting the user decoding order of NOMA.
	In particular, the authors of \cite{NOMA_RIS_OMA} pursued a theoretical performance comparison between NOMA and OMA schemes in the SF-RIS assisted downlink communication networks.
	The research was further extended to multi-cell NOMA networks \cite{Ni_NOMA_RIS_TWC}, where user association, subchannel assignment, power allocation, phase shifts, and decoding order were jointly designed to maximize the sum rate.
	Under various RIS elements assumptions, the authors of \cite{Mu_NOMA_RIS_TWC} proposed different efficient beamforming designs to maximize the achievable sum rate of SF-RIS aided networks.
	Different from the sum-rate	maximization\cite{Ni_NOMA_RIS_TWC,Mu_NOMA_RIS_TWC} and total transmit power minimization\cite{NOMA_RIS_OMA}, energy-efficient algorithms were developed in \cite{SEE_TVT} and \cite{NOMA-EE-deployment}, which yielded a tradeoff between the sum-rate maximization and the total power consumption minimization.
	
	\emph{2) Studies on SF-RIS Assisted PLS:}
	The existing research on leveraging SF-RIS to improve the PLS started from a simple network	model with only one Bob and one Eve\cite{RIS-secure-GC,RIS-secure-CL,RIS-secure-WCL}.
	In this network, the active transmit beamforming at the BS and the passive reflect beamforming at the RIS were jointly optimized to maximize the secrecy rate.
	Based  on the above works, the studies in \cite{NOMA_RIS_secure_EL,NOMA_RIS_secure_robust,NOMA_RIS_secure_AN,NOMA_RIS_secure_CL,Eve-SIC} considered the use of RIS to enhance the PLS of NOMA networks.
	Specifically, by deriving the secrecy outage probability (SOP), the authors of \cite{NOMA_RIS_secure_EL} revealed that SF-RISs can help guarantee the secure NOMA transmission.
	To proceed, the work in \cite{NOMA_RIS_secure_CL} extended the SOP analysis to a more practical scenario where both the direct link and reflected links exist.
	In \cite{NOMA_RIS_secure_AN}, artificial noise was introduced to reduce information leakage to Eves while minimizing the effect on the reception quality of Bobs.
	Additionally, considering imperfect eavesdropping channel state information (CSI), the authors of \cite{NOMA_RIS_secure_robust} rigorously exploited a robust beamforming scheme in a SF-RIS assisted NOMA network.
	Recently, the authors of \cite{Eve-SIC} investigated the case with distributed SF-RISs.
	Simulation results showed that for NOMA network assisted by multiple SF-RISs, the maximum	secrecy rate is achieved when the SF-RISs share equal number of elements.

	\emph{2) Studies on DF-RIS Assisted Networks:}
	This topic has attracted an	increasing amount of research attention in wireless communications\cite{CL_IOS_secure,STAR-IOS,STAR-Liu,STAR-RIS-NOMA-CL,STAR-RIS-NOMA-Hou,STAR-RIS-NOMA-Zuo,STAR-RIS-NOMA-uplink,STAR-RIS-TWC,IOS-Zhang}.
	Particularly, the feasibility of implementing DF-RISs was discussed in \cite{STAR-IOS}, through pointing out several possible implementation options and providing solid theoretical evidence.
	The authors of \cite{STAR-Liu} introduced a basic signal model of STAR-RISs and presented three practical operating protocols, namely energy splitting (ES), mode switching (MS), and time switching (TS).
	A joint active-passive beamforming optimization problem was considered in \cite{STAR-RIS-TWC} for the STAR-RIS assisted downlink networks, where the unicast and multicast transmission cases and three operation protocols were discussed. 
	Furthermore, the potential of STAR-RISs in improving security was exploited in \cite{CL_IOS_secure}.
	In addition, the working principle and hybrid beamforming scheme for IOSs were proposed in \cite{IOS-Zhang}.
	It is verified that the proposed scheme can effectively control the angles of the reflected and transmitted beams.
	When it comes to the adoption of DF-RISs in NOMA systems, the authors of \cite{STAR-RIS-NOMA-CL} studied the fundamental coverage range of STAR-RIS aided two-user networks for both NOMA and OMA.
	The resource allocation at the BS and the transmission and reflection coefficients at the STAR-RIS were jointly optimized in \cite{STAR-RIS-NOMA-CL} to satisfy the communication requirements of users.
	Moreover, the authors of \cite{STAR-RIS-NOMA-Zuo,STAR-RIS-NOMA-Hou,STAR-RIS-NOMA-uplink} investigated the key benefits of STAR-RIS assisted NOMA systems with respect to the sum rate maximization, interference mitigation, and power consumption minimization, respectively.		
	
	\begin{table}[t]\small
		\centering
		\renewcommand{\arraystretch}{1.2}
		\caption{Comparison of this work with other representative works}
		\vspace{-2mm}
		\label{Comparison}
		\scalebox{0.88}{
			\begin{tabular}{|c|c|c|c|c|c|c|}
				\hline  
				\multirow{2}*{\textbf{Ref.}}  & \multicolumn{2}{c|}{\textbf{Scenarios}} & \multicolumn{2}{c|}{\textbf{Performance}} &{\textbf{Imperfect}}  \\
				\cline{2-5}
				& DF-RIS & NOMA& Secrecy & Energy & \textbf{CSI}\\
				\hline
				\cite{NOMA_RIS_OMA},\cite{SEE_TVT}&  & $\surd$ & &   $\surd$&  \\
				\hline
				\cite{RIS-secure-GC,RIS-secure-CL,RIS-secure-WCL}&  & & $\surd$ &  &  \\
				\hline 
				\cite{NOMA_RIS_secure_robust}&   &  $\surd$  &  $\surd$  &  $\surd$  &  $\surd$  \\
				\hline
				\cite{NOMA_RIS_secure_AN},\cite{Eve-SIC}&  & $\surd$  & $\surd$ & & $\surd$   \\
				\hline
				\cite{CL_IOS_secure} & $\surd$  &  & $\surd$  & &   \\
				\hline
				\cite{STAR-RIS-NOMA-CL,STAR-RIS-NOMA-Zuo,STAR-RIS-NOMA-Hou} &  $\surd$ &  $\surd$  &  &  & \\
				\hline
				\cite{STAR-RIS-NOMA-uplink} &  $\surd$ &  $\surd$  &  & $\surd$  & \\
				\hline
				This work & $\surd$  & $\surd$ & $\surd$  & $\surd$  &  $\surd$ \\
				\hline
		\end{tabular}}
		\vspace{-1mm}
	\end{table}
	
	\vspace{-2mm}
	\subsection{Motivations and Contributions}
	Table \ref{Comparison} illustrates the comparison of this work with other representative works. 
	As shown in Table \ref{Comparison}, although few recent efforts have been devoted to DF-RISs assisted NOMA networks\cite{STAR-RIS-NOMA-CL,STAR-RIS-NOMA-Zuo,STAR-RIS-NOMA-Hou,STAR-RIS-NOMA-uplink}, the tradeoff between the secrecy and energy consumption remains an open issue.
	The existing work in \cite{CL_IOS_secure} just considered a simplified case with one Bob on each side of the STAR-RIS.
	Besides, most of the literature assumed that the perfect CSI of all channels is available to the BS.
	In practice, since DF-RISs are passive and can neither send nor receive pilot symbols, perfect CSI is challenging to obtain.
	Simply treating the estimated channels as perfect ones will inevitably lead to system performance loss.
	Therefore, these overly optimistic assumptions greatly weaken the generality and practicality of the considered system models and the proposed algorithms in \cite{CL_IOS_secure} and \cite{STAR-RIS-NOMA-CL,STAR-RIS-NOMA-Zuo,STAR-RIS-NOMA-Hou,STAR-RIS-NOMA-uplink}.
	
	Thus, we propose a DF-RIS assisted NOMA network under imperfect CSI and explore a more robust beamforming design for secure green communications.
	The main novelty of this paper is that, to the best of our knowledge, this is the first work to exploit the potential of STAR-RISs in improving secrecy energy efficiency (SEE), and also the first work to design the robust transmission strategy for DF-RIS assisted NOMA networks.
	In summary, we face the following challenges:
	1) The newly introduced transmission and reflection coefficients at the DF-RIS make the coupling between the optimization variables complicated;
	2) The intra- and inter-cluster interference leads to the problem of interference management;
	3) The infinitely many inequality constraints caused by imperfect CSI increase the difficulty of the robust beamforming design.
	Hence, the contributions of this paper can be summarized as follows:
	\begin{itemize}
		\item	
		We first propose a DF-RIS aided secure communication network.
		To be specific, we consider a STAR-RIS assisted NOMA network to provide full-space transmission security.
		The STAR-RIS is deployed to enhance the desired transmission while mitigating inter-user interference and restraining full-space eavesdropping.
		From a practical point of view, all Bobs are randomly distributed in the whole space around the STAR-RIS in the presence of potential Eves, while all channel information is assumed to be imperfect.
		To maximize the SEE for the ES protocol, we jointly optimize the power allocation coefficients and the active-passive beamforming schemes.
		\item 
		These coupled optimization variables and imperfect CSI render the formulated problem non-convex and difficult to solve directly.
		To this end, we first leverage the sequential convex approximation (SCA) technique to transform the original problem into a tractable form.
		Then, the $\mathcal{S}$-procedure and general sign-definiteness are invoked to approximate the semi-infinite constraints and convert the infinite possibilities into finite ones.
		Eventually, using the alternating optimization (AO) framework, we adopt the SCA and a penalty convex-concave procedure (PCCP) to deal with this challenging problem.
		\item
		Furthermore, the formulated SEE maximization problem and the proposed algorithm are extended to the MS and TS protocols.
		For MS, the existence of the binary amplitude coefficients yields a mixed-integer non-convex optimization problem.
		Through transforming the binary constraint into equivalent forms and adding them into the objective function as penalties, the AO algorithm for ES solves this problem effectively.
		While for TS, a two-layer optimization method is developed, where the outer-layer iteration determines the time allocation using bisection search, and the inner-layer iteration updates the remaining variables using the previously designed algorithm.
		\item
		Simulation results are provided to verify the convergence and effectiveness of the proposed algorithms. 
		Particularly, the following observations can be made from the numerical results:
			1) \emph{Secrecy superiority}: 
			In terms of SEE performance, our proposed system exhibits significant DF gain and NOMA gain in comparison with conventional SF-RIS assisted NOMA systems and STAR-RIS assisted OMA systems, respectively;
			2) \emph{Protocol selection}: TS is suitable for low power domain, while ES achieves the best performance when downlink power is sufficient.
			Additionally, MS always performs worse than ES as it can be regarded as a special case of ES;
			3) \emph{System design}: For high CSI uncertainty (CSIU) or large bit resolution power consumption, the increment of STAR-RIS elements may hamper the improvement of SEE.
			This indicates that there exists an optimal deployment scale for STAR-RIS to maximize the SEE.
		%
	\end{itemize}
	
	The rest of this paper is organized as follows. 
	Section \ref{Preliminaries} provides a brief introduction to the signal models of SF-RISs and DF-RISs, and their applications to secure communications.
	Section \ref{System Model and Problem Formulation} elaborates the system model and problem formulation for a STAR-RIS aided secrecy NOMA network.
	Then, Section \ref{Solution} proposes an efficient algorithm to achieve a suboptimal solution.
	Next, Section \ref{Extension} extends our studies to the MS and TS cases. 
	Finally, numerical results are presented in Section \ref{Simulation Results}, and Section \ref{Conclusion} concludes this paper.
	
	\emph{Notations:}
	Column vectors and matrices are denoted by boldface lowercase and uppercase boldface letters, respectively.
		$\mathbb{C}^{M\times N}$ denotes the complex space with $M\times N$ dimensions, and $\mathbf{I}_N$ denotes an $N\times N$ identity matrix.
	$\mathbb{H}^{N}$ denotes the set of all $N$-dimensional complex Hermitian matrices.
	$\mathbf{X}^{\ast}$, $\mathbf{X}^{\rm T}$, $\mathbf{X}^{\rm H}$, $\lVert\mathbf{X} \lVert_F$, and ${\rm vec}(\mathbf{X})$ denote the conjugate, transpose, Hermitian, Frobenius norm and vectorization operator of matrix $\mathbf{X}$, respectively.
	The symbols $\lVert\mathbf{x} \lVert_1$ and $\lVert\mathbf{x} \lVert_2$ denote 1-norm and 2-norm of vector $\mathbf{x}$, respectively.
	${\rm Re}\{\cdot\}$ denotes the real part of a complex number.
	${\rm diag}(\mathbf{x})$ is a diagonal matrix with the entries of $\mathbf{x}$ on its main diagonal.
	$[\mathbf{x}]_m$ denotes the $m$-th element of vector $\mathbf{x}$.
	$\otimes$ denotes the Kronecker product and $\mathbf{X}\succeq \mathbf{0}$ indicates that matrix $\mathbf{X}$ is positive semi-definite.
	$\nabla f(x)$ denotes the gradient of the function $f(x)$.
	For two given sets $\mathcal{A}$ and $\mathcal{B}$, $\mathcal{A} / \mathcal{B} \triangleq\{x \mid x \in \mathcal{A}, x \notin \mathcal{B}\}$.
	
	\section{Preliminaries}\label{Preliminaries}
	In this section, SF-RISs, DF-RISs and their corresponding signal models are firstly reviewed.
	On this basis,  we propose a DF-RIS assisted secure communication network,
	while illustrating its advantages and disadvantages.
	
	\subsection{SF-RIS}
	As shown in Table \ref{three_surfaces}, according to the type of wireless signal manipulation, SF-RISs can be divided into two categories: reflecting-only RISs and transmitting-only RISs. 
	All elements of the former are operated in the reflection mode, while the latter are in the transmission mode.
	For a SF-RIS equipped with $M$ elements, let $s_m$ denote the signal incident on the $m$-th element, $\forall m\!\in\!\mathcal{M}\!=\!\{1,2,\ldots,M\}$.
	Then the signal reflected by the $m$-th element of reflecting-only RISs is modeled by
	$y_m^r\!=\!\sqrt{\beta_{m}^{r}}e^{j\theta_{m}^{r}}s_m$, $\forall m$, where $\beta_{m}^{r}\!\in\![0,1]$ and $\theta_{m}^{r}\!\in\![0,2 \pi)$ characterize the reflective amplitude and phase shift response of the $m$-th reflecting element, respectively.
	In contrast, for transmitting-only RISs, the signal transmitted by the $m$-th element is given by
	$y_m^t\!=\!\sqrt{\beta_{m}^{t}}e^{j\theta_{m}^{t}}s_m$, where $\beta_{m}^{t}\in[0,1]$ and $\theta_{m}^{t}\in[0,2 \pi)$ denote the transmissive (refractive) amplitude and phase shift of the $m$-th transmitting element, respectively.
	
	\begin{table*}[t]\small
		\caption{Summary of single-functional RISs and dual-functional RISs and their applications to secure communications} 
		\vspace{-0.2cm}
		\centering
		\includegraphics[width=7in,height=2.4in]{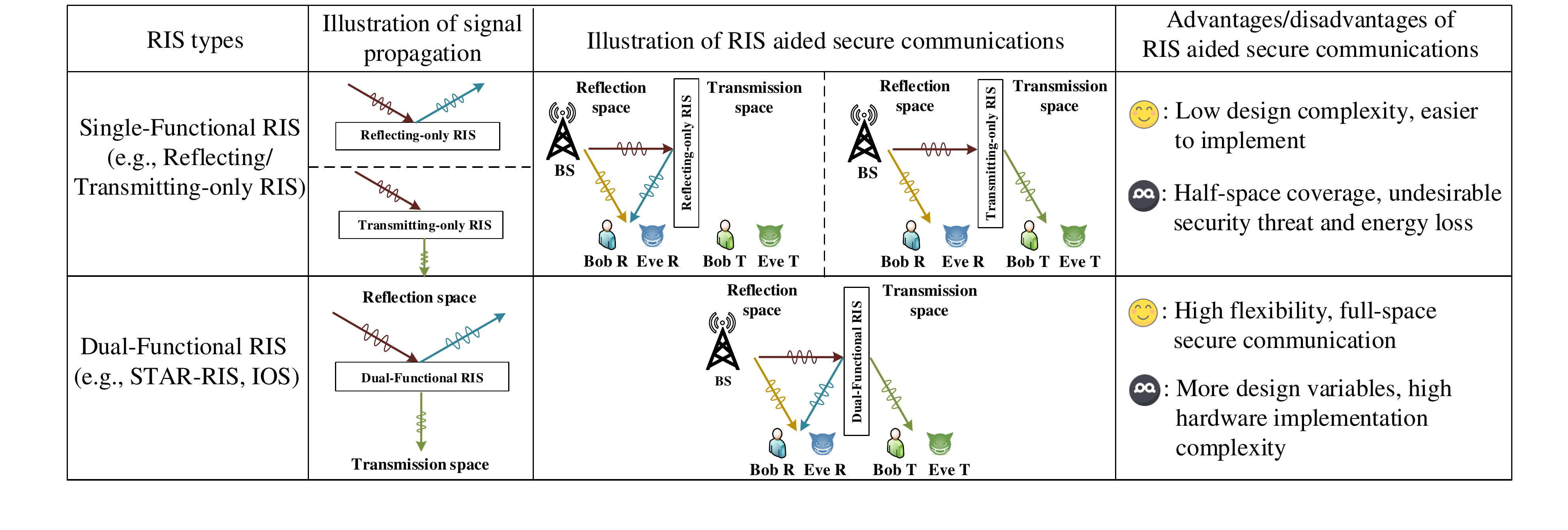}
		\label{three_surfaces}
		\vspace{-5mm}
	\end{table*}
	\subsection{DF-RIS}
	As illustrated in Table \ref{three_surfaces}, different from SF-RISs, the wireless signals impinging on a DF-RIS element are divided into two parts.
	Part of signals are transmitted into the transmission space ($k\!=\!t$), while the remaining part are  reflected into the reflection space ($k\!=\!r$), where $k\!\in\!\mathcal{K}\!=\!\{r,t\}$, with $\mathcal{K}$ denoting the set of spaces.
	This DF feature means that the $m$-th element of DF-RISs has both reflection $\{\beta_m^r, \theta_m^r\}$ and transmission coefficients $\{\beta_m^t, \theta_m^t\}$\footnote{
	To characterize the theoretical performance upper bound, we assume that the transmission and reflection phase shifts can be adjusted independently.
	The realistic coupled phase shift model has been studied in the literature \cite{STAR-Coupled}, and we will extend our results to this model in future research.
	}.
	After being reconfigured by these coefficients, the transmitted ($k\!=\!t$) or reflected ($k\!=\!r$) signal from the $m$-th element is given by $y_m^k\!=\!\sqrt{\beta_{m}^{k}}e^{j\theta_{m}^{k}}s_m$, $\forall k,m$.
	Due to the law of energy conservation, the transmitting-reflecting amplitude coefficients are coupled with each other, i.e., $\beta_m^r+\beta_m^t=1$, $\forall m$\footnote{
	To investigate the maximum performance gain of DF-RISs, we assume that the DF-RIS does not introduce performance loss.
	It is worth mentioning that our proposed solutions are also applicable to the case with performance loss, i.e., $\beta_{m}^r+\beta_{m}^t=c, \forall m$, where $0< c< 1$.}.
	
	In terms of the hardware implementations, we can loosely classify DF-RISs into two categories: metasurface based and patch-array based DF-RISs.
	As a popular representative of the former, STAR-RISs achieve independent phase shift controls with the aid of surface magnetic currents, thus providing more DoFs for system design.
	Nevertheless, this implementation has ultra-small periodic cells, requiring higher manipulation accuracy.
	Generally, the practical protocols for operating STAR-RISs include the following three types\cite{STAR-Liu}:
	
	\subsubsection{Energy Splitting}
	For ES, all elements are operated in the simultaneous transmission and reflection mode.
	Since the total radiation energy is split into two parts, ES can be presented as $\beta_{m}^{k}\!\in\![0,1]$, $\forall k,m$, and $\beta_{m}^{t}\!+\!\beta_{m}^r\!=\!1$, $\forall m$.
	
	\subsubsection{Mode Switching} 
	For MS, the elements are partitioned into two groups.
	One group is operated in the transmission mode, while the other group is exploited to be operated in the reflection mode.
	In this case, $\{\beta_{m}^k\}$ are restricted to binary values, i.e., $\beta_{m}^{k}\!\in\!\{0,1\}$, $\forall k,m$, and $\beta_{m}^{t}\!+\!\beta_{m}^r\!=\!1$, $\forall m$.
	
	\subsubsection{Time Switching}
	For TS, all elements are operated in the transmission mode or the reflection mode during the orthogonal time slots.
	Let $\{\tau_k\}$ be the percentages of time allocated to the two periods, where $\tau_k\!\in\![0,1]$, $\sum_{k} \tau_{k}\!=\!1$.
	Then, for time slots with the transmission period, we have $\beta_{m}^t\!=\!1$ and $\beta_{m}^r\!=\!0$, otherwise $\beta_{m}^t\!=\!0$ and $\beta_{m}^r\!=\!1$, $\forall m$.
	
	Different from STAR-RISs, patch-array based IOSs consist of periodic cells about a few centimeters in size.
	Their relatively large sizes allow the patches to be tuned using positive intrinsic negative (PIN) diodes\cite{STAR-IOS}.
	Since the PIN diodes have only two states,  ``ON" or ``OFF", IOSs are easier to implement, but may incur a certain performance loss.
	Specifically, this PIN diodes implementation makes each IOS element only support a finite-cardinality reflection and transmission coefficient set.
	Besides, for a given state of all PIN diodes, the signals transmitted and reflected by an element are adjusted via a common phase shift, i.e., $\theta_{m}^t\!=\!\theta_{m}^r$, $\forall m$.
	
	\subsection{DF-RIS Aided Secure Communication}
	The prototypes of SF-RIS and DF-RIS aided secure communication networks are also presented in Table \ref{three_surfaces}.
	Through comparison, we find that deploying DF-RISs has the following advantages:
	1) DF-RIS is capable of forwarding the received signals to Bobs in either a reflective or a transmissive manner, thus facilitating the full-space signal enhancement.
		2) By properly configuring the reflection and transmission coefficients, DF-RIS can effectively deteriorate the reception of Eves on both sides.
	As a result, transmission security can be improved in the whole space, regardless of the locations of Bobs and Eves.
	However, the newly introduced optimization variables and complex operating mechanisms lead to higher complexity of software design and hardware implementation for DF-RISs.
	
	In the rest parts of this paper, we take STAR-RIS as a typical example of DF-RISs to develop robust beamforming schemes in practical secure communication networks.
	It is worth mentioning that the obtained results of this work can be easily extended to the IOS-based scenarios by considering the identical phase shift and coupled amplitude response\footnote{For the phase shifts under IOS-based scenarios, we have $\theta_{m}^t=\theta_{m}^r,\forall m$. 
	However, the reflective and transmissive amplitude responses are different since they are related to the direction of the re-emitted signals.
	In addition, these amplitude coefficients are coupled due to the law of energy conservation.
	More details can be found in our previous work\cite{Wen-IOS-TVT}.}.
	%
	\begin{figure}[t]
		\setlength{\abovecaptionskip}{-0.1cm} 
		\centering
		\includegraphics[width=3.6in]{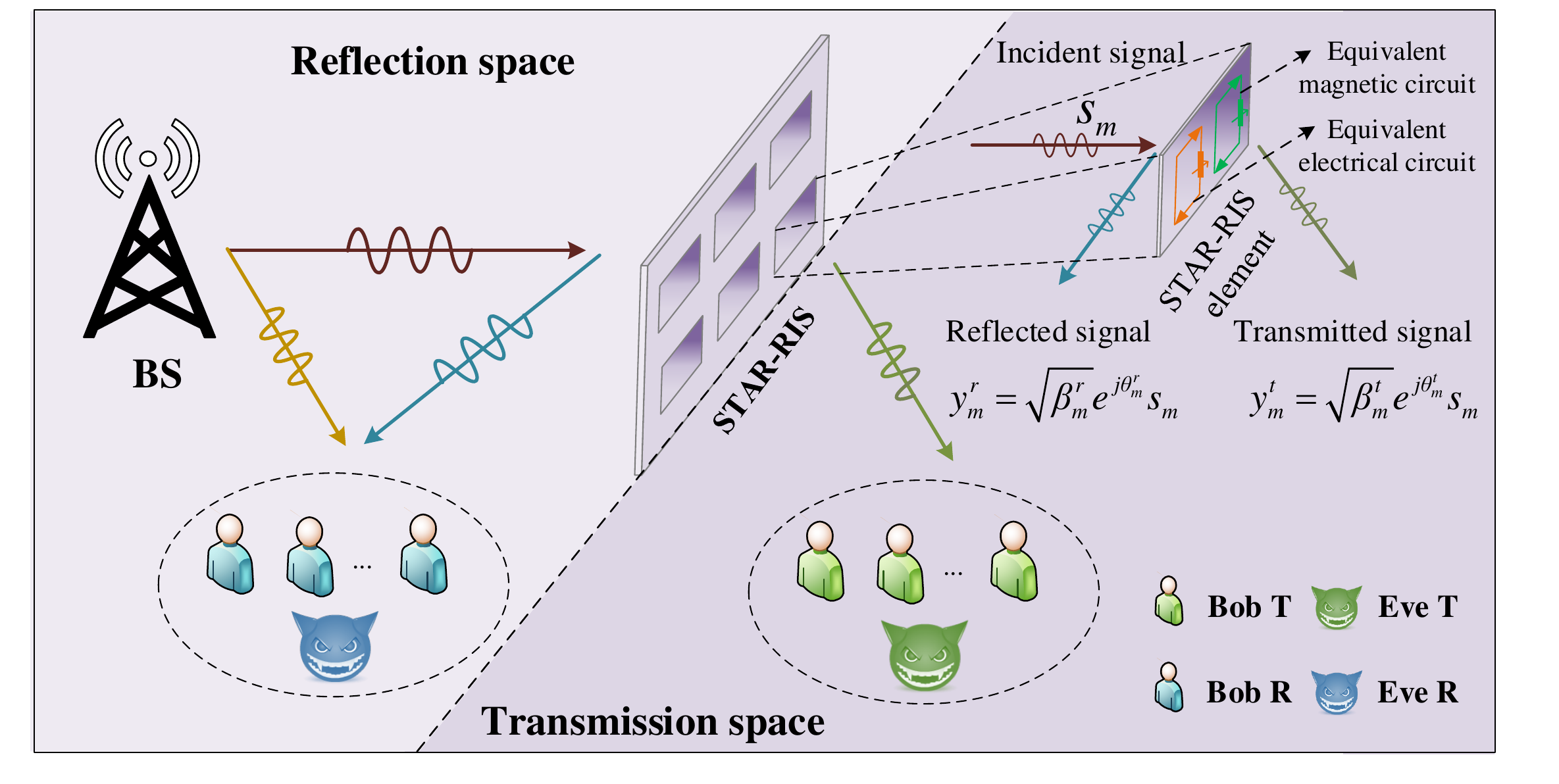}
		\caption{System model of a STAR-RIS assisted secrecy NOMA network.}
		\label{system_model}
	\end{figure}
	\section{System Model and Problem Formulation}\label{System Model and Problem Formulation}
	\subsection{System Model}
	As shown in Fig. \ref{system_model}, we consider a STAR-RIS assisted downlink NOMA network, where the BS sends information to $J$ Bobs in the presence of two Eves.
	The sets of Bobs and Eves are denoted by $\mathcal{J}=\{1,2,\ldots,J\}$ and $\mathcal{E}=\{E_r,E_t\}$, respectively.
	The set of Bobs in space $k$ is denoted by $\mathcal{J}_k=\{1,2,\ldots,J_k\}$, where $J_k+J_{\bar{k}}=J$, $\forall k,\bar{k}\in\mathcal{K}, k\neq \bar{k}$.
	The STAR-RIS composes of $M$ elements and the BS is equipped with $N$ antennas, while Bobs and Eves are all single-antenna.
	
	Let $\mathbf{u}_{k}\!=\!\big[\sqrt{\beta_{1}^{k}}e^{j \theta_{1}^{k}}\!,\sqrt{\beta_{2}^{k}}e^{j \theta_{2}^{k}}\!,\ldots,\sqrt{\beta_{M}^{k}}e^{j \theta_{M}^{k}}\big]^{\mathrm T}\!\in\!\mathbb{C}^{M\times1}$, $\forall k, m$, be the transmission or reflection beamforming vector.
	Then the diagonal passive beamforming matrix is given by $\boldsymbol{\Theta}_{k}\!=\!{\rm diag}(\mathbf{u}_{k})$.
	If Bob or Eve is located at the reflection space, we have $\mathbf{u}_{k}\!=\!\mathbf{u}_{r}$; otherwise $\mathbf{u}_{k}\!=\!\mathbf{u}_{t}$.
	Following the ES protocol, the set of constraints to the transmission and reflection coefficients is denoted by
	\setlength{\abovedisplayskip}{3pt}
	\setlength{\belowdisplayskip}{3pt}
	\begin{eqnarray}
		\label{C_ES_IOS}
		\mathbb{R}_{\beta, \theta}\!=\!\big\{\beta_{m}^{k}, \theta_{m}^{k} | \sum \nolimits_{k}\beta_{m}^{k}\!=\!1;\beta_{m}^{k}\!\in\![0,1]; \theta_{m}^{k}\!\in\![0,2 \pi)\big\}.
	\end{eqnarray}
	
	For notation simplicity, we index the $j$-th Bob in space $k$ by $U_{k,j}$.
	Let $\alpha_{k,j}$ and $s_{k,j}$ be the power allocation factor and the desired signal of $U_{k,j}$, where $\mathbb{E}[|s_{k,j}|^{2}]=1$.
	Accordingly, the received signals at $U_{k,j}$ and Eve $e$ are given by
	\begin{subequations}
		\setlength{\abovedisplayskip}{3pt}
		\setlength{\belowdisplayskip}{3pt}
		\begin{eqnarray}
			\label{received signal_user}
			\nonumber
			\!\!\!\!\!y_{k,j}&{}&\!\!\!\!\!\!\!\!\!\!=\underbrace{\bar{\mathbf{h}}_{k,j}\mathbf{f}_k\alpha_{k,j}s_{k,j}}_{\rm desired \ signal}
			+\underbrace{\bar{\mathbf{h}}_{k,j}\mathbf{f}_k\sum\nolimits_{i\in\{\mathcal{J}_{k}/j\}}\alpha_{k,i}s_{k,i}}_{\rm intra-cluster \ interference}\\
			\!\!\!\!\!&{}&\!\!\!\!\!\!\!\!\!\!+\underbrace{\bar{\mathbf{h}}_{k,j}\mathbf{f}_{\bar{k}}\sum\nolimits_{i\in\mathcal{J}_{\bar{k}} }\alpha_{\bar{k},i}s_{\bar{k},i}}_{\rm inter-cluster \ interference}+\underbrace{n_{k,j}}_{\rm noise},  ~\forall k,\forall j\!\in\!\mathcal{J}_k,\\
			\label{received signal_Eve}
			\!\!\!\!\!y_{e}&{}&\!\!\!\!\!\!\!\!\!\!\!=\bar{\mathbf{h}}_{e}\sum\nolimits_{k}\mathbf{f}_k\sum\nolimits_{j}\alpha_{k,j}s_{k,j}+n_{e}, ~\forall e\in\mathcal{E},
		\end{eqnarray}
	\end{subequations}
	where $\mathbf{f}_k$ is the active beamforming vector for $U_{k,j}$, $n_{k,j} \sim \mathcal{C} \mathcal{N}(0, \sigma^{2})$ and $n_{e} \sim \mathcal{C} \mathcal{N}(0, \sigma^{2})$ denote the noises  at $U_{k,j}$ and Eve $e$ with zero mean and variance $\sigma^2$, respectively.
	In addition, $\bar{\mathbf{h}}_{k,j}$ and $\bar{\mathbf{h}}_{e}$ denote the combined channel vectors from the BS to $U_{k,j}$ and Eve $e$, given by
	\begin{subequations}
		\begin{eqnarray}
			\label{channel gain}
			\bar{\mathbf{h}}_{k,j}\!\!\!\!&=\!\!\!\!&\mathbf{h}_{k,j}^{\mathrm H}+\mathbf{g}_{k,j}^{\mathrm H}\boldsymbol{\Theta}_{k}\mathbf{H}_{b}, ~\forall k, \forall j\in \mathcal{J}_k,\\ \bar{\mathbf{h}}_{e}\!\!\!\!&=\!\!\!\!&\mathbf{h}_{e}^{\mathrm H}+\mathbf{g}_{e}^{\mathrm H}\boldsymbol{\Theta}_{k}\mathbf{H}_{b}, ~\forall e, 
		\end{eqnarray}
	\end{subequations}
	where $\mathbf{h}_{k,j}^{\mathrm H}\in\mathbb{C}^{1\times N}$, $\mathbf{g}_{k,j}^{\mathrm H}\in\mathbb{C}^{1\times M}$, $\mathbf{H}_{b}\in\mathbb{C}^{M\times N}$, $\mathbf{h}_{e}^{\mathrm H}\in\mathbb{C}^{1\times N}$ and $\mathbf{g}_{e}^{\mathrm H}\in\mathbb{C}^{1\times M}$ represent the channel vectors from the BS to $U_{k,j}$, from the STAR-RIS to $U_{k,j}$, from the BS to STAR-RIS, from the BS to Eve $e$, and from the STAR-RIS to Eve $e$, respectively.
	
	Without loss of generality, we assume that the $J_k$ Bobs in space $k$ are indexed in the descending order as follows\cite{TVT-NOMA-decoding}\footnote{Determining the optimal SIC decoding order is an NP-hard problem, which relies on exhaustive search, branch and bound methods or heuristic approaches to solve\cite{EE_SCA}. For simplicity, we assume that the decoding order is given and all optimization variables need to be designed with satisfying decoding order constraint (\ref{descending order})\cite{TVT-NOMA-decoding}.}.
	\begin{eqnarray}
		\label{descending order}
		| \bar{\mathbf{h}}_{k,j}\mathbf{w}_{k,j} |^2\geq | \bar{\mathbf{h}}_{k,j+1}\mathbf{w}_{k,j+1} |^2,  ~\forall k,\forall j\!\in\!\mathcal{J}_k^{'}\!=\!\{\mathcal{J}_k/J_k\},
	\end{eqnarray}
	where $\mathbf{w}_{k,j}\!=\!\alpha_{k,j}\mathbf{f}_{k}$.
	As per the principle of NOMA, Bobs employ the SIC to eliminate the intra-cluster interference.
	By defining $\mathbf{w}_{k,-j}\!=\![\alpha_{k,1}\mathbf{f}_k, \alpha_{k,2}\mathbf{f}_k,\ldots,\alpha_{k,j-1}\mathbf{f}_k, \mathbf{f}_{\bar{k}}]$,
	the achievable data rate of decoding $s_{k,j}$ at $U_{k,l}$ is given by
	\setlength{\abovedisplayskip}{2pt}
	\setlength{\belowdisplayskip}{2pt}
	\begin{eqnarray}
		\label{SIC-User}
		R_{k,j}^{l}\!=\!\log_2\left(1\!+\!\frac{|\bar{\mathbf{h}}_{k,l}\mathbf{w}_{k,j}|^2}
		{\lVert \bar{\mathbf{h}}_{k,l} \mathbf{w}_{k,-j}\lVert_2^2+ \sigma^2}\right),\forall k, \forall j\!\in\!\mathcal{J}_k, \forall l\!\in\!\mathcal{L}_k,\!\!\!\!\!
	\end{eqnarray}
	where $\mathcal{L}_k=\{l|l\leq j, \forall l\in\mathcal{J}_k, \forall j\in\mathcal{J}_k, \forall k\}$.
	Moreover, to guarantee that SIC performs successfully, the achievable date rate of $U_{k,j}$
	should satisfy\cite{RIS-TWC-Rate}
	\begin{eqnarray}
		\label{R-User}
		R_{k,j}={\rm min}\big\{R_{k,j}^{l}|\forall l\in\mathcal{L}_k\big\}, ~\forall k, \forall j\in\mathcal{J}_k.
	\end{eqnarray}
	
	As for Eves, we adopt a worst-case assumption in PLS.
	First, all Bobs are their intended users.
	Besides, Eves perfectly know the decoding order of Bobs and the precoding vector information, so that they can carry out SIC decoding technique to detect the target signals the same as Bobs\cite{Eve-SIC}.
	Therefore, the eavesdropping rate of Eve $e$ to decode $s_{k,j}$ is given by
	\begin{eqnarray}
		\label{R_E}
		R_{k,j}^{e}=\log_2\left(1+\frac{|\bar{\mathbf{h}}_{e} \mathbf{w}_{k,j}|^2}	{\lVert \bar{\mathbf{h}}_{e} \mathbf{w}_{k,-j}\lVert_2^2 + \sigma^2}  \right),
		~\forall k,e,\forall j\!\in\!\mathcal{J}_k.
	\end{eqnarray}
	Combining (\ref{R-User}) with (\ref{R_E}), the achievable secrecy rate of $U_{k,j}$ is expressed as $R_{k,j}^{s}\!=\!\big[R_{k,j}\!-\!\sum_e R_{k,j}^{e}\big]^{+}$, where $[x]^+ \!=\! \max \{x,0\}$.
	On this basis, the sum secrecy rate (SSR) of the system is given by $R_s=\sum_k\sum_{j}R_{k,j}^{s}$, $\forall k,\forall j\!\in\!\mathcal{J}_k$.
	
	The considered system is composed of the direct links $\mathbf{h}_{k,j}$ and $\mathbf{h}_{e}$, as well as the cascaded links $\mathbf{G}_{k,j}\!=\!{\rm diag}(\mathbf{g}_{k,j}^{\mathrm H})\mathbf{H}_b$ and $\mathbf{G}_{e}\!=\!{\rm diag}(\mathbf{g}_{e}^{\mathrm H})\mathbf{H}_b$.
	Unfortunately, various factors such as channel estimation and quantization errors would lead to outdated and coarse CSI\cite{TSP-A framework}\footnote{For TS, the CSI of users in the transmission and reflection spaces can be consecutively estimated using the mature low-complexity channel estimation methods\cite{CSI-1,CSI-2,CSI-3}.
	However, the channel estimation design for STAR-RIS under the ES and MS protocols needs further research, which will be left for our future work.}.
	To describe this effect, we adopt the bounded CSI model to characterize the uncertainties of CSI, given by\cite{NOMA_RIS_secure_robust}
	\begin{subequations}
		\begin{eqnarray}
			\label{imperfect CSI_user}
			&\!\!\!\!\!\!\!\!\!\!\!\mathbf{h}_{k,j}\!\!=\!\hat{\mathbf{h}}_{k,j}\!+\!\triangle\mathbf{h}_{k,j},  \mathbf{G}_{k,j}\!=\!\hat{\mathbf{G}}_{k,j}\!+\!\triangle\mathbf{G}_{k,j},~\forall k,\!\forall j\!\in\!\mathcal{J}_k,\\
			&\!\!\!\!\!\!\!\!\Omega_{k,j}\!\!=\!\{ \lVert \triangle\mathbf{h}_{k,j} \lVert_2\leq \xi_{k,j},  \lVert\triangle\mathbf{G}_{k,j} \lVert_F\leq \zeta_{k,j} \!\},~\forall k,\!\forall j\!\in\!\mathcal{J}_k,\\
			\label{imperfect CSI_Eve}
			&\!\!\!\!\!\!\!\!\!\!\!\!\!\!\!\!\!\!\!\!\!\!\!\!\!\!\!\!\!\!\!\!\!\!\!\!\!\!\!\!\!\!\!\!\!\!\!\!\!\mathbf{h}_{e}\!=\!\hat{\mathbf{h}}_{e}\!+\!\triangle\mathbf{h}_{e}, \mathbf{G}_{e}\!=\!\hat{\mathbf{G}}_{e}\!+\!\triangle\mathbf{G}_{e}, ~\forall e,\\
			&\!\!\!\!\!\!\!\!\!\!\!\!\!\!\!\!\!\!\!\!\!\!\!\!\!\!\!\!\!\!\!\!\!\!\!\!\!\!\!\!\!\Omega_{e}\!=\!\{ \lVert \triangle\mathbf{h}_{e} \lVert_2\leq \xi_{e}, \lVert \triangle\mathbf{G}_{e} \lVert_F\leq \zeta_{e}  \} ,~\forall e,
		\end{eqnarray}
	\end{subequations}
	where $\hat{\mathbf{h}}_{k,j}$ and $\hat{\mathbf{G}}_{k,j}$ denote the estimations of the corresponding channels $\mathbf{h}_{k,j}$ and $\mathbf{G}_{k,j}$, respectively.
	$\triangle\mathbf{h}_{k,j}$ and $\triangle\mathbf{G}_{k,j}$ represent the channel estimation errors of $\mathbf{h}_{k,j}$ and $\mathbf{G}_{k,j}$, respectively.
	Moreover, the continuous set $\Omega_{k,j}$ collects all possible channel estimation errors, with $\xi_{k,j}$ and $\zeta_{k,j}$ denoting the corresponding radii of the uncertainty regions known at the BS.
	In addition, $\hat{\mathbf{h}}_{e}$, $\hat{\mathbf{G}}_{e}$, $\triangle\mathbf{h}_{e}$ and $\triangle\mathbf{G}_{e}$ are defined similarly.
	
	\subsection{Problem Formulation}
	Our objective is to maximize the SEE while satisfying the secrecy and power constraints.
	Based on the secrecy capacity, the SEE is defined as the ratio of the achievable SSR over the total power consumption\cite{SEE_TVT}.
	Here, the total power dissipated to operate the considered system includes the BS transmit power and the hardware static power $P_0$, given by\cite{RIS-review-3}
	\begin{eqnarray}
		\label{total power}
		\mathcal{P}=\varrho \sum\nolimits_{k}\lVert \mathbf{f}_k\lVert_2^2+P_0,
	\end{eqnarray}
	where $\varrho $ is the power amplifier efficiency, and $P_0=P_{\rm B}+JP_{\rm U}+MP_r(b)$, with $P_{\rm B}$, $P_{\rm U}$ and $P_r(b)$ denoting the static power consumed by the BS, each Bob and each phase shifter having $b$-bit resolution, respectively.
	
	With these definitions in place, by jointly designing the power allocation coefficients $\boldsymbol{\alpha}\!=\!\{\alpha_{k,j}|\forall k,\forall j\!\in\!\mathcal{J}_k\}$, the active beamforming $\mathbf{F}\!=\!\{\mathbf{f}_{k}|\forall k\}$, the transmission and reflection coefficients $\boldsymbol{\Phi
	}\!=\!\{\boldsymbol{\Theta}_k|\forall k\}$, the SEE maximization problem is formulated as
	\begin{subequations}
		\label{P0}
		\begin{eqnarray}
			\label{P0_function}
			&\!\!\!\!\!\!\!\!\!\!\!\!\!\!\!\!\!\!\!\!\underset{\boldsymbol{\alpha},\mathbf{F},\boldsymbol{\Phi} }{\max}&\!\!\!\!\!\! \frac{\sum_k\sum_{j}\big(R_{k,j}-\sum_e R_{k,j}^{e}\big)}
			{ \varrho  \sum\nolimits_{k}\lVert \mathbf{f}_k\lVert_2^2+P_0 }\\
			\label{C_max_power}
			&\!\!\!\!\!\!\!\!\!\!\!\!\!\!\!\operatorname{s.t.}&\!\!\!\!\!\!
			\sum\nolimits_{k}\lVert \mathbf{f}_k\lVert_2^2  \leq P_{\rm max}, \\
			\label{C_power allocation}
			&&\!\!\!\!\!\!\sum\nolimits_{j}\alpha_{k,j}^2=1, ~\forall k, \forall j\!\in\!\mathcal{J}_k,\\
			\label{C_STAR-RIS}
			&&\!\!\!\!\!\!\beta_{m}^{k}, \theta_{m}^{k} \in \mathbb{R}_{\beta, \theta}, ~\forall k,  m, \\
			\label{C_User}
			&&\!\!\!\!\!\!R_{k,j}   \geq C_{k,j}, ~\Omega_{k,l},\forall k, \forall j\!\in\!\mathcal{J}_k, \forall l\!\in\!\mathcal{L}_k,\\
			\label{C_Eve_k}
			&&\!\!\!\!\!\!R_{k,j}^{e}   \leq C_{k,j}^{e}, ~\Omega_{e},\forall k, e,\forall j\!\in\!\mathcal{J}_k,\\
			\label{C_decoding_order}
			&&\!\!\!\!\!\!	| \bar{\mathbf{h}}_{k,j}\mathbf{w}_{k,j} |^2\!\geq\!  | \bar{\mathbf{h}}_{k,j+1}\mathbf{w}_{k,j+1} |^2, \Omega_{k,j}, \!\forall k,\! \forall j\!\in\!\mathcal{J}_k^{'},
		\end{eqnarray}
	\end{subequations}
	where $P_{\rm max}$ denotes the maximum transmit power of the BS.
	The power constraints (\ref{C_max_power}) and (\ref{C_power allocation}) represent the total transmission power constraint and power allocation restriction, respectively.
	Constraint (\ref{C_STAR-RIS}) specifies the range of the reflection and transmission coefficients.
	Moreover, constraint (\ref{C_User}) guarantees that the minimum rate requirement $C_{k,j}$ is satisfied at $U_{k,j}$, and $C_{k,j}^{e}$ in (\ref{C_Eve_k}) denotes the maximum tolerable information leakage to Eve $e$ for eavesdropping $s_{k,j}$.
	In particular, the secrecy constraints (\ref{C_User}) and (\ref{C_Eve_k}) ensure that the SSR is lower bounded by $R_s\!\geq\! \sum_k\sum_{j}\big[C_{k,j}\!-\!\sum_e C_{k,j}^{e}\big]^{+}$, $\forall k,e,\forall j\!\in\!\mathcal{J}_k$.
	Note that the operator $[\cdot]^{+}$ is omitted in the objective function (\ref{P0_function})  since we can set $C_{k,j}\geq \sum_eC_{k,j}^e$ such that $R_{k,j}$ is always greater or equal to $\sum_e R_{k,j}^e$.
	Additionally, constraint (\ref{C_decoding_order}) determines the SIC decoding order.
	
	We notice that problem (\ref{P0}) is hard to be solved directly due to the following reasons:
	1) Compared with the conventional SF-RISs, the newly introduced transmission and reflection coefficients are intricately coupled with the other variables;
	2) Constraint (\ref{C_STAR-RIS}) is highly non-convex as each phase shifter is limited to the unit magnitude;
	3) The CSI estimation error is considered in all involved channels, resulting in infinitely many non-
	convex constraints (\ref{C_User})-(\ref{C_decoding_order}).
	To sum up, problem (\ref{P0}) is a non-linear and non-convex problem, and is non-trivial to solve optimally.
	In the next sections, we propose an effective algorithm to solve problem (\ref{P0}) sub-optimally, and extend it to the MS and TS protocols.
	
	\section{Proposed Solutions} \label{Solution}
	To solve problem (\ref{P0}) efficiently, we firstly reformulate problem (\ref{P0}) to an equivalent problem through introducing slack variables, which paves the way for decomposing the coupling variables.
	Next, relying on the $\mathcal{S}$-procedure and the general sign-definiteness, we transform the semi-definite constraints into the tractable forms.
	Finally, the AO strategy is utilized to divide problem (\ref{P3}) into three sub-problems.
	
	Inspired by the SCA framework\cite{SEE_TVT}, we introduce the slack variables $\psi$ and $\rho$ and rewrite problem (\ref{P0}) as
	\begin{subequations}
		\label{P1}
		\begin{eqnarray}
			&\underset{\boldsymbol{\alpha},\mathbf{F},\boldsymbol{\Phi},\psi,\rho}{\max}& \psi \\
			&\operatorname{s.t.}&
			\label{P1_C_1}
			R_s\geq \psi \rho,\\
			\label{P1_C_2}
			&&\mathcal{P}\leq \rho,\\
			&&{\rm (\ref{C_max_power})-(\ref{C_decoding_order})}.
		\end{eqnarray}
	\end{subequations}
	Note that problem (\ref{P1}) is equivalent to problem (\ref{P0}) since constraints (\ref{P1_C_1}) and (\ref{P1_C_2}) are active at the optimum.
	Obviously, constraint (\ref{P1_C_2}) is a convex set because it can be expressed as a second-order cone (SOC) representation:
	\begin{eqnarray}
		\label{P1_C_2_1}
		\frac{\rho-P_0+\varrho}{2\varrho}\geq \left\|  \left[  \frac{\rho-P_0-\varrho}{2\varrho}, \mathbf{f}_t^{\mathrm T},\mathbf{f}_r^{\mathrm T} \right] ^{\mathrm T}\right\| _2.
	\end{eqnarray}
	
	In order to track the convexity of constraint (\ref{P1_C_1}), we introduce the slack variable set $\mathbf{r}\!=\!\{r_{k,j},r_{k,j}^{e}|\forall k,e,\forall j\!\in\!\mathcal{J}_k\}$, which satisfies $R_{k,j}\!=\! r_{k,j}$ and $R_{k,j}^{e}\!=\! r_{k,j}^{e}$.
	Furthermore, we employ the convex upper bound approximation to address the non-convex term $\psi \rho$.
		Define $g(\psi,\rho)\!=\!\psi \rho$ and $G(\psi,\rho,t)\!=\!\frac{t}{2}\psi^2\!+\!\frac{\rho^2}{2t} (t>0)$,
		then we have $G(\psi,\rho,t)\geq g(\psi,\rho)$\cite{Ni_NOMA_RIS_TWC}.
		Moreover, when $t=\rho/\psi$, we can notice that $G(\psi,\rho,t)=g(\psi,\rho)$ and
		$\nabla G(\psi,\rho,t)=\nabla g(\psi,\rho)$.
		Accordingly, a convex upper bounded of $\psi \rho$ can be obtained as $(\psi \rho)^{\rm ub}= \frac{t}{2}\psi^2\!+\!\frac{\rho^2}{2 t}$,
		where the fixed point $t$ is updated in the $\ell$-th iteration by $t^{(\ell)}\!=\! \rho^{(\ell-1)}/\psi^{(\ell-1)}$.
	Then problem (\ref{P1}) is reformulated as
	\begin{subequations}
		\label{P2}
		\begin{eqnarray}
			&\!\!\!\!\!\!\!\!\!\!\underset{\boldsymbol{\alpha},\mathbf{F},\boldsymbol{\Phi},\psi,\rho,\mathbf{r}}{\max}& \psi \\
			\label{P2_C_1}
			&\operatorname{s.t.}& R_{k,j}\geq r_{k,j}, ~\Omega_{k,l},\forall k, \forall j\!\in\!\mathcal{J}_k, \forall l\!\in\!\mathcal{L}_k,\\
			\label{P2_C_2}
			&&R_{k,j}^{e}\leq r_{k,j}^{e}, ~\Omega_{e},\forall k,e,\forall j\!\in\!\mathcal{J}_k,\\
			\label{P2_C_3}
			&&r_{k,j}\geq C_{k,j}, ~\forall k,\forall j\!\in\!\mathcal{J}_k,\\
			\label{P2_C_4}
			&&r_{k,j}^{e}\leq C_{k,j}^{e},~\forall k,e,\forall j\!\in\!\mathcal{J}_k,\\
			\nonumber
			&&\sum\nolimits_k\sum\nolimits_{j}(r_{k,j}-\sum\nolimits_e r_{k,j}^{e})\geq (\psi \rho)^{\rm ub}, \\
			\label{P1_C_1_3}
			&&\forall k,e,\forall j\!\in\!\mathcal{J}_k,\\
			&&{\rm (\ref{C_max_power})-(\ref{C_STAR-RIS}), (\ref{C_decoding_order}), (\ref{P1_C_2_1})}.
		\end{eqnarray}
	\end{subequations}
	
	We can observe that the main difficulty in solving problem (\ref{P2}) comes from the semi-definite constraints (\ref{P2_C_1}), (\ref{P2_C_2}) and (\ref{C_decoding_order}).
	To this end, we construct finite linear matrix inequalities (LMIs) equivalent to them as follows.
	
	\vspace{-3mm}
	\subsection{Semi-Infinite Constraint Transformation} 
	With the assistance of the auxiliary variable sets $\boldsymbol{\eta}=\{\eta_{k,j}^l,\eta_{k,j}^e|\forall k,e,\forall j\in\mathcal{J}_k,\forall l\in\mathcal{L}_k\}$ and $\boldsymbol{\varsigma}=\{\varsigma_{k,j}|\forall k,\forall j\in\mathcal{J}_{k}^{''}\}$, where $\mathcal{J}_k^{''}=\{\mathcal{J}_{k}^{'}/{J_{k-1}}\}$, constraints (\ref{P2_C_1}), (\ref{P2_C_2}) and (\ref{C_decoding_order}) are respectively transformed into
	\begin{subequations}
		\label{P2_C_1_2}
		\begin{eqnarray}
			\label{P2_C_1.1}
			&&\!\!\!\!\!\!\!\!\!\!\!\!\!\!\!\!\!\!\!\!\!\!\!|\bar{\mathbf{h}}_{k,l} \mathbf{w}_{k,j}|^2\!\geq \eta_{k,j}^l (2^{r_{k,j}}-1), ~\Omega_{k,l}, \forall k, \forall j\!\in\!\mathcal{J}_k, \forall l\!\in\!\mathcal{L}_k,\\
			\label{P2_C_1.2}
			&&\!\!\!\!\!\!\!\!\!\!\!\!\!\!\!\!\!\!\!\!\!\!\!\lVert \bar{\mathbf{h}}_{k,l} \mathbf{w}_{k,-j}\lVert_2^2\ + \ \sigma^2\leq \eta_{k,j}^l, ~\Omega_{k,l}, \forall k, \forall j\!\in\!\mathcal{J}_k, \forall l\!\in\!\mathcal{L}_k,\\
			\label{P2_C_2.1}
			&&\!\!\!\!\!\!\!\!\!\!\!\!\!\!\!\!\!\!\!\!\!\!\!|\bar{\mathbf{h}}_{e} \mathbf{w}_{k,j}|^2\leq \eta_{k,j}^{e} (2^{r_{k,j}^{e}}-1), ~\Omega_{e}, \forall k,e,\forall j\!\in\!\mathcal{J}_k,\\
			\label{P2_C_2.2}
			&&\!\!\!\!\!\!\!\!\!\!\!\!\!\!\!\!\!\!\!\!\!\!\!\lVert \bar{\mathbf{h}}_{e} \mathbf{w}_{k,-j} \lVert_2^2\ + \ \sigma^2\geq \eta_{k,j}^{e} , ~\Omega_{e}, \forall k,e,\forall j\!\in\!\mathcal{J}_k,\\
			\label{P2_C_5.1}
			&&\!\!\!\!\!\!\!\!\!\!\!\!\!\!\!\!\!\!\!\!\!\!\!| \bar{\mathbf{h}}_{k,j}\mathbf{w}_{k,j} |^2 \geq \varsigma_{k,j}, ~\Omega_{k,j}, \forall k, \forall j\!\in\!\mathcal{J}_k^{'}, \\
			\label{P2_C_5.2}
			&&\!\!\!\!\!\!\!\!\!\!\!\!\!\!\!\!\!\!\!\!\!\!\!| \bar{\mathbf{h}}_{k,j+1}\mathbf{w}_{k,j+1} |^2 \leq \varsigma_{k,j}, ~\Omega_{k,j+1}, \forall k, \forall j\!\in\!\mathcal{J}_k^{'}, \\
			\label{P2_C_5.3}
			&&\!\!\!\!\!\!\!\!\!\!\!\!\!\!\!\!\!\!\!\!\!\!\!\varsigma_{k,j}\geq \varsigma_{k, j+1}, ~\forall k, \forall j\!\in\!\mathcal{J}_k^{''}.
		\end{eqnarray}
	\end{subequations}
	
	Due to the continuity of CSI uncertainty sets, except for the linear constraint (\ref{P2_C_5.3}), other constraints in (\ref{P2_C_1_2}) all have infinite possibilities.
	Fortunately, it can be observed that since the forms of constraints (\ref{P2_C_1.1}) and (\ref{P2_C_5.1}) are similar, they can be handled in the same way with the help of $\mathcal{S}$-procedure lemma.
	In addition, constraints (\ref{P2_C_1.2})-(\ref{P2_C_2.2}), and (\ref{P2_C_5.2}) can be solved by using the general sign-definiteness.
	To tackle constraint (\ref{P2_C_1.1}), we first derive its linear approximation in the following proposition.
	\begin{proposition}
		\label{proposition1}
		\emph{
			By denoting $(\boldsymbol{\alpha}^{(\ell)}, \mathbf{F}^{(\ell)}, \boldsymbol{\Phi}^{(\ell)})$ as the optimal solutions obtained in the $\ell$-th iteration, constraint (\ref{P2_C_1.1}) can be equivalently linearized by
			\begin{eqnarray}
				\nonumber		   
				&{}&\!\!\!\!\!\!\!\!\!\!(\mathbf{x}_{k,j}^l)^{\mathrm H}\mathbf{A}_{k,j}\mathbf{x}_{k,j}^l+2{\mathrm Re}\{(\mathbf{a}_{k,j}^l)^{\mathrm H}\mathbf{x}_{k,j}^l\}+a_{k,j}^l \\
				\label{C_proposition1}
				&{}&\!\!\!\!\!\!\!\!\!\!\geq \eta_{k,j}^l (2^{r_{k,j}}-1), ~\Omega_{k,l}, \forall k, \forall j\!\in\!\mathcal{J}_k, \forall l\in\!\mathcal{L}_k,
			\end{eqnarray}
			where the introduced coefficients $\mathbf{x}_{k,j}^l$, $\mathbf{A}_{k,j}$, $\mathbf{a}_{k,j}^l$ and $a_{k,j}^l$ are given as (\ref{proposition1_coefficients}) at the bottom of this page.
		}
		\begin{figure*}[bp]
			\hrulefill
			\begin{subequations}
				\label{proposition1_coefficients}
				\begin{eqnarray}
					\label{rho}
					\!\!\!\!\!\!\!\!\!\!\!\!\!\!\!\mathbf{x}_{k,j}^l\!\!\!\!\!\!\!\!\!\!&{}&=\Big[ \triangle\mathbf{h}_{k,l}^{\mathrm H}\ {\rm vec}^{\mathrm H}(\triangle \mathbf{G}_{k,l}^{\ast})\Big]^{\mathrm H},
					~\mathbf{A}_{k,j}=\mathbf{\widetilde{A}}_{k,j}+(\mathbf{\widetilde{A}}_{k,j})^{\mathrm H}-\mathbf{\widehat{A}}_{k,j},
					~\mathbf{a}_{k,j}^l=\mathbf{\widetilde{a}}_{k,j}^l+\mathbf{\widehat{a}}_{k,j}^l-\mathbf{\bar{a}}_{k,j}^l,
					~a_{k,j}^l=2{\rm Re}\{\widetilde{a}_{k,j}^l\}\!-\!\widehat{a}_{k,j}^l,\\
					\!\!\!\!\!\!\!\!\!\!\!\!\mathbf{\widetilde{A}}_{k,j}\!\!\!\!\!\!\!\!\!\!&{}&=\left[\begin{array}{c}
						\mathbf{w}_{k,j}^{(\ell)} \\
						\mathbf{w}_{k,j}^{(\ell)} \otimes (\mathbf{u}_{k}^{(\ell)})^{\ast}
					\end{array}\right]\Big[\mathbf{w}_{k,j}^{\mathrm H} \  \mathbf{w}_{k,j}^{\mathrm H} \otimes \mathbf{u}_{k}^{\mathrm T}\Big],
					~\mathbf{\widehat{A}}_{k,j}=\left[\begin{array}{c}
						\mathbf{w}_{k,j}^{(\ell)} \\
						\mathbf{w}_{k,j}^{(\ell)} \otimes (\mathbf{u}_{k}^{(\ell)})^{\ast}
					\end{array}\right]\Big[(\mathbf{w}_{k,j}^{(\ell)})^{\mathrm H} \  (\mathbf{w}_{k,j}^{(\ell)})^{\mathrm H} \otimes (\mathbf{u}_{k}^{(\ell)})^{\mathrm T}\Big],\\
					\!\!\!\!\!\!\!\!\!\!\!\!\mathbf{\widetilde{a}}_{k,j}^l\!\!\!\!\!\!\!\!\!\!&{}&=\left[\begin{array}{c}
						\mathbf{w}_{k,j}(\mathbf{w}_{k,j}^{(\ell)})^{\mathrm H}(\hat{\mathbf{h}}_{k,l}+\hat{\mathbf{G}}_{k,l}^{\mathrm H}\mathbf{u}_{k}^{(\ell)}) \\
						{\rm vec}^{\ast}(\mathbf{u}_{k}(\hat{\mathbf{h}}_{k,l}^{\mathrm H}+(\mathbf{u}_{k}^{(\ell)})^{\mathrm H}\hat{\mathbf{G}}_{k,l})\mathbf{w}_{k,j}^{(\ell)}\mathbf{w}_{k,j}^{\mathrm H}
						)\end{array}\right],
					~\mathbf{\widehat{a}}_{k,j}^l=\left[\begin{array}{c}
						\mathbf{w}_{k,j}^{(\ell)}\mathbf{w}_{k,j}^{\mathrm H}(\hat{\mathbf{h}}_{k,l}+\hat{\mathbf{G}}_{k,l}^{\mathrm H}\mathbf{u}_{k}) \\
						{\rm vec}^{\ast}(\mathbf{u}_{k}^{(\ell)}(\hat{\mathbf{h}}_{k,l}^{\mathrm H}+\mathbf{u}_{k}^{\mathrm H}\hat{\mathbf{G}}_{k,l})\mathbf{w}_{k,j}(\mathbf{w}_{k,j}^{(\ell)})^{\mathrm H}
						),\\
					\end{array}\right],\\
					\!\!\!\!\!\!\!\!\!\!\!\!\mathbf{\bar{a}}_{k,j}^l\!\!\!\!\!\!\!\!\!\!&{}&=\left[\begin{array}{c}
						\mathbf{w}_{k,j}^{(\ell)}(\mathbf{w}_{k,j}^{(\ell)})^{\mathrm H}(\hat{\mathbf{h}}_{k,l}+\hat{\mathbf{G}}_{k,l}^{\mathrm H}\mathbf{u}_{k}^{(\ell)}) \\
						{\rm vec}^{\ast}(\mathbf{u}_{k}^{(\ell)}(\hat{\mathbf{h}}_{k,l}^{\mathrm H}+(\mathbf{u}_{k}^{(\ell)})^{\mathrm H}\hat{\mathbf{G}}_{k,l})\mathbf{w}_{k,j}^{(\ell)}(\mathbf{w}_{k,j}^{(\ell)})^{\mathrm H}
						),\\
					\end{array}\right],\\
					\!\!\!\!\!\!\!\!\!\widetilde{a}_{k,j}^l\!\!\!\!\!\!\!\!\!\!&{}&=(\hat{\mathbf{h}}_{k,l}^{\mathrm H}+(\mathbf{u}_{k}^{(\ell)})^{\mathrm H}\hat{\mathbf{G}}_{k,l})\mathbf{w}_{k,j}^{(\ell)}\mathbf{w}_{k,j}^{\mathrm H}(\hat{\mathbf{h}}_{k,l}+\hat{\mathbf{G}}_{k,l}^{\mathrm H}\mathbf{u}_{k}),
					~\widehat{a}_{k,j}^l=(\hat{\mathbf{h}}_{k,l}^{\mathrm H}+(\mathbf{u}_{k}^{(\ell)})^{\mathrm H}\hat{\mathbf{G}}_{k,l})\mathbf{w}_{k,j}^{(\ell)}(\mathbf{w}_{k,j}^{(\ell)})^{\mathrm H}(\hat{\mathbf{h}}_{k,l}\!+\!\hat{\mathbf{G}}_{k,l}^{\mathrm H}\mathbf{u}_{k}^{(\ell)}).
				\end{eqnarray}
			\end{subequations}
		\end{figure*}
	\end{proposition}
	\begin{IEEEproof}
			See Appendix A.
	\end{IEEEproof}
	
	Although constraint (\ref{P2_C_1.1}) is transformed into the more tractable linear form in (\ref{C_proposition1}), there are still an infinite number of such linear forms.
	To facilitate derivation, we resort to the $\mathcal{S}$-procedure to further convert them into a manageable form.
	\begin{lemma}
		\label{lemma1}
		\emph{($\mathcal{S}$-procedure \cite{s_procedure})
			Let a quadratic function $f_i(\mathbf{x})$, $\mathbf{x}\in\mathbb{C}^{N\times 1}$, $i\in\mathcal{I}=\{0,1,\ldots,I\}$, be defined as
		}
		\begin{eqnarray}
			\label{C_Lemma1}
			f_i(\mathbf{x})=\mathbf{x}^{\mathrm H}\mathbf{A}_i\mathbf{x}+2{\mathrm Re}\{\mathbf{a}_i^{\mathrm H}\mathbf{x}\}+a_i, 
		\end{eqnarray}
		\emph{
			where $\mathbf{A}_i\in\mathbb{H}^N$, $\mathbf{a}_i\in\mathbb{C}^{N\times 1}$, and $a_i\in\mathbb{R}$.
			Then the condition $\{f_i(\mathbf{x})\geq 0\}_{i=1}^I\Rightarrow f_0(\mathbf{x})\geq 0$ holds if and only if there exist $\upsilon_i\geq 0,\forall i\in\mathcal{I}$ such that
		}
		\begin{eqnarray}
			\left[\begin{array}{cc}
				\mathbf{A}_0 & \mathbf{a}_0 \\
				\mathbf{a}_0^{\mathrm H} & a_0
			\end{array}\right]-\sum\nolimits_{i=1}^{I}\upsilon_i\left[\begin{array}{cc}
				\mathbf{A}_i & \mathbf{a}_i \\
				\mathbf{a}_i^{\mathrm H} & a_i
			\end{array}\right] \succeq \mathbf{0}.
		\end{eqnarray}
	\end{lemma}
	
	Since $\lVert \triangle\mathbf{h}_{k,l} \lVert_2\leq \xi_{k,l}$ is equivalent to $ \triangle\mathbf{h}_{k,l}^{\mathrm H}\triangle\mathbf{h}_{k,l}\leq \xi_{k,l}^2$, 
	we can express $\Omega_{k,l}$ in terms of the quadratic expression as
	\begin{eqnarray}
		\label{Omega_expansion}	
		\Omega_{k,l}\triangleq\!\left\{\begin{array}{c}\!\!\!
			(\mathbf{x}_{k,j}^l)^{\mathrm{H}}\mathbf{C}_1 \mathbf{x}_{k,j}^l\!-\!\xi_{k,l}^2\leq 0, \\
			\!\!\!	(\mathbf{x}_{k,j}^l)^{\mathrm{H}}\mathbf{C}_2 \mathbf{x}_{k,j}^l\!-\!\zeta_{k,l}^{2} \leq 0.
		\end{array}\right.  \forall k, \forall j\!\in\!\mathcal{J}_k, \forall l\!\in\!\mathcal{L}_k,
	\end{eqnarray}
	where	$\mathbf{C}_1\!=\!\Big[\begin{array}{cc}
		\mathbf{I}_{N} & \mathbf{0} \\
		\mathbf{0} & \mathbf{0}
	\end{array}\Big] ~{\text{and}}~\mathbf{C}_2\!=\!\Big[\begin{array}{cc}
		\mathbf{0} & \mathbf{0} \\
		\mathbf{0} & \mathbf{I}_{M N}
	\end{array}\Big]$.
	Then according to Lemma \ref{lemma1}, the implication
	(\ref{Omega_expansion})$\Rightarrow$(\ref{C_proposition1}) holds if and only if there exist $\upsilon_{k, j}^{l,h}\geq 0$ and $\upsilon_{k, j}^{l,G}\geq 0$
	such that 
	\begin{eqnarray}
		\label{P2_C_1.1_LMIs}
		\left[\!\!\!\begin{array}{cc}
			\mathbf{A}_{k,j}\!\!+\!\upsilon_{k, j}^{l,h}\mathbf{C}_1\!\!+\! \upsilon_{k, j}^{l,G} \mathbf{C}_2 &\!\!\! \mathbf{a}_{k,j}^l \\
			(\mathbf{a}_{k,j}^l)^{\mathrm{H}} &\!\!\! Q_{k,j}^l
		\end{array}\!\!\!\right] \succeq \mathbf{0}, \forall k, \forall j\!\in\!\mathcal{J}_k, \forall l\!\in\!\mathcal{L}_k,\!\!
	\end{eqnarray}
	where $Q_{k,j}^l=a_{k,j}^l-\eta_{k,j}^l (2^{r_{k,j}}\!-\!1)-\upsilon_{k, j}^{l,h}\xi_{k,l}^2-\upsilon_{k, j}^{l,G}\zeta_{k,l}^2$.
	
	Using the same method, constraint (\ref{P2_C_5.1}) is rewritten as
	\begin{eqnarray}
		\label{P2_C_5.1_LMIs}
		\left[\!\!\!\begin{array}{cc}
			\mathbf{A}_{k,j}+\upsilon_{k, j}^{h}\mathbf{C}_1+ \upsilon_{k, j}^{G} \mathbf{C}_2 & \mathbf{a}_{k,j} \\
			\mathbf{a}_{k,j}^{\mathrm{H}} & Q_{k,j}
		\end{array}\!\!\!\right] \succeq \mathbf{0}, ~\forall k,\forall j\!\in\!\mathcal{J}_k^{'},
	\end{eqnarray}
	where $\upsilon_{k, j}^{h},\upsilon_{k, j}^{G}\!\geq \!0$ and $Q_{k,j}\!=\!a_{k,j}\!-\!\varsigma_{k,j}\!-\!\upsilon_{k, j}^h\xi_{k,j}^2\!-\!\upsilon_{k, j}^G\zeta_{k,j}^2$, $\forall k, \forall j\in\mathcal{J}_k^{'}$.
	$\mathbf{a}_{k,j}$ and $a_{k,j}$ are obtained by replacing $\hat{\mathbf{h}}_{k,l}$ with $\hat{\mathbf{h}}_{k,j}$ and $\hat{\mathbf{G}}_{k,l}$ with $\hat{\mathbf{G}}_{k,j}$ in $\mathbf{a}_{k,j}^l$ and $a_{k,j}^l$, respectively.
	
	Then, to deal with constraints (\ref{P2_C_1.2})-(\ref{P2_C_2.2}), and (\ref{P2_C_5.2}), we first provide the following proposition.
		\begin{proposition}
			\label{proposition2}
			\emph{Let $\boldsymbol{\pi}_{k,j}^l\!=\!((\hat{\mathbf{h}}_{k,l}^{\mathrm H}+\mathbf{u}_{k}^{\mathrm H}\hat{\mathbf{G}}_{k,l})\mathbf{w}_{k,-j})^{\mathrm H}$, constraint (\ref{P2_C_1.2}) can be equivalently transformed into the matrix inequality as follows
				\begin{eqnarray}
					\nonumber
					&{}&\!\!\!\!\!\!\!\!\!\!\mathbf{0} \preceq\left[\!\begin{array}{c}
						\mathbf{0} \\
						\mathbf{w}_{k,-j}^{\mathrm{H}}
					\end{array}\!\right]\left[\!\begin{array}{ll}
						\triangle \mathbf{h}_{k,l} & \!\! \mathbf{0}
					\end{array}\!\right]+\left[\!\begin{array}{c}
						\triangle \mathbf{h}_{k,l}^{\mathrm{H}} \\
						\mathbf{0}
					\end{array}\!\right]\left[\!\begin{array}{ll}
						\mathbf{0} & \!\!\mathbf{w}_{k,-j}
					\end{array}\!\right]	\\
					\nonumber
					&{}&\!+\left[\!\!\begin{array}{c}
						\mathbf{0} \\
						\mathbf{w}_{k,-j}^{\mathrm{H}}
					\end{array}\!\!\right] \triangle \mathbf{G}_{k,l}^{\mathrm{H}}\left[\!\!\begin{array}{cc}\!
						\mathbf{u}_{k} & \!\!\mathbf{0}
					\end{array}\!\!\right]+\left[\!\!\begin{array}{c}
						\mathbf{u}_{k}^{\mathrm{H}} \\
						\mathbf{0}
					\end{array}\!\!\right] \triangle \mathbf{G}_{k,l}\left[\!\!\begin{array}{cc}
						\mathbf{0} & \!\!\mathbf{w}_{k,-j}
					\end{array}\!\!\right]	\\
					\label{P2_C_1.3.1}
					&{}&\!+\left[\!\begin{array}{cc}
						\eta_{k,j}^l-\sigma^{2} &\! (\boldsymbol{\pi}_{k,j}^l)^{\mathrm{H}} \\
						\boldsymbol{\pi}_{k,j}^l&\!\mathbf{I}
					\end{array}\!\right], \forall k, \forall j\in\mathcal{J}_k, \forall l\in\mathcal{L}_k.
				\end{eqnarray}
			}	
		\end{proposition}
		\begin{IEEEproof}
			See Appendix B.
		\end{IEEEproof}
		
		It can be observed that the resulting constraint (\ref{P2_C_1.3.1}) contains multiple complex valued uncertainties.
		In order to tackle it, we introduce the general sign-definiteness lemma as follows.
	\begin{lemma}
		\label{lemma2}
		\emph{(General sign-definiteness\cite{general sign_definiteness})
			Given matrices $\mathbf{A}$ and $\{\mathbf{E}_i,\mathbf{F}_i\}_{i=1}^{I}$ with $\mathbf{A}=\mathbf{A}^{\rm H}$, the semi-infinite LMI
		}
		\begin{eqnarray}
			\label{C_lemma2}
			\mathbf{A} \succeq \sum\nolimits_{i=1}^{I}\left(\mathbf{E}_i^{\mathrm{H}} \mathbf{G}_i \mathbf{F}_i+\mathbf{F}_i^{\mathrm{H}} \mathbf{G}_i^{\mathrm{H}} \mathbf{E}_i\right),  ~\forall i,\left\|\mathbf{G}_i\right\|_{F} \leq \xi_i,
		\end{eqnarray}
		\emph{
			holds if and only if there exist $\varpi_i\geq 0,\forall i$, such that
		}
		\begin{eqnarray}
			\left[\begin{array}{cccc}
				\mathbf{A}-\sum_{i=1}^{I} \varpi_{i} \mathbf{F}_{i}^{\mathrm{H}} \mathbf{F}_{i} & -\xi_{1} \mathbf{E}_{1}^{\mathrm{H}} & \cdots & -\xi_{I} \mathbf{E}_{I}^{\mathrm{H}} \\
				-\xi_{1} \mathbf{E}_{1} & \varpi_{1} \mathbf{I} & \cdots & \mathbf{0}\\
				\vdots & \vdots & \ddots & \vdots \\
				-\xi_{I} \mathbf{E}_{P} & \mathbf{0} & \cdots & \varpi_{I} \mathbf{I}
			\end{array}\right] \succeq \mathbf{0}.
		\end{eqnarray}
	\end{lemma}
	
	Please refer to \cite{general sign_definiteness_proof} for the detailed proof of Lemma \ref{lemma2}.
	
	Applying Lemma \ref{lemma2}, defining slack variables $\varpi_{k, j}^{l,h},\varpi_{k, j}^{l,G}\geq 0$, and denoting $T_{k,j}^l=\eta_{k,j}^l-\sigma^2-\varpi_{k, j}^{l,h}-\varpi_{k, j}^{l,G}\sum\nolimits_{m} \beta_{m}^{k}$, the equivalent LMIs of constraint (\ref{P2_C_1.2}) are given by
	\begin{eqnarray}
		\label{P2_C_1.2_LMIs}
		\nonumber
		&	\left[\!\!\!\begin{array}{cccc}
			T_{k,j}^l &\!\! (\boldsymbol{\pi}_{k,j}^l)^{\mathrm{H}} &\!\! \mathbf{0}_{1 \times N} &\!\! \mathbf{0}_{1 \times N} \\
			\boldsymbol{\pi}_{k,j}^l &\!\! \mathbf{I}_{j} &\!\! \xi_{k,l} \mathbf{w}_{k,-j}^{\mathrm{H}} &\!\! \zeta_{k,l} \mathbf{w}_{k,-j}^{\mathrm{H}} \\
			\mathbf{0}_{N \times 1} &\!\! \xi_{k,l} \mathbf{w}_{k,-j} &\!\!\varpi_{k, j}^{l,h} \mathbf{I}_{N} &\!\! \mathbf{0}_{N} \\
			\mathbf{0}_{N \times 1} &\!\! \zeta_{k,l} \mathbf{w}_{k,-j} &\!\! \mathbf{0}_{N} &\!\! \varpi_{k, j}^{l,G} \mathbf{I}_{N}
		\end{array}\!\!\!\right] \!\succeq\! \mathbf{0},  \\
		&\!\!\!\!\!\!\!\!\!\!\!\!\!\!\!\!\!\!\!\!\!\!\!\!\!\!\!\!\!\!\!\!\!\!\!\!\!\!\!\!\!\!\!\!\!\!\!\!\!\!\!\!\!\!\!\!\!\!\!\!\!\forall k, \forall j\in\mathcal{J}_k, \forall l\in\mathcal{L}_k.
	\end{eqnarray}
	
	Similarly, it is possible to reformulate constraints (\ref{P2_C_2.1}), (\ref{P2_C_2.2}) and (\ref{P2_C_5.2}) as the finite LMIs in (\ref{C_rest_LMI}) at the bottom of this page.
	The corresponding parameters are given in (\ref{LMI_coefficients}) at the bottom of this page, where $\omega_{k, j}^{e,h}$, $\omega_{k, j}^{e,G}$, $\widetilde{\omega}_{k, j}^{e,h}$, $\widetilde{\omega}_{k, j}^{e,G}\geq 0$, $\forall k,e,\forall j\in\mathcal{J}_k$, and $\varpi_{k,j}^h$, $\varpi_{k,j}^G\geq 0$, $\forall k,\forall j\in\mathcal{J}_k^{'}$.
	\begin{figure*}[bp]
		\hrulefill
		\begin{subequations}
			\label{C_rest_LMI}
			\begin{eqnarray}
				&\!\!\!\!\!\!\!\!\!\!\!\!\!		{\rm (\overline {\ref{P2_C_2.1}})}\!\! :\!\!\left[\!\!\begin{array}{cccc}
					T_{k,j}^{e} &\!\! \!(\pi_{k,j}^{e})^{\ast} &\!\!\! \mathbf{0}_{1 \times N} &\!\!\! \mathbf{0}_{1 \times N} \\
					\pi_{k,j}^{e} &\!\! \!1 &\!\!\! \xi_{e} \mathbf{w}_{k,j}^{\mathrm{H}} &\!\!\! \zeta_{e} \mathbf{w}_{k,j}^{\mathrm{H}} \\
					\mathbf{0}_{N \times 1} &\!\! \!\xi_{e} \mathbf{w}_{k,j} &\!\!\! \omega_{k, j}^{e,h} \mathbf{I}_{N} &\!\!\! \mathbf{0}_{N} \\
					\mathbf{0}_{N \times 1} &\!\!\! \zeta_{e} \mathbf{w}_{k,j} &\!\!\! \mathbf{0}_{N} &\!\!\! \omega_{k, j}^{e,G} \mathbf{I}_{N}
				\end{array}\!\!\right] \!\succeq \!\mathbf{0}, 
				~	{\rm (\overline {\ref{P2_C_2.2}})}\!\! :\!\!
				\left[\!\!\!\begin{array}{cccc}
					\widetilde{T}_{k,j}^{e} &\!\!\!- (\boldsymbol{\widetilde{\pi}}_{k,j}^{e})^{\mathrm{H}} &\!\!\! \mathbf{0}_{1 \times N} &\!\!\! \mathbf{0}_{1 \times N} \\
					-\boldsymbol{\widetilde{\pi}}_{k,j}^{e} &\!\!\!- \mathbf{I}_{j} &\!\!\! -\xi_{k,e} \mathbf{w}_{k,-j}^{\mathrm{H}} &\!\!\! -\zeta_{e} \mathbf{w}_{k,-j}^{\mathrm{H}} \\
					\mathbf{0}_{N \times 1} &\!\!\! -\xi_{e} \mathbf{w}_{k,-j} &\!\!\!\widetilde{\omega}_{k, j}^{e,h} \mathbf{I}_{N} &\!\!\! \mathbf{0}_{N} \\
					\mathbf{0}_{N \times 1} &\!\!\! -\zeta_{e} \mathbf{w}_{k,-j} &\!\!\! \mathbf{0}_{N} &\!\!\! \widetilde{\omega}_{k, j}^{e,G} \mathbf{I}_{N}
				\end{array}\!\!\!\right] \!\succeq\! \mathbf{0}, ~\forall k, e,\forall j\!\in\!\mathcal{J}_k, \\
				&\!\!\!\!\!\!\!\!\!\!\!\!\!\!\!\!	\!\!\!\!\!\!\!\!\!\!\!\!\!	\!\!\!\!\!\!\!\!\!\!\!\!\!\!\!\!\!\!\!\!\!\!\!\!\!\!\!\!\!\!\!\!\!\!\!\!\!\!\!	\!\!\!\!\!\!\!\!\!\!\!\!\!\!\!\!\!\!\!\!\!\!\!\!\!\!\!\!\!\!\!\!\!\!\!\!\!\!\!{\rm (\overline {\ref{P2_C_5.2}})}\!\! :\!\!	\left[\!\!\begin{array}{cccc}
					T_{k,j}&\!\! (\pi_{k,j+1})^{\ast} &\!\! \mathbf{0}_{1 \times N} &\!\! \mathbf{0}_{1 \times N} \\
					\pi_{k,j+1}&\!\! 1 &\!\! {\xi}_{k,j+1} \mathbf{w}_{k,j+1}^{\mathrm{H}} &\!\! {\zeta}_{k,j+1} \mathbf{w}_{k,j+1}^{\mathrm{H}} \\
					\mathbf{0}_{N \times 1} &\!\! {\xi}_{k,j+1} \mathbf{w}_{k,j+1} &\!\! \varpi_{k, j}^h \mathbf{I}_{N} &\!\! \mathbf{0}_{N} \\
					\mathbf{0}_{N \times 1} &\!\! {\zeta}_{k,j+1} \mathbf{w}_{k,j+1} &\!\! \mathbf{0}_{N} &\!\! \varpi_{k, j}^G \mathbf{I}_{N}
				\end{array}\!\!\right] \succeq\mathbf{0}, ~\forall k, \forall j\in\mathcal{J}_k^{'}. 
			\end{eqnarray}
		\end{subequations}
		\begin{subequations}
			\label{LMI_coefficients}
			\begin{eqnarray}
				&{}&\!\!\!\!\!\!\!\!\!\!\!\!\!\!\!\!\!\!\!	\!\!\!\!\!\!\!\!\!\!\!\!\!\!\!\!\!\!\!	\!\!\!\!\!\!\!\!\!\!\!\!\!	\!\!\!\!\!\!\!\!\!\!\!\!\!\!\!\!\!T_{k,j}^{e}=\eta_{k,j}^{e} (2^{r_{k,j}^{e}}-1)-\omega_{k, j}^{e,h}-\omega_{k, j}^{e,G}\sum\nolimits_m\beta_m^k,
				~~~~\pi_{k,j}^{e}=((\hat{\mathbf{h}}_{k,e}^{\mathrm H}+\mathbf{u}_{k}^{\mathrm H}\hat{\mathbf{G}}_{k,e})\mathbf{w}_{k,j})^{\ast}, 
				~\forall k, e,\forall j\in\mathcal{J}_k,\\
				&{}&\!\!\!\!\!\!\!\!\!\!\!\!\!\!\!\!\!\!\!	\!\!\!\!\!\!\!\!\!\!\!\!\!\!\!\!\!\!\!	\!\!\!\!\!\!\!\!\!\!\!\!\!	\!\!\!\!\!\!\!\!\!\!\!\!\!\!\!\!	\widetilde{T}_{k,j}^{e}=\sigma^2-\eta_{k,j}^{e}-\widetilde{\omega}_{k, j}^{e,h}-\widetilde{\omega}_{k, j}^{e,G}\sum\nolimits_m\beta_m^k,
				~~~~\boldsymbol{\widetilde{\pi}}_{k,j}^{e}=((\hat{\mathbf{h}}_{k,e}^{\mathrm H}+\mathbf{u}_{k}^{\mathrm H}\hat{\mathbf{G}}_{k,e})\mathbf{w}_{k,-j})^{\mathrm H}, ~\forall k,e, \forall j\in\mathcal{J}_k,\\
				&{}&\!\!\!\!\!\!\!\!\!\!\!\!\!\!\!\!\!\!\!	\!\!\!\!\!\!\!\!\!\!\!\!\!\!\!\!\!\!\!	\!\!\!\!\!\!\!\!\!\!\!\!\!	\!\!\!\!\!\!\!\!\!\!\!\!\!\!\!\!	T_{k,j}=\varsigma_{k,j}-\varpi_{k, j}^h-\varpi_{k, j}^G\sum\nolimits_m\beta_m^k,
				~~~~\pi_{k,j+1}=((\hat{\mathbf{h}}_{k,j+1}^{\mathrm H}+\mathbf{u}_{k}^{\mathrm H}\hat{\mathbf{G}}_{k,j+1})\mathbf{w}_{k,j+1})^{\ast}, 
				~\forall k, \forall j\in\mathcal{J}_k^{'}.
			\end{eqnarray}
		\end{subequations}
	\end{figure*}	
	
	Eventually, reformulating problem (\ref{P2}) by replacing constraints (\ref{P2_C_1}), (\ref{P2_C_2}) and (\ref{C_decoding_order}) 
	with the LMIs (\ref{P2_C_1.1_LMIs}), (\ref{P2_C_5.1_LMIs}), (\ref{P2_C_1.2_LMIs}) and (\ref{C_rest_LMI}), and the linear constraint (\ref{P2_C_5.3}) yields the following optimization problem
	\begin{subequations}
		\label{P3}
		\begin{eqnarray}
			&\!\!\!\!\!\!\!\!\!\!\!\!\!\!\!\!\underset{\boldsymbol{\alpha},\mathbf{F},\boldsymbol{\Phi},\mathbf{\Delta}}{\max}&\!\!\!\!\! \psi \\
			&\!\!\!\!\!\!\!\!\!\!\operatorname{s.t.}&\!\!\!\!\!{\rm (\ref{C_max_power})-(\ref{C_STAR-RIS}), (\ref{P1_C_2_1}),(\ref{P2_C_3})-(\ref{P1_C_1_3}) }, \\
			&&\!\!\!\!\!{\rm (\ref{P2_C_5.3}) , (\ref{P2_C_1.1_LMIs}),  (\ref{P2_C_5.1_LMIs}), (\ref{P2_C_1.2_LMIs}), (\ref{C_rest_LMI})},\\
			\label{P3_C_coefficient1}
			&&\!\!\!\!\!\upsilon_{k, j}^{l,h},\upsilon_{k,j}^{l,G},\varpi_{k, j}^{l,h},\varpi_{k,j}^{l,G}\geq 0, \forall k,\forall j\!\in\!\mathcal{J}_k,\forall l\!\in\!\mathcal{L}_k,\\
			&&\!\!\!\!\!\omega_{k, j}^{e,h},\omega_{k,j}^{e,G},\widetilde{\omega}_{k, j}^{e,h},\widetilde{\omega}_{k,j}^{e,G} \geq 0, \forall k,e,\forall j\!\in\!\mathcal{J}_k,\\
			\label{P3_C_coefficient4}
			&&\!\!\!\!\!\upsilon_{k, j}^h,\upsilon_{k,j}^G, \varpi_{k, j}^h,\varpi_{k,j}^G \geq 0, \forall k,\forall j\!\in\!\mathcal{J}_k^{'}, 
		\end{eqnarray}
	\end{subequations}
	where $\mathbf{\Delta}\!=\!\{\psi,\rho,\mathbf{r}, \boldsymbol{\eta},\boldsymbol{\varsigma},\boldsymbol{\upsilon}_1, \boldsymbol{\varpi}_1,\boldsymbol{\omega}_1,\boldsymbol{\omega}_2,\boldsymbol{\upsilon}_2,\boldsymbol{\varpi}_2\}$ represents the set of all introduced auxiliary variables, with 
	\begin{subequations}
		\label{P3_coefficients}
		\begin{eqnarray}
			\boldsymbol{\varpi}_1\!\!\!\!\!\!\!\!\!\!&{}&=\{\varpi_{k, j}^{l,h},\!\varpi_{k,j}^{l,G}|~\forall k,\forall j\in\mathcal{J}_k,\forall l\in\mathcal{L}_k\},\\
			\boldsymbol{\omega}_1\!\!\!\!\!\!\!\!\!\!&{}&=\{\omega_{k, j}^{e,h},\omega_{k,j}^{e,G}|~\forall k,e,\forall j\in\mathcal{J}_k\},\\
			\boldsymbol{\omega}_2\!\!\!\!\!\!\!\!\!\!&{}&=\{\widetilde{\omega}_{k, j}^{e,h},\widetilde{\omega}_{k,j}^{e,G}| ~\forall k,e,\forall j\in\mathcal{J}_k\},\\
			\boldsymbol{\upsilon}_2\!\!\!\!\!\!\!\!\!\!&{}&=\{\upsilon_{k, j}^h,\upsilon_{k,j}^G|~\forall k,\forall j\in\mathcal{J}_k^{'}\},\\
			\boldsymbol{\varpi}_2\!\!\!\!\!\!\!\!\!\!&{}&=\{\varpi_{k, j}^h,\varpi_{k,j}^G|~\forall k,\forall j\in\mathcal{J}_k^{'}\}.
		\end{eqnarray}
	\end{subequations}
	
	We notice that problem (\ref{P3}) is still non-convex and difficult to optimize $\boldsymbol{\alpha}$, $\mathbf{F}$ and $\boldsymbol{\Phi}$ simultaneously, since they are highly coupled in parameters such as $\mathbf{a}_{k,j}^l$ and $\boldsymbol{\pi}_{k,j}^l$.
	To solve it effectively, we employ the AO strategy to decouple these variables and transform problem (\ref{P3}) into three subproblems.
	
	\vspace{-2mm}
	\subsection{Power Allocation}
	For given $\mathbf{F}$ and $\boldsymbol{\Phi}$, the optimization problem for the design of power allocation coefficients $\boldsymbol{\alpha}$ can be formulated as
	\begin{subequations}
		\label{P4}
		\begin{eqnarray}
			&\underset{\boldsymbol{\alpha},\mathbf{\Delta} }{\max}&\psi \\
			&\operatorname{s.t.}&{\rm (\ref{C_power allocation}), (\ref{P1_C_2_1}), (\ref{P2_C_3})-(\ref{P1_C_1_3}), (\ref{P2_C_5.3})},\\
			&&{\rm (\ref{P2_C_1.1_LMIs}),  (\ref{P2_C_5.1_LMIs}), (\ref{P2_C_1.2_LMIs}), (\ref{C_rest_LMI}), (\ref{P3_C_coefficient1})-(\ref{P3_C_coefficient4})}.
		\end{eqnarray}
	\end{subequations}
	At this point, it can be observed that all the constraints in problem (\ref{P4}) are convex except constraint (\ref{P2_C_1.1_LMIs}).
	Moreover, its non-convexity arises from the term $\eta_{k,j}^l2^{r_{k,j}}$ in $Q_{k,j}^l$.
	To proceed, we adopt the SCA approach to obtaining approximation.
	Specifically, for a given feasible point $((\eta_{k,j}^l)^{(\ell)},r_{k,j}^{(\ell)})$ in the $\ell$-th iteration, $\eta_{k,j}^l2^{r_{k,j}}$ is upper bounded by
	\begin{eqnarray}
		\label{P4-SCA}
		(\eta_{k,j}^l2^{r_{k,j}})^{\rm ub}=\big( (r_{k,j}-r_{k,j}^{(\ell)}) (\eta_{k,j}^l)^{(\ell) }{\rm ln}2+\eta_{k,j}^l\big) 2^{r_{k,j}^{(\ell)}}.
	\end{eqnarray}
	
	Substituting (\ref{P4-SCA}) into (\ref{P2_C_1.1_LMIs}), we can recast problem (\ref{P4}) as
	\begin{subequations}
		\label{P4-1}
		\begin{eqnarray}
			&\underset{\boldsymbol{\alpha},\mathbf{\Delta}}{\max}&\psi \\
			&\operatorname{s.t.}&{\rm (\ref{C_power allocation}), (\ref{P1_C_2_1}), (\ref{P2_C_3})-(\ref{P1_C_1_3}), (\ref{P2_C_5.3})},\\
			&&{\rm (\ref{P2_C_5.1_LMIs}), (\ref{P2_C_1.2_LMIs}), (\ref{C_rest_LMI}), (\ref{P3_C_coefficient1})-(\ref{P3_C_coefficient4})},\\
			\nonumber
			&&\left[\begin{array}{cc}
				\mathbf{A}_{k,j}+\upsilon_{k, j}^{l,h}\mathbf{C}_1+\upsilon_{k, j}^{l,G} \mathbf{C}_2 & \mathbf{a}_{k,j}^l \\
				(\mathbf{a}_{k,j}^l)^{\mathrm{H}} & \widehat{Q}_{k,j}^l
			\end{array}\right] \succeq \mathbf{0}, \\
			\label{P4_1_C_1}
			&&~\forall k, \forall j\in\mathcal{J}_k, \forall l\in\mathcal{L}_k,
		\end{eqnarray}
	\end{subequations}
	where $\widehat{Q}_{k,j}^l=a_{k,j}^l+\eta_{k,j}^l-	(\eta_{k,j}^l2^{r_{k,j}})^{\rm ub}-\upsilon_{k, j}^{l,h}\xi_{k,l}^2-\upsilon_{k, j}^{l,G}\zeta_{k,l}^2$.
	Obviously, problem (\ref{P4-1}) is a semi-definite programming (SDP) problem, and hence can be efficiently solved by standard convex program solvers such as the CVX.
	
	\vspace{-2mm}
	\subsection{Active Beamforming Design}
	With fixed $\boldsymbol{\alpha}$ and $\boldsymbol{\Phi}$, we arrive at the optimization of the active beamforming $\mathbf{F}$.
	As the non-convex constraint (\ref{P2_C_1.1_LMIs}) is also an obstacle to this subproblem, we can employ the similar procedure devised above to address it.
	Armed with (\ref{P4_1_C_1}), the optimization problem for $\mathbf{F}$ is given by
	\begin{subequations}
		\label{P5}
		\begin{eqnarray}
			&\underset{\mathbf{F},\mathbf{\Delta} }{\max}&\psi \\
			&\operatorname{s.t.}&{\rm (\ref{C_max_power}), (\ref{P1_C_2_1}), (\ref{P2_C_3})-(\ref{P1_C_1_3}), (\ref{P2_C_5.3})},\\
			&&{\rm (\ref{P2_C_5.1_LMIs}), (\ref{P2_C_1.2_LMIs}), (\ref{C_rest_LMI}), (\ref{P3_C_coefficient1})-(\ref{P3_C_coefficient4}), (\ref{P4_1_C_1}) }.
		\end{eqnarray}
	\end{subequations}
	It is straightforward to observe that problem (\ref{P5}) is a SDP problem, which can be solved efficiently via the CVX.
	
	\vspace{-2mm}
	\subsection{Passive Beamforming Design}
	Now we focus on the design of passive beamforming.
	For given $\boldsymbol{\alpha}$ and $\mathbf{F}$, we have the following optimization problem
	\begin{subequations}
		\label{P6}
		\begin{eqnarray}
			\label{P6_OJ}
			&\underset{\boldsymbol{\Phi}, \mathbf{\Delta} }{\max}&\!\!\psi \\
			\label{P6_C_00}
			&\operatorname{s.t.}&\!\!{\rm (\ref{P1_C_2_1}), (\ref{P2_C_3})-(\ref{P1_C_1_3}), (\ref{P2_C_5.3})},\\
			\label{P6_C_01}
			&&\!\!{\rm (\ref{P2_C_1.1_LMIs}),  (\ref{P2_C_5.1_LMIs}), (\ref{P2_C_1.2_LMIs}), (\ref{C_rest_LMI}), (\ref{P3_C_coefficient1})-(\ref{P3_C_coefficient4})},\\
			\label{P6_C_02}
			&&\!\!\sum\nolimits_{k} \beta_{m}^{k}=1, ~\beta_{m}^{k}\in[0,1],  \forall k,m,\\
			\label{P6_C_03}
			&&\!\![\mathbf{u}_{k}]_m=\sqrt{\beta_{m}^{k}}e^{j \theta_{m}^{k}}, ~\theta_{m}^{k}\in [0,2\pi),  \forall k,m.
		\end{eqnarray}
	\end{subequations}
	Note that problem (\ref{P6}) is intractable due to the non-convex constraint (\ref{P2_C_1.1_LMIs}) and the unit-modulus constraint (\ref{P6_C_03}).
	The former can be replaced by the approximate form (\ref{P4_1_C_1}), while the latter can be handled by the PCCP method\cite{CL_IOS_secure}.
	
	\begin{algorithm}[tbp]
		\caption{PCCP-Based Algorithm for Solving (\ref{P6})}
		\label{Algorithm1}
		\begin{algorithmic}[1] 
			\STATE \textbf{Initialize} $\mathbf{u}_{k}^{(0)}$, the scaling factor $\varepsilon>1$ and the iteration index $\ell=0$.
			Set the allowable tolerances $\epsilon_1$ and $\epsilon_2$, the maximum number of iterations $T_{\rm max}$, and the maximum value $\lambda_{\max}$.
			\REPEAT
			\IF {$n\leq T_{\rm max}$  }
			\STATE Update $\mathbf{u}_{k}^{(\ell+1)}$ by solving problem (\ref{P6-1});
			\STATE Update $\lambda^{(\ell+1)}={\min}\{ \varepsilon \lambda^{(\ell)}, \lambda_{\max} \}$;
			\STATE Update $\ell=\ell+1$;
			\ELSE
			\STATE Reinitialize with a new $\mathbf{u}_{k}^{(0)}$, set $\varepsilon>1$ and $\ell=0$.
			\ENDIF
			\UNTIL $\lVert\mathbf{u}_{k}^{(\ell)}-\mathbf{u}_{k}^{(\ell-1)}\lVert_1 \leq \epsilon_1$ and $C\leq \epsilon_2$.
			\STATE \textbf{Output} the converged solution $\mathbf{u}_{k}^{\star}=\mathbf{u}_{k}^{(\ell)}$.
		\end{algorithmic}  
	\end{algorithm}

	In particular, the auxiliary variable set $\mathbf{b}\!=\!\{b_{k,m}|\forall k,m\}$ is introduced to linearize (\ref{P6_C_03}), satisfying $b_{k,m}\!=\![\mathbf{u}_{k}]_m^\ast[\mathbf{u}_{k}]_m$. 
	Then we can rewrite $b_{k,m}\!=\![\mathbf{u}_{k}]_m^\ast[\mathbf{u}_{k}]_m$ as 
	$b_{k,m}\!\leq\![\mathbf{u}_{k}]_m^\ast[\mathbf{u}_{k}]_m\!\leq\! b_{k,m}$.
	Based on the FTS, the non-convex part $b_{k,m}\!\leq\![\mathbf{u}_{k}]_m^\ast[\mathbf{u}_{k}]_m$ can be approximated by $b_{k,m}\!\leq \!2{\rm Re}\big\{ [\mathbf{u}_{k}]_m^\ast[\mathbf{u}_{k}^{(\ell)}]_m\big\}-[\mathbf{u}_{k}^{(\ell)}]_m^{\ast}[\mathbf{u}_{k}^{(\ell)}]_m$.
	Following the PCCP framework, we penalize these terms which are included in the objective function (\ref{P6_OJ}), and recast problem (\ref{P6}) as
	\begin{subequations}
		\label{P6-1}
		\begin{eqnarray}
			\label{P6-1-function}
			&\!\!\!\underset{\boldsymbol{\Phi},\mathbf{\Delta},\mathbf{b},\mathbf{c} }{\max}&\psi-\lambda^{(\ell)} C \\
			\label{P6-1-00}
			&\operatorname{s.t.}&{\rm (\ref{P1_C_2_1}),(\ref{P2_C_3})-(\ref{P1_C_1_3}), (\ref{P2_C_5.3})},\\
			\label{P6-1-01}
			&&{\rm (\ref{P2_C_5.1_LMIs}), (\ref{P2_C_1.2_LMIs}), (\ref{C_rest_LMI}), (\ref{P3_C_coefficient1})-(\ref{P3_C_coefficient4}), (\ref{P4_1_C_1}) },\\
			\label{C_P6-1_1}
			&& [\mathbf{u}_{k}]_m^\ast[\mathbf{u}_{k}]_m\leq b_{k,m}+c_{k,m},  ~\forall k,m,\\
			\nonumber
			&&2{\rm Re}\big\{ [\mathbf{u}_{k}]_m^\ast[\mathbf{u}_{k}^{(\ell)}]_m\big\}-[\mathbf{u}_{k}^{(\ell)}]_m^{\ast}[\mathbf{u}_{k}^{(\ell)}]_m\\
			\label{C_P6-1_2}
			&&\geq b_{k,m}-\hat{c}_{k,m}, ~\forall k, m,\\
			\label{C_P6-1_3}
			&&\sum\nolimits_{k} b_{k,m}=1, ~b_{k,m}\geq 0, ~\forall k,m,
		\end{eqnarray}
	\end{subequations} 
	where $\mathbf{c}\!=\!\{c_{k,m},\hat{c}_{k,m}|\forall k, m\}$ is the slack variable set imposed over the modulus constraint (\ref{P6_C_03}).
	Also, $C\!=\!\sum\nolimits_k\sum\nolimits_m (c_{k,m}+\hat{c}_{k,m})$ is the penalty term, and is scaled by the penalty factor $\lambda^{(\ell)}$, which controls the feasibility of the constraint.
	
	Now problem (\ref{P6-1}) is a SDP problem and can be solved via the CVX.
	The conventional approach to tackle the unit-modulus constraint is applying the semi-definite relaxation (SDR).
	However, the SDR method usually generates a passive beamforming matrix, and needs a rank-one method to obtain the final passive beamforming vector. 
	The resulting solution may not be feasible for the original problem. 
	In contrast, one advantage of PCCP is that the passive beamforming vector can be directly obtained, which reduces the performance loss.
	The steps of finding a feasible $\boldsymbol{\Phi}$ to problem (\ref{P6}) are summarized in Algorithm \ref{Algorithm1}.
	The main points are as follows:
	1) The unit-modulus constraint (\ref{P6_C_03}) can be guaranteed by the stopping criteria $C \leq \epsilon_2$ when $\epsilon_2$ is sufficiently small\cite{TSP-A framework};
	2) The maximum value $\lambda_{\max}$ is imposed to avoid numerical	problems.
	Specifically, a feasible solution satisfying $C \leq \epsilon_2$ may not be found when the iteration converges under increasing large values of $\lambda^{(\ell)}$\cite{PCCP};
	3) The convergence of Algorithm \ref{Algorithm1} can be controlled by the stopping criteria $\lVert\mathbf{u}_{k}^{(\ell)}-\mathbf{u}_{k}^{(\ell-1)}\lVert_1 \leq \epsilon_1$.
	
	\subsection{Complexity and Convergence Analysis}
	\begin{table}[tbp]\small
		\renewcommand{\arraystretch}{1.7}
		\caption{LMI constraints description}
		\label{P5_constraints}
		\centering
		\scalebox{1}{
		\begin{tabular}{|c|c|c|c|c|c|}
			\hline  
			constraint &size & number & constraint & size & number    \\
			\hline 
			(\ref{P2_C_5.1_LMIs}) &$a_1$ & $J-2$ 
			&(\ref{P2_C_1.2_LMIs}) & $a_2$ & $\sum$ \\
			\hline
			${\rm (\overline {\ref{P2_C_2.1}})}$& $a_3$ &$2J$ 
			&${\rm (\overline {\ref{P2_C_2.2}})}$& $a_2$ &$2J$ \\
			\hline
			${\rm (\overline {\ref{P2_C_5.2}})}$ & $a_3$ &$J-2$
			&(\ref{P4_1_C_1})& $a_1$ &$\sum$ \\
			\hline
		\end{tabular}}
	\end{table}
	
	\begin{figure}[tbp]
		\centering
		\includegraphics[width=3.6in,height=2.2in]{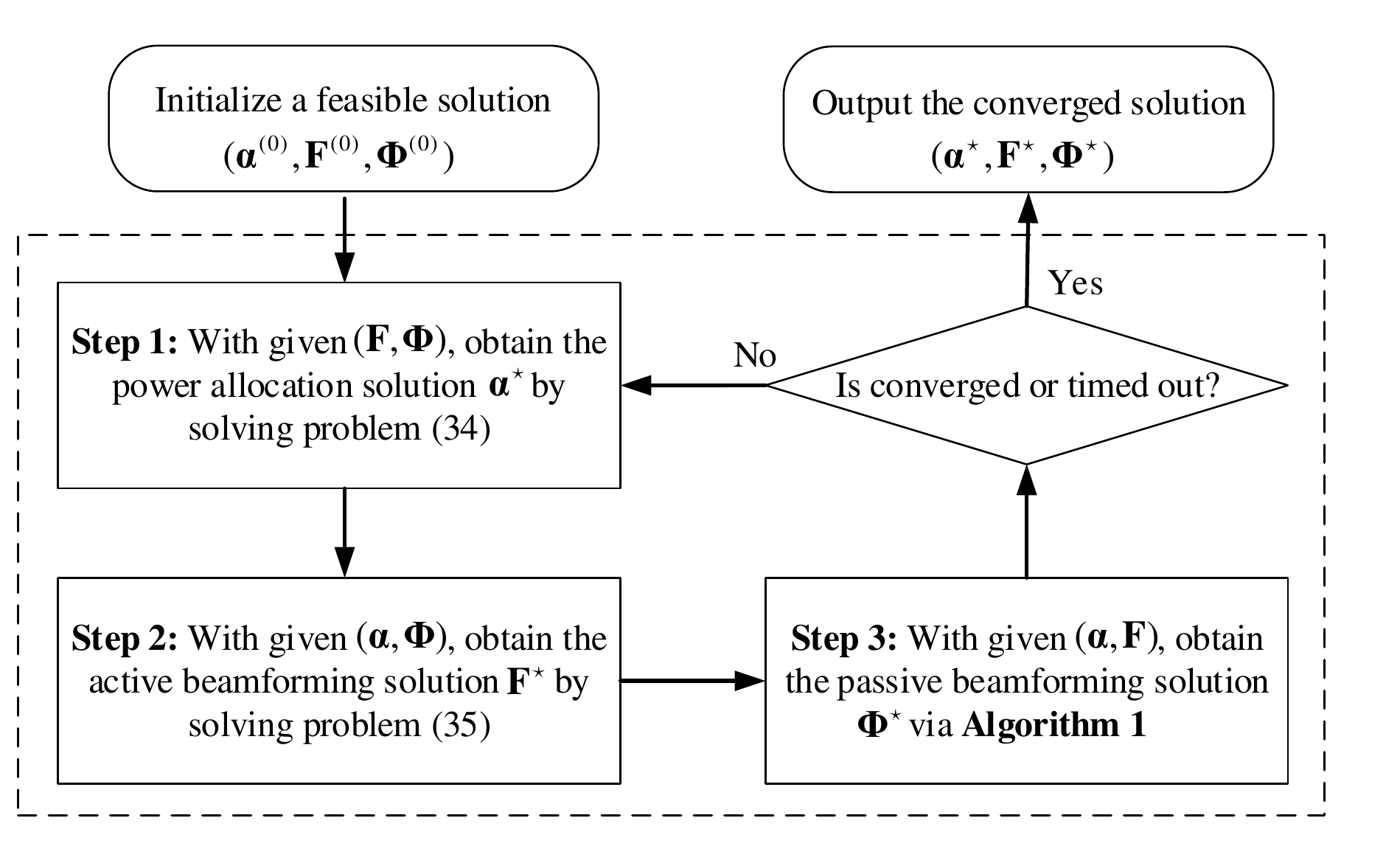}
		\caption{A flowchart of the proposed AO algorithm.}
		\label{final_flowchart}
	\end{figure}

	The final flowchart of the proposed AO algorithm is given in Fig. \ref{final_flowchart}, in which problems (\ref{P4-1}), (\ref{P5}) and (\ref{P6-1}) are alternatively solved until the convergence metric is triggered.
	It can be observed that these resulting convex problems involve the LMI, SOC constraints and linear constraints.
	Thus, all problems can be solved by the interior point method\cite{interior point}, and the total complexity of the AO algorithm can be estimated in terms of their worst-case runtime.
	Generally, by ignoring the non-dominated linear constraints, the upper bound of the number of operations required to solve such problem is provided in \cite{TSP-A framework}.
	Taking problem (\ref{P5}) as an example, the number of variables is $n_2=2N$.
	The number and size of LMI constraints contained in (\ref{P5}) are presented in Table \ref{P5_constraints}, where $a_1=MN+N+1$, $a_2=2N+j+1$, $a_3=2N+2$ and $\sum=\sum_k\sum_{j}j$, $\forall k,\forall j\in\mathcal{J}_k$.
	Moreover, the number of SOC in (\ref{P1_C_2_1}) is $1$ with the size of $2N+1$.
	The complexity of solving problems (\ref{P4-1}), (\ref{P5}) and (\ref{P6-1}) is denoted by $\mathcal{O}_{\boldsymbol{\alpha}}$, $\mathcal{O}_{\mathbf{F}}$, and $\mathcal{O}_{\boldsymbol{\Phi}}$, respectively.
	The exact expressions of $\mathcal{O}_{\boldsymbol{\alpha}}$, $\mathcal{O}_{\mathbf{F}}$, and $\mathcal{O}_{\boldsymbol{\Phi}}$ are given by (\ref{complexity}) at the bottom of the next page, where $n_1=J$ and $n_3=2M$.
	On the other hand, since each sub-algorithm converges to their individual suboptimal solution\cite{SEE_TVT,TSP-A framework}, the objective value of problem (\ref{P0}) is non-decreasing in each AO iteration, and is upper bounded by a finite value. 
	Hence, the proposed AO algorithm is guaranteed to converge.
	\begin{figure*}[bp]
		\hrulefill
		\begin{subequations}
			\label{complexity}
			\begin{eqnarray}
				&{}&\!\!\!\!\!\!\!\!\!\!\!\!\mathcal{O}_{\boldsymbol{\alpha}}=\mathcal{O}\big\{ \sqrt{f_1} 
				n_1 
				\big(n_1^{2}+n_1f_2+f_3\big) \big\},
				~~ \mathcal{O}_{\mathbf{F}}=\mathcal{O}\big\{ \sqrt{f_1+2}
				n_2 
				\big(n_2 ^{2}+n_2f_2+f_3+(2N+1)^2n_2\big) \big\},\\
				&{}&\!\!\!\!\!\!\!\!\!\!\!\!\mathcal{O}_{\boldsymbol{\Phi}}=\mathcal{O}\big\{ \sqrt{f_1+4M} 
				n_3 
				\big(n_3^{2}+n_3f_2+f_3+2Mn_3\big) \big\},
				~~f_1=\sum(a_1+a_2)+2J(a_2+a_3)+(J-2)(a_1+a_3),\\
				&{}&\!\!\!\!\!\!\!\!f_2=\sum (a_1^2+a_2^2)+2J(a_2^2+a_3^2)+(J-2)(a_1^2+a_3^2),
				~~f_3=\sum (a_1^3+a_2^3)+2J(a_2^3+a_3^3)+(J-2)(a_1^3+a_3^3).
			\end{eqnarray}
		\end{subequations}
	\end{figure*}	
	\section{Extension to the MS and TS Protocols} \label{Extension}
	\subsection{Optimization for the MS Protocol}
	Defining $\mathbb{R}_{\beta, \theta}^{\rm MS}=\big\{\beta_{m}^{k}, \theta_{m}^{k} | \sum \nolimits_{k}\beta_{m}^{k}=1;\beta_{m}^{k}\in\{0,1\}; \theta_{m}^{k}\in[0,2 \pi)\big\}$,
	the SEE maximization problem for the MS protocol can be formulated as
	\begin{subequations}
		\label{P0_MS}
		\begin{eqnarray}
			&\underset{\boldsymbol{\alpha},\mathbf{F},\boldsymbol{\Phi} }{\max}&\frac{\sum_k\sum_{j}\big(R_{k,j}-\sum_e R_{k,j}^{e}\big)}
			{ \varrho \sum\nolimits_{k}\lVert \mathbf{f}_k\lVert_2^2+P_0 }\\
			&\operatorname{s.t.}&\!\!{\rm (\ref{C_max_power}), (\ref{C_power allocation}), (\ref{C_User})-(\ref{C_decoding_order}) },\\
			\label{C_MS_IOS}
			&&\beta_{m}^{k}, \theta_{m}^{k} \in \mathbb{R}_{\beta, \theta}^{\rm MS}, ~\forall k,  m.
		\end{eqnarray}
	\end{subequations}
	It can be observed that problem (\ref{P0_MS}) is a mixed-integer non-convex problem due to the binary constraint $\beta_{m}^{k}\in\{0,1\}$ in (\ref{C_MS_IOS}).
	Since constraint (\ref{C_MS_IOS}) does not affect the optimizations of $\boldsymbol{\alpha}$ and $\mathbf{F}$, the corresponding power allocation and active beamforming design subproblems can be solved by the methods proposed above.
	By contrast, with fixed $\boldsymbol{\alpha}$ and $\mathbf{F}$, the subproblem of the passive beamforming is recast to
	\begin{subequations}
		\label{PMS_0}
		\begin{eqnarray}
			&\!\!\!\!\!\!\!\!\!\!\!\!\underset{\boldsymbol{\Phi}, \mathbf{\Delta} }{\max}&\psi \\
			&\!\!\!\!\!\!\!\!\!\!\!\!\operatorname{s.t.}&{\rm (\ref{P6_C_00}), (\ref{P6_C_01}), (\ref{P6_C_03})},\\
			\label{C_PMS_inter00}
			&&\sum\nolimits_k \beta_{m}^{k}=1, ~\forall k,m,\\
			\label{C_PMS_inter}
			&& \beta_{m}^{k}\in\{0,1\}, ~\forall k,m.
		\end{eqnarray}
	\end{subequations}
	The only difference between problem (\ref{P6}) and problem (\ref{PMS_0}) is the additional non-convex binary constraint (\ref{C_PMS_inter}).
	Therefore, we only need to focus on how to tackle this new obstacle.
	By introducing the slack variable set $\mathbf{d}\!=\!\{d_{k,m}|\forall k,m\}$, constraint (\ref{C_PMS_inter}) can be equivalently rewritten as 
	\begin{eqnarray}
		\label{C_PMS_inter1}
		\beta_{m}^{k}=d_{k,m} ~\text{and}~\beta_{m}^{k}(1-d_{k,m})=0, ~\forall k,m.
	\end{eqnarray}
	
	Furthermore, by adding (\ref{C_PMS_inter1}) as penalty terms into the objective function (\ref{P6-1-function}),  one can obtain that
	\begin{subequations}
		\label{PMS_1}
		\begin{eqnarray}
			&\!\!\!\!\!\!\underset{\boldsymbol{\Phi},\mathbf{\Delta},\mathbf{b},\mathbf{c},\mathbf{d} }{\max}&\psi-\lambda^{(\ell)} C - \widetilde{\lambda}^{(\ell)}\widetilde{C} \\
			&\!\!\!\operatorname{s.t.}&{\rm (\ref{P6-1-00})-(\ref{C_P6-1_3})},
		\end{eqnarray}
	\end{subequations} 
	where $\widetilde{C}\!=\!\sum_{k}\sum_m(|b_{k,m}-d_{k,m}|^2+|b_{k,m}(1-d_{k,m})|^2)$.
	Moreover, $\widetilde{\lambda}^{(\ell)}$ is the penalty factor in the $\ell$-th iteration, which can be updated in the same way as $\lambda^{(\ell)}$.
	
	For given $\{\boldsymbol{\Phi},\mathbf{\Delta},\mathbf{b},\mathbf{c}\}$, the optimal $\mathbf{d}$ can be obtained by leveraging the first-order optimal condition\cite{CL_IOS_secure}, expressed as
	\begin{eqnarray}
		d_{k,m}=\frac{b_{k,m}+b_{k,m}^2 }{1+b_{k,m}^2 }, ~\forall k,m.
	\end{eqnarray}
	With a fixed $\mathbf{d}$, problem (\ref{PMS_1}) is a convex SDP problem, and thus can be efficiently solved by the existing toolbox.

	\subsection{Optimization for the TS Protocol}
	When employing TS, the BS consecutively sends information to the two spaces in the transmission and reflection periods.
	This time domain exploration enables Bobs to successively eliminate the inter-cluster interference.
	Therefore, with the time allocation variable set $\boldsymbol{\tau}=\{\tau_{k}|\forall k\}$, the achievable rates of decoding $s_{k,j}$ at $U_{k,l}$ and Eve $e$ are recast to
	\begin{subequations}
		\begin{eqnarray}
			\label{TS_SIC-User}
			\nonumber
			&{}&R_{k,j}^{l,{\rm TS}}=\tau_{k}\log_2\left(1+\frac{|\bar{\mathbf{h}}_{k,l}\mathbf{w}_{k,j}|^2}
			{\lVert \bar{\mathbf{h}}_{k,l} \mathbf{w}_{k,-j}^{\rm TS}\lVert_2^2+ \tau_{k}\sigma^2}\right),\\
			&{}&\forall k,\forall j\in\mathcal{J}_k, \forall l\in\mathcal{L}_k,\\
			\nonumber
			&{}&R_{k,j}^{e,{\rm TS}}=\tau_{k}\log_2\left(1+\frac{|\bar{\mathbf{h}}_{e} \mathbf{w}_{k,j}|^2}	{\lVert \bar{\mathbf{h}}_{e} \mathbf{w}_{k,-j}^{\rm TS}\lVert_2^2+  \tau_{k} \sigma^2}  \right),\\
			&{}&\forall k,e,\forall j\in\mathcal{J}_k,
		\end{eqnarray}
	\end{subequations}
	where $\mathbf{w}_{k,-j}^{\rm TS}\!=\![\alpha_{k,1}\mathbf{f}_k, \alpha_{k,2}\mathbf{f}_k,\ldots,\alpha_{k,j-1}\mathbf{f}_k]$.
	Note that the transmit power of $U_{k,j}$ is increased by $1/\tau_{k}$ to ensure a fair comparison with ES and MS.
	At this point, the passive beamforming vector is given by  $\mathbf{u}_{k}^{\rm TS}\!=\!\big[e^{j \theta_{1}^{k}}\!,e^{j \theta_{2}^{k}}\!,\ldots,e^{j \theta_{M}^{k}}\big]^{\mathrm T}$, $\forall k, m$, and the constraint on the STAR-RIS coefficients reduces to $ \{\theta_{m}^{k}\!\in\![0,2 \pi)\}$.
	As a result, the SEE maximization problem for TS is formulated as
	\begin{subequations}
		\label{PTS_0}
		\begin{eqnarray}
			&\!\!\!\!\!\!\!\!\underset{\boldsymbol{\alpha},\mathbf{F},\boldsymbol{\Phi},\boldsymbol{\tau} }{\max}\!\!\!\!& \frac{\sum_k\sum_{j}\big(R_{k,j}^{\rm TS}-\sum\nolimits_{e}R_{k,j}^{e,{\rm TS}} \big)}
			{ \varrho \sum\nolimits_{k} \lVert\mathbf{f}_k\lVert_2^2+P_0 }\\
			&\operatorname{s.t.}&\theta_{m}^{k}\in[0,2 \pi), ~\forall  k, m, \\
			&&R_{k,j}^{\rm TS}   \geq C_{k,j}, ~\Omega_{k,l},\forall k, \forall j\!\in\!\mathcal{J}_k, \forall l\!\in\!\mathcal{L}_k,\\
			&&R_{k,j}^{e,{\rm TS}}   \leq C_{k,j}^{e}, ~\Omega_{e},\forall k,e, \forall j\!\in\!\mathcal{J}_k, \\
			&&\sum\nolimits_{k} \tau_{k}=1, ~\tau_{k}\in[0,1],  ~\forall k,\\
			&&{\rm (\ref{C_max_power}), (\ref{C_power allocation}),(\ref{C_decoding_order})},
		\end{eqnarray}
	\end{subequations}
	where $R_{k,j}^{\rm TS}={\rm min}\big\{R_{k,j}^{l,{\rm TS}}|\forall l\in\mathcal{L}_k\big\}$, $\forall k, \forall j\in\mathcal{J}_k$.
	Here, a two-layer iterative algorithm is proposed to solve problem (\ref{PTS_0}).
	Specifically, the outer-layer iteration is designed to determine the time allocation coefficients $\boldsymbol{\tau}$.
	Given that $\tau_{k}$ is a one-dimensional variable, bisection search is a straightforward and basic method to obtain its optimal solution $\boldsymbol{\tau}^{\star}$.
	Then, with $\boldsymbol{\tau}^{\star}$, the remaining variables can be solved by the inner-layer iteration, which updates $\{\boldsymbol{\alpha},\mathbf{F},\boldsymbol{\Phi}\}$ alternatively using the similar methods described in Section \ref{Solution}.
	 \begin{figure}[tp]
		\setlength{\abovecaptionskip}{-1.2cm} 
		\centering
		\includegraphics[width=3.2in,height=2.1in]{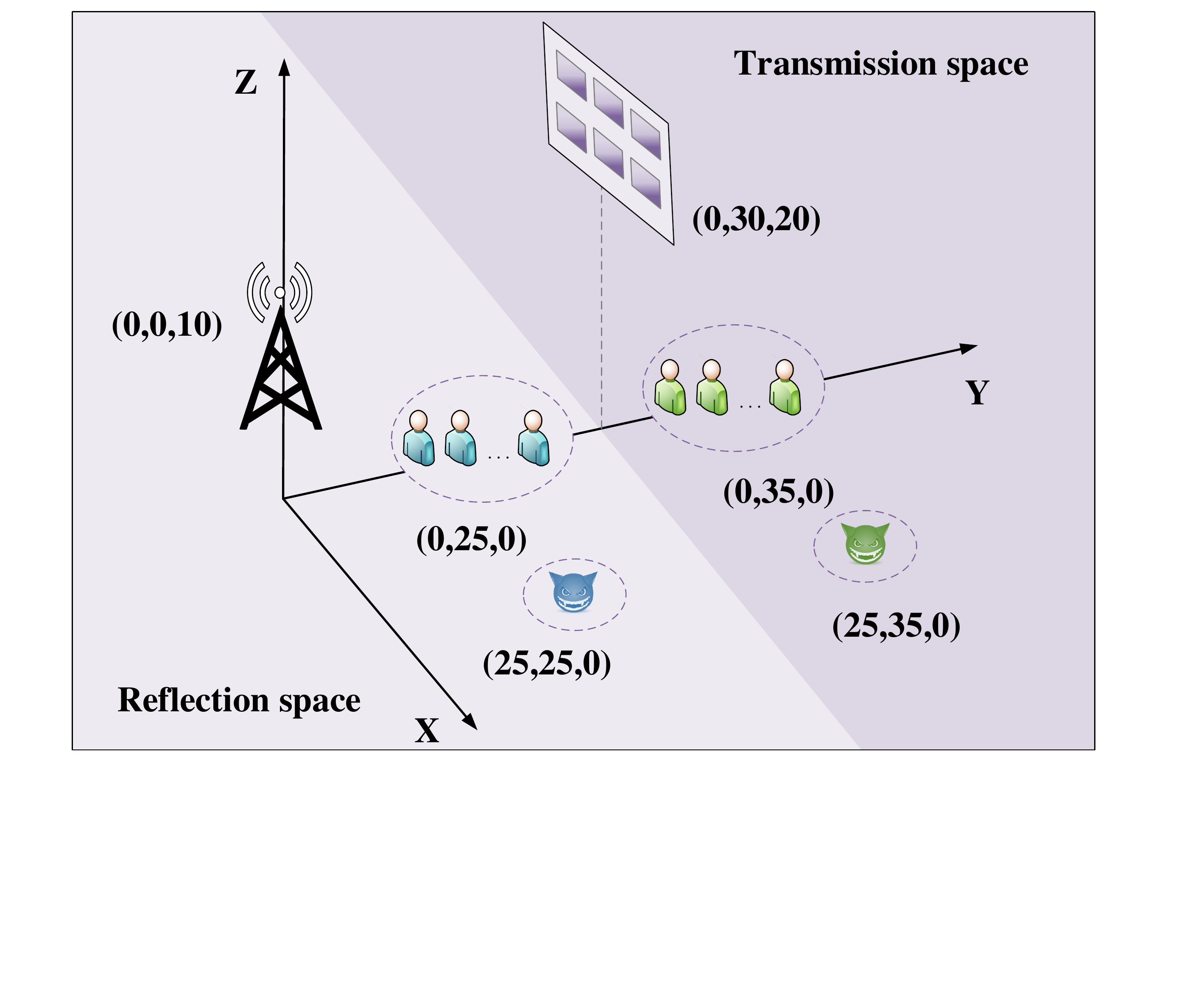}
		\caption{Simulation scenario of the STAR-RIS assisted NOMA system.}
		\label{simulation_model}
	\end{figure}
	\section{Simulation Results}\label{Simulation Results}
	We concentrate on a three-dimensional coordinate network as shown in Fig. \ref{simulation_model}, where the BS and the STRA-RIS are located at $(0,0,10)$ and $(0,30,20)$ m, respectively.
	Moreover, Bobs and Eves of each space are randomly and uniformed distributed in their own circle with the radius of 4 m.
	The clusters of Bobs and Eve in the reflection space are centered on $(0,25,0)$ and $(25,25,0)$ m, respectively, while the corresponding centers in the transmission space are $(0,35,0)$ and $(25,35,0)$ m, respectively.
	Since the BS and the STAR-RIS usually have fixed positions, the line-of-sight (LoS) transmission can be exploited\cite{STAR-RIS-TWC,CL_IOS_secure}.
	In particular, we adopt Rician fading for all channels.
	For instance, the channel matrix $\mathbf{H}_b$ between the BS and the STAR-RIS is given by
		\begin{eqnarray}
			\label{H_b}
			\mathbf{H}_{b}\!=\!\sqrt{\varepsilon_0d_b^{-\alpha_b}}\left(\sqrt{\frac{\nu_b}{\nu_b+1}}\mathbf{H}_{b}^{\rm{LoS}}\!+\!\sqrt{\frac{1}{\nu_b+1}}\mathbf{H}_{b}^{\rm{NLoS}}\right),
		\end{eqnarray}
		where $\varepsilon_0$ is the path loss at the reference distance of 1 m, $d_b$ is the BS-STAR-RIS link distance, and $\alpha_b$ is the corresponding path loss exponent.
		As for small-scale fading, $\nu_b$ is the Rician factor.
		Moreover, $\mathbf{H}_{b}^{\rm{LoS}}$ and $\mathbf{H}_{b}^{\rm{NLoS}}$ denote the LoS and non-LoS components, which are modeled as the product of the array response vectors of the transceivers and Rayleigh fading, respectively.
		Other channels are generated in the same way.
	To facilitate the presentation, we define $\kappa_{k,j}^h$, $\kappa_{k,j}^G$, $\kappa_{e}^h$, and $\kappa_{e}^G$ as the maximum normalized estimation errors for the channels $\mathbf{h}_{k,j}$, $\mathbf{G}_{k,j}$, $\mathbf{h}_{e}$, and $\mathbf{G}_{e}$, i.e., $\kappa_{k,j}^h\!=\!\frac{\xi_{k,j}}{\lVert \hat{\mathbf{h}}_{k,j}\lVert_2}$, $\kappa_{k,j}^G\!=\!\frac{\zeta_{k,j}}{\lVert \hat{\mathbf{G}}_{k,j}\lVert_F}$, $\kappa_{e}^h\!=\!\frac{\xi_{e}}{\lVert \hat{\mathbf{h}}_{e}\lVert_2}$, and $\kappa_{e}^G\!=\!\frac{\zeta_{e}}{\lVert \hat{\mathbf{G}}_{e}\lVert_F}$.
	Unless specified otherwise, we set $(\kappa_{k,j}^h)^2\!=\!(\kappa_{k,j}^G)^2\!=\!(\kappa_{e}^h)^2\!=\!(\kappa_{e}^G)^2\!=\!0.1$, $\forall k,e,\forall j\in\mathcal{J}_k$.
	Other parameters are summarized in Table \ref{parameter}.
	
	\begin{table}[t]\small
		\vspace{-0.3cm}
		\centering
		\renewcommand{\arraystretch}{1.3}
		\caption{Simulation parameters}
		\vspace{-0.2cm}
		\label{parameter}
		\scalebox{0.86}{
			\begin{tabular}{|c|c|}
				\hline  
				\bfseries Parameter & \bfseries Value\\
				\hline
				Path loss at 1 m &$\varepsilon_0= -30$ dB  \\
				\hline
				\tabincell{c}
				{Path loss exponents of BS-STAR-RIS, \\ 
					BS-Bob/Eve, STAR-RIS-Bob/Eve links} & 2.2, 3.2, 2.6\\
				\hline
				\tabincell{c}
				{Rician factors of BS-STAR-RIS, \\ 
					BS-Bob/Eve, STAR-RIS-Bob/Eve links}  & 3 dB \\
				\hline
				Noise power at Bobs and Eves & $\sigma^2=-80$ dBm \\ 
				\hline
				Power amplifier efficiency & $\varrho =1$  \\
				\hline
				Hardware static power  \cite{RIS-review-3} & $P_{\rm B}\!=\!10$ dBW, $P_{\rm U}\!=\!10$ dBm \\
				\hline
				Minimum rate requirement for Bobs & $C_{k,j}=1.5$ bits/s/Hz \\
				\hline 
				Maximum tolerable information leakage & $C_{k,j}^{e}=0.6$  bits/s/Hz \\
				\hline
				Initialized penalty factors\cite{STAR-RIS-TWC} & $\lambda^{(0)}=\widetilde{\lambda}^{(0)}=10^{-3}$ \\
				\hline
				Scaling factor & $\varepsilon=10$ \\
				\hline
				Convergence tolerance & $10^{-4}$ \\
				\hline
		\end{tabular}}
	\end{table}
	
	\begin{figure}[t]
		\centering
		\setlength{\abovecaptionskip}{-0.1mm} 
		\includegraphics[width=3in]{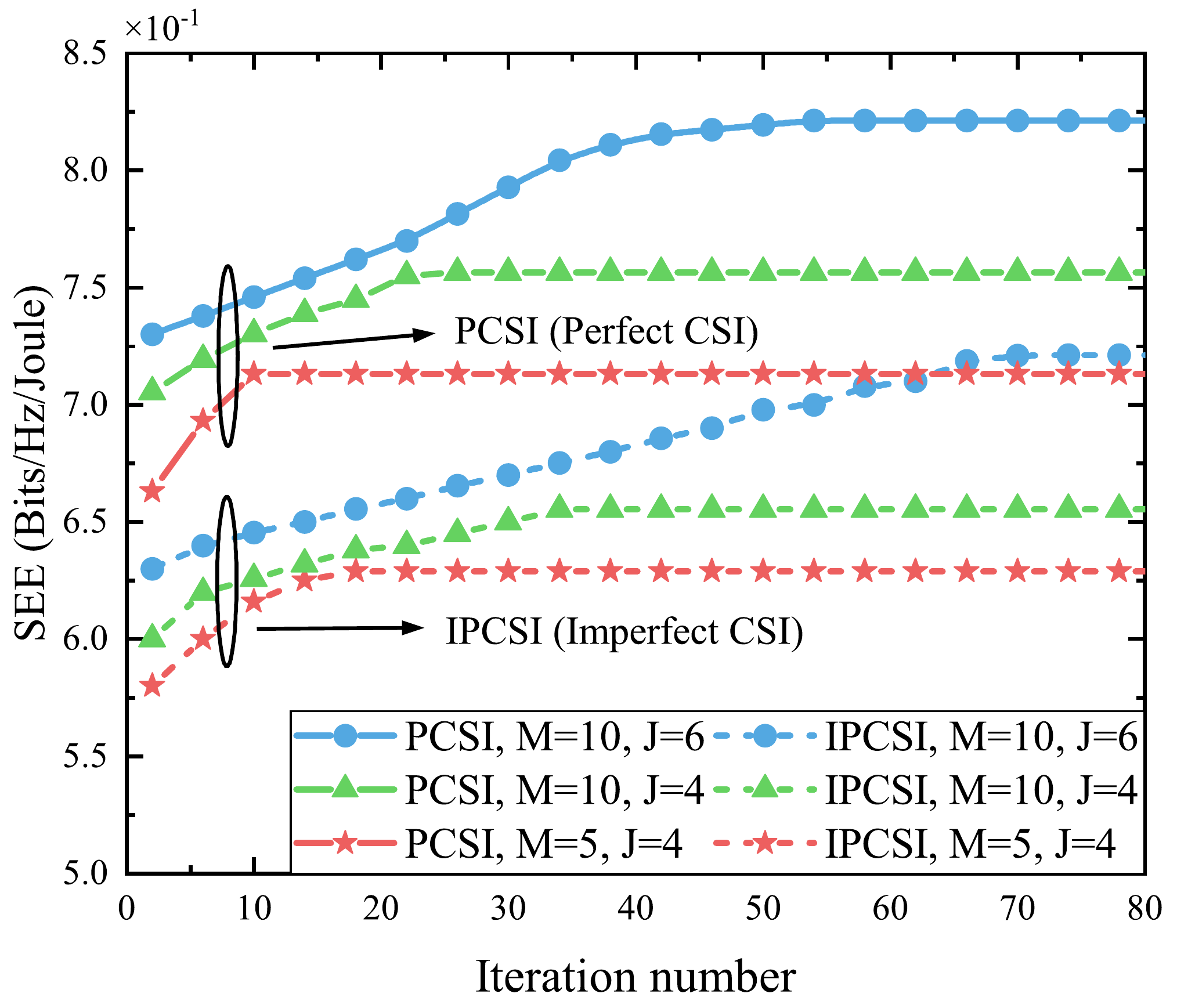}
		\caption{Convergence behaviors of the proposed algorithms for different values of $M,J$, and different CSIU levels. 
			For perfect CSI, we set $(\kappa_{k,j}^h)^2\!=\!(\kappa_{k,j}^G)^2\!=\!(\kappa_{e}^h)^2\!=\!(\kappa_{e}^G)^2\!=\!0$, $\forall k,e,\forall j\in\mathcal{J}_k$.
			For $J\!=\!4$, we set $J_t\!=\!J_r\!=\!2$. 
			While for $J\!=\!6$, we set $J_t\!=\!J_r\!=\!3$.
			The other system parameters are set as $N\!=\!5$, $P_{\max}\!=\!40$ dBm, and $P_r(b)=10$ mW\cite{Prb_reference}.}
		\label{convergence}
	\end{figure}
	The convergence behaviors of the proposed algorithms with different number of STAR-RIS elements $M$, different number of Bobs $J$, and different settings of CSI are investigated in Fig. \ref{convergence}.
	Considering that the proposed AO algorithms in Fig. \ref{Algorithm1} can be easily extended to the MS and TS protocols, here we focus on depicting the obtained SEE under ES against the number of iterations.
	It can be observed that all curves gradually increase and exhibit trend of convergence after a finite number of iterations.
	Meanwhile, the proposed algorithms under imperfect CSI require more iterations than those with perfect CSI to convergence since the CSIU increases the dimensions of LMIs.
	Particularly, Fig. \ref{convergence} shows that a larger $M$ or $J$ leads to slower convergence, and the increase of $J$ needs more additional iterations.
	The reason is that the dimension of the solution space and the dimensions of LMIs scale with $M$, while the number of optimization variables, the dimensions and number of LMIs constraints all increase with $J$.
	In summary, compared to the number of elements, the convergence of the AO algorithms seems more sensitive to the number of Bobs.
	This phenomenon is consistent with Fig. 3 in \cite{convergence} and the complexity analysis results given in (\ref{complexity}).
	

	To evaluate the performance of the proposed system, the STAR-RIS assisted OMA systems with three operating protocols are considered as benchmark schemes.
	Under the OMA scheme, the BS serves Bobs in orthogonal time slots.
	Additionally, the NOMA and OMA systems with the aid of SF-RISs are also considered, where one conventional reflecting-only RIS and one transmitting-only RIS are deployed adjacent to each other, and each RIS has $M/2$ elements.
	
\begin{figure}[t]
		\centering
		\setlength{\abovecaptionskip}{-0.1mm} 
		\includegraphics[width=3in]{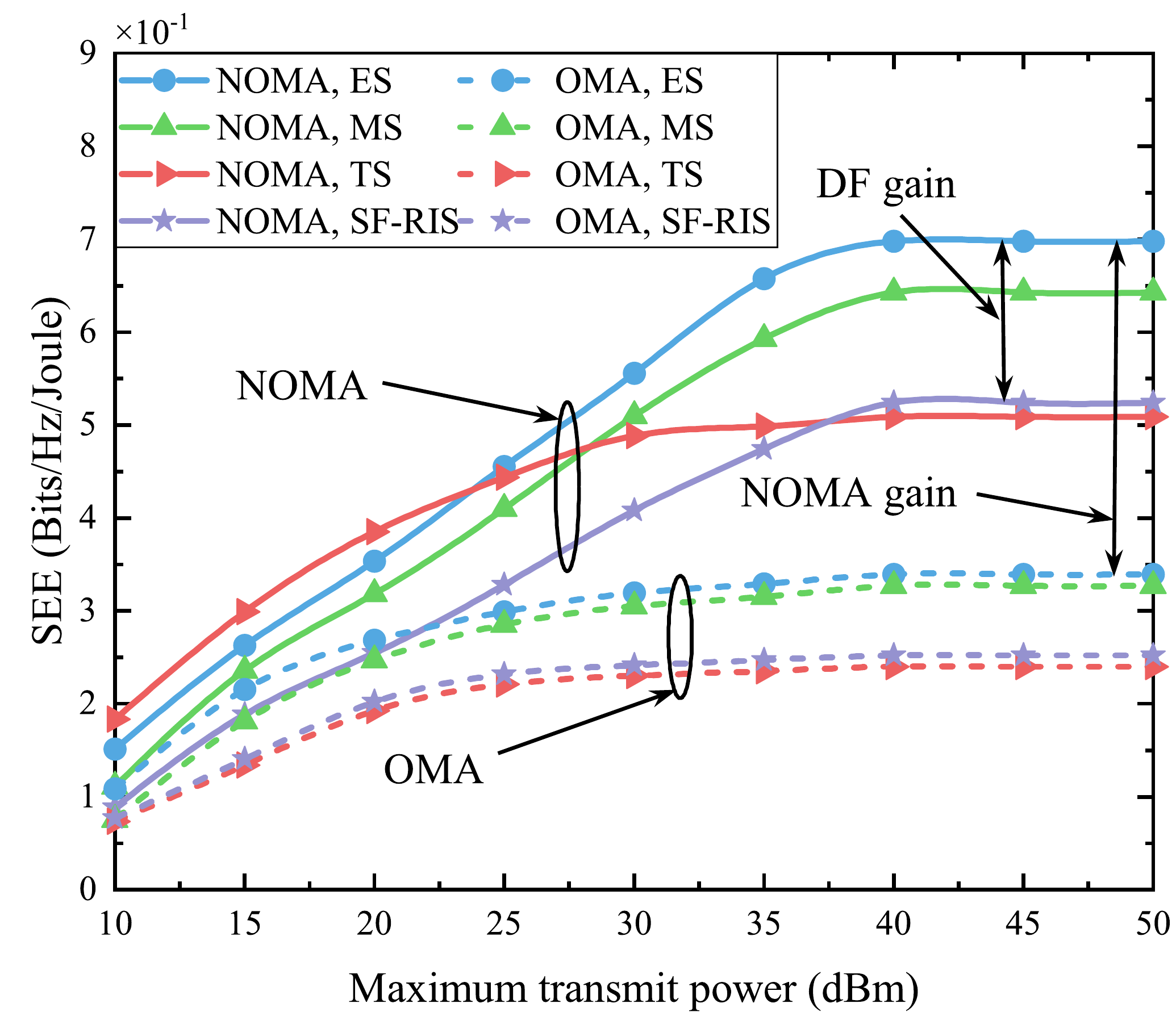}
		\caption{SEE versus the maximum transmit power $P_{\max}$.
			The other system parameters are set as $J_t\!=\!J_r\!=\!2$, $M\!=\!20$, $N\!=\!5$, and $P_r(b)=10$ mW.}
		\label{P_max_SEE}
	\end{figure}

	Fig. \ref{P_max_SEE} depicts the achievable SEE versus the maximum transmit power $P_{\max}$.
	It is interesting to remark that the SEE of all the schemes first increase with $P_{\max}$ and eventually saturate at large $P_{\max}$.
	The reason for this behavior is that the SEE maximization reduces to the SSR maximization for low $P_{\max}$, and using full transmit power is optimal.
	However, when the power availability is sufficient, further increase of the SSR generates the repaid elevation of the energy consumption.
	In this case, for large $P_{\max}$, the optimal allocation strategy is 
	maintaining the power constantly equal to the finite maximizer of the SEE. 
	Regarding the performance of the three operating protocols for the STAR-RIS, when adopting the NOMA scheme, TS is preferable for low $P_{\max}$, whereas ES achieves the best performance for high $P_{\max}$.
	The reason is that by eliminating the inter-cluster interference, TS can effectively facilitate the rate improvement of Bobs.
	While for larger $P_{\max}$, the impact of intra and inter-cluster interference brought by ES on the SSR increases with $P_{\max}$. 
	When properly designed, these interferences can be leveraged to deteriorate Eves' reception, thereby enhancing the SEE.
	
	Also, the effects of $P_{\max}$ on the SEE of OMA schemes are investigated in Fig. \ref{P_max_SEE}.
	Different from NOMA, we can observe from Fig. \ref{P_max_SEE} that TS under OMA yields a smaller value than ES and MS even at the limited power.
	This can be explained as follows.
	Since the OMA scheme results in interference-free communications, both the rates of Bobs and Eves are dominated by the signal reception.
	At this time, for TS, there is always an Eve who is not constrained by the STAR-RIS, resulting in serious information leakage.
	Consequently, TS suffers from more SEE performance loss than ES and MS.
	Moreover, it is illustrated that the proposed NOMA schemes perform better than the OMA schemes.
	To be specific, when $P_{\max}\!=\!40$ dBm, the NOMA systems under ES, MS, TS and SF-RIS are capable of enjoying 106.5\%, 96.3\%, 111.6\%, and 107.9\% higher SEE than OMA counterparts, respectively.
	The reason behind this is twofold.
	On the one hand, the interference-free transmission implemented by the OMA also means that the suppression of information leakage by interference is weakened, which is not conducive to boost the SEE.
	On the other hand, the NOMA scheme serves all Bobs in the whole transmission phase, thus obtaining higher spectrum efficiency.

	Regarding the performance comparison between ES, MS and SF-RIS, as illustrated in Fig. \ref{P_max_SEE}, ES always performs better than the two, while SF-RIS looks the worst.
	More particularly, when $P_{\max}\!=\!40$ dBm, ES STAR-RIS assisted NOMA and OMA networks are able to provide up to 33.6\% and 34.5\% higher SEE than the conventional RISs counterparts, respectively.
	This is because, compared with ES which employs flexible amplitude and phase shift models, MS and SF-RIS have limited control of the reflection and transmission coefficients.
	Specifically, the amplitude coefficients of MS and SF-RIS are limited to binary variables.
	Moreover, SF-RIS employs fixed element-based mode selection.
	By contrast, ES can adjust the phase shift coefficients and the continuous amplitude coefficients in an element-wise manner.
	This modulated form allows ES to take full advantage of the DoFs available at each element to enhance the desired signal, mitigate inter-user interference and inhibit bilateral eavesdropping.
	Additionally, this phenomenon can be explained by the fact that MS is a special case of ES and SF-RIS is a special case of MS.
	
	
	\begin{figure}[t]
		\centering
		\subfigure[$P_r(b)=10$ mW]
		{
			\begin{minipage}[t]{0.5\textwidth}
				\centering
				\includegraphics[width=3in]{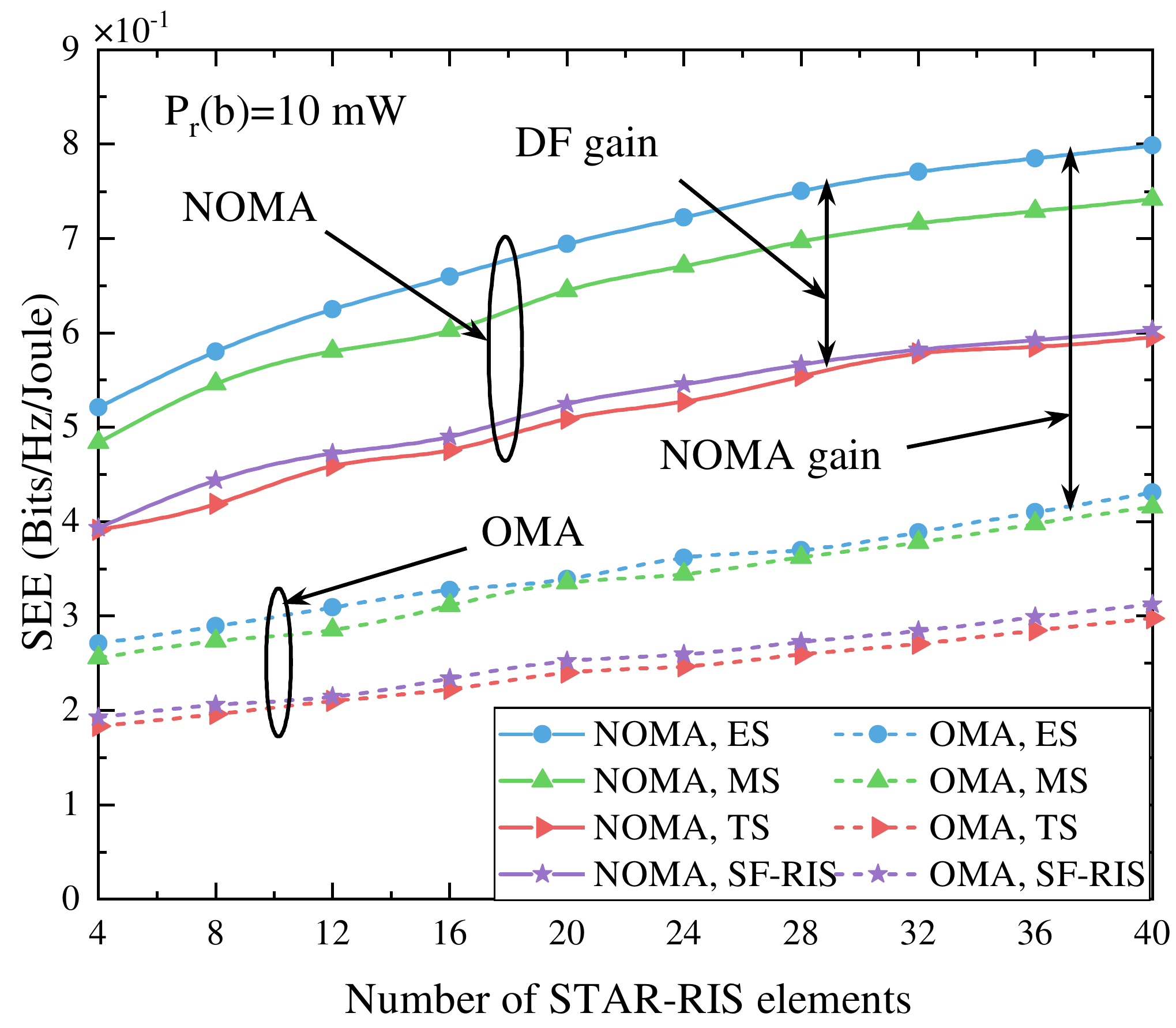}
				\label{M_SEE_1}
			\end{minipage}
		}
		\subfigure[$P_r(b)=1$ W]
		{
			\begin{minipage}[t]{0.5\textwidth}
				\centering
				\includegraphics[width=3in]{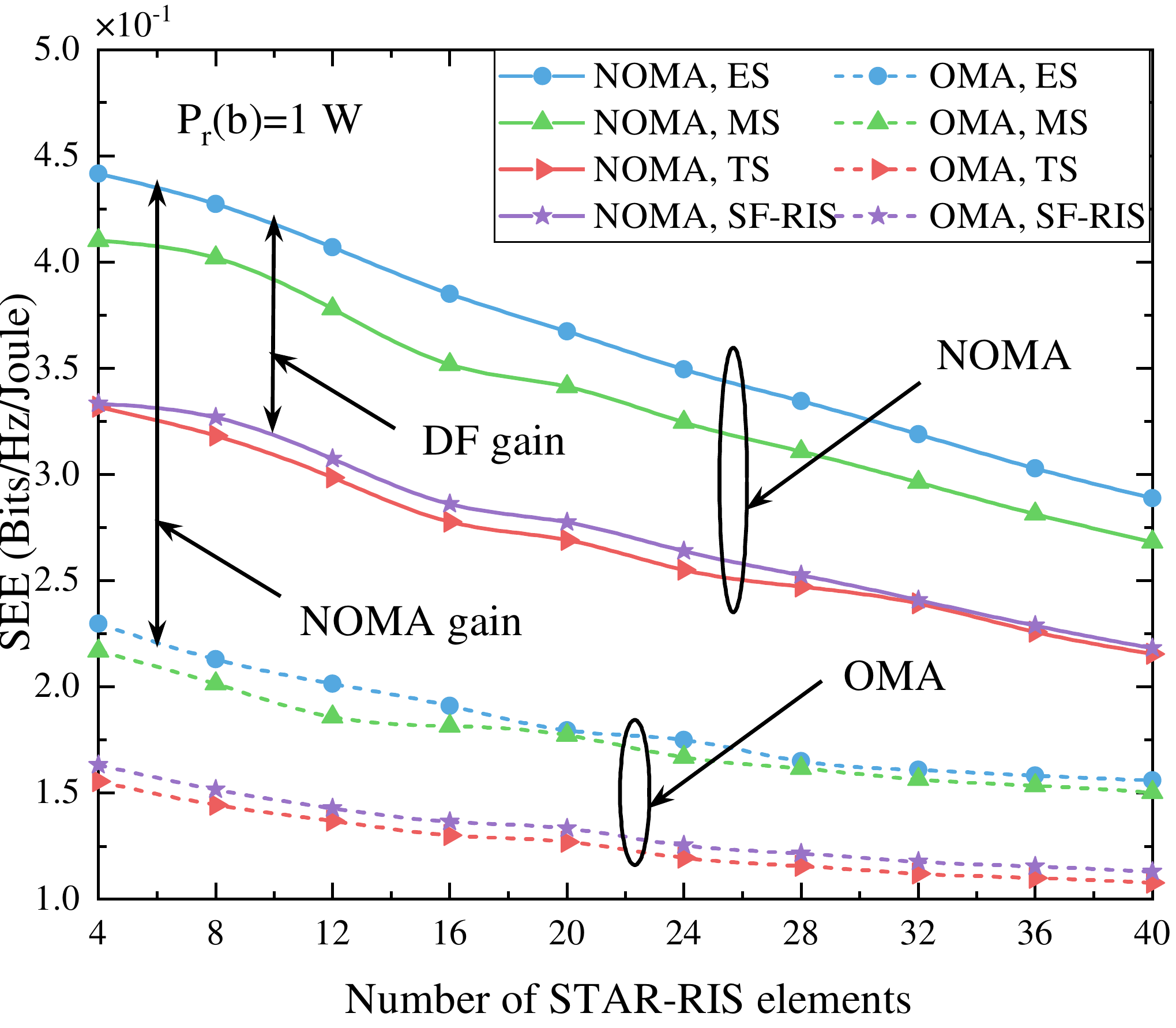}
				\label{M_SEE_2}
			\end{minipage}
		}	
		\caption{SEE versus the number of STAR-RIS elements $M$ for different bit resolution power consumption $P_r(b)$.
			The other system parameters are set as $J_t\!=\!J_r\!=\!2$, $P_{\max}=40$ dBm, and $N\!=\!5$.
		} 
		\label{M_SEE}
	\end{figure}
	
	Fig. \ref{M_SEE} characterizes the SEE versus the number of STAR-RIS elements $M$ for different bit resolution power consumption $P_r(b)$.
	As clearly shown in Fig. \ref{M_SEE_1}, all designs exhibit the same trend when $P_r(b)\!=\!10$ mW.
	Specifically, the SEE performance continues to increase for a wide range of $M$.
	This indicates that this value of $P_r(b)$ is quite small, and the increase of $M$ will not cause significant energy consumption cost.
	Furthermore, the extra spatial DoFs offered by the large-scale STAR-RIS can provide stronger cascaded channels for Bobs but substantially deteriorate the information reception at Eves.
	Unfortunately, for higher value of $P_r(b)$ in Fig. \ref{M_SEE_2}, the STAR-RIS static power consumption becomes one of the dominant items of energy efficiency.
	As a result, the SEE starts decreasing even for a small-to-moderate $M$ value.
	In other words, there exists a trade-off between the secure performance benefit of deploying the large scale STAR-RIS and its power consumption.
	In addition, it is evident that the STAR-RIS assisted NOMA systems achieve significant DF gain and NOMA gain in terms of SEE, which again demonstrates the superiority of the proposed system design.
	
	\begin{figure}[t]
		\centering
		\includegraphics[width=3in]{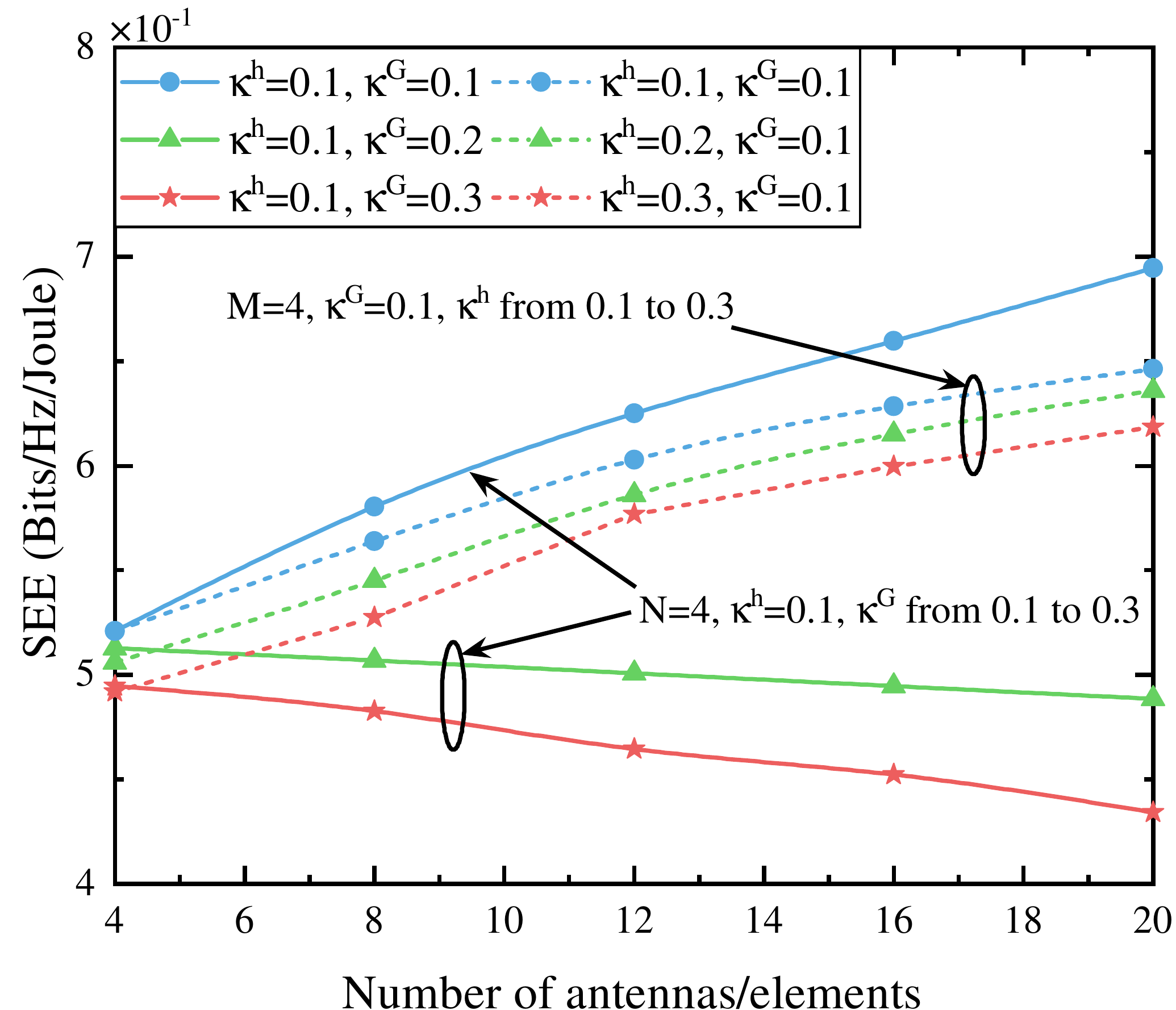}
		\caption{SEE versus the number of the BS transmit antennas $N$ and STAR-RIS elements $M$, and different CSIU levels.
			The other system parameters are set as $J_r=J_t=2$, $P_{\max}\!=\!40$ dBm, and $P_r(b)\!=\!10$ mW. }
		\label{CSI_SEE}
	\end{figure}
	\begin{figure}[t]
	\centering
	\includegraphics[width=3in]{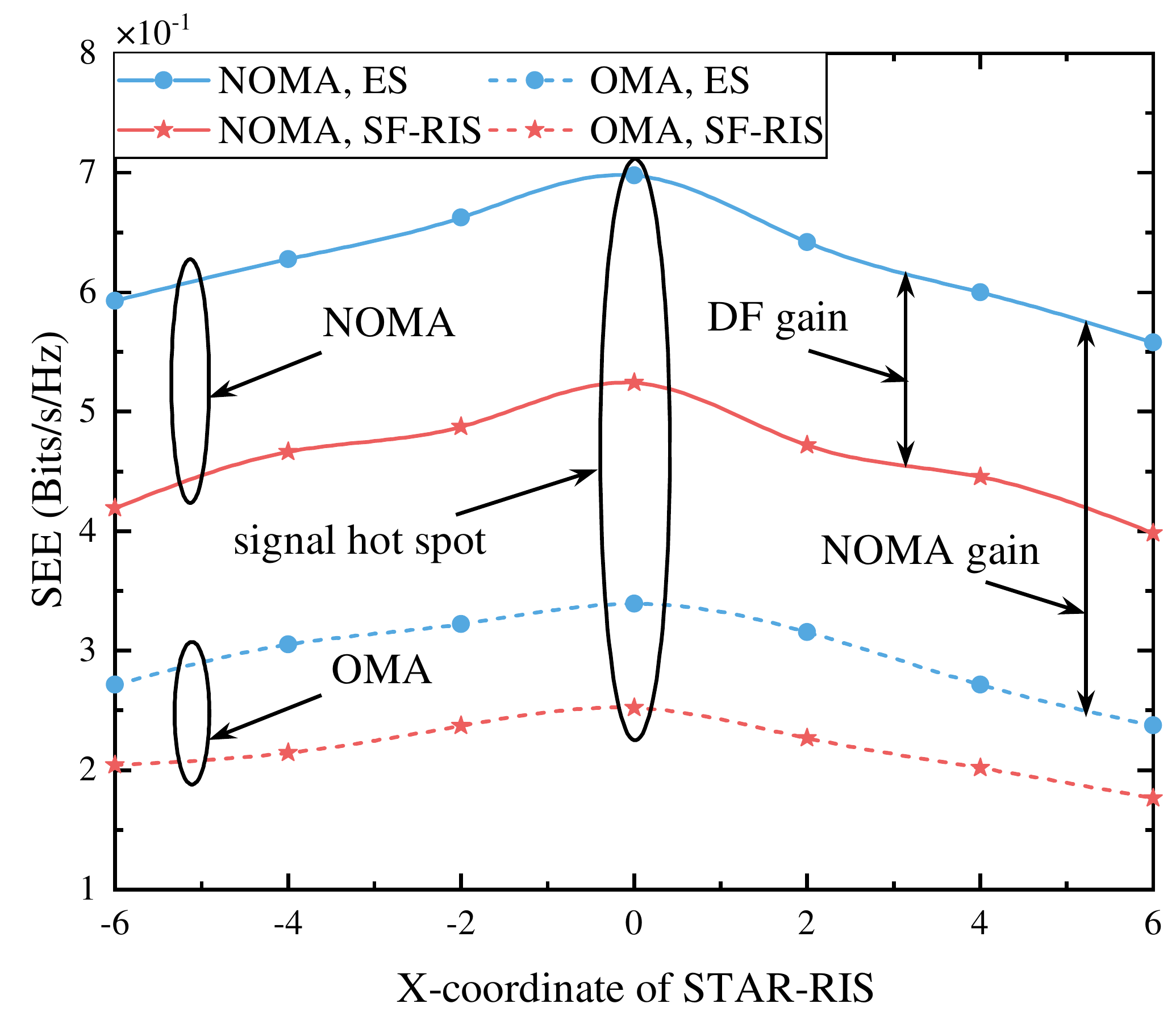}
	\caption{SEE versus the $X$-coordinate of STAR-RIS, $x_r$. The other system parameters are set as $J_r=J_t=2$, $P_{\max}=40$ dBm, $M=20$, $N=5$, and $P_r(b)=10$ mW.}
	\label{Y-coordinate_SEE}
  \end{figure}
	Fig. \ref{CSI_SEE} illustrates the SEE versus the number of the BS transmit antennas $N$ and the number of STAR-RIS elements $M$ under different CSIUs.
	Here, we choose the (NOMA, ES) scheme, as well as set $\kappa^h\!=\!(\kappa_{k,j}^h)^2\!\!=\!(\kappa_{e}^h)^2$ and $\kappa^G\!=\!(\kappa_{k,j}^G)^2\!\!=\!(\kappa_{e}^G)^2$, $\forall k,e,\forall j\in\mathcal{J}_k$.
	In Fig. \ref{CSI_SEE}, for the cases with fixed $N$, the SEE under low CSIU, e.g., $\kappa^h=\kappa^G=0.1$, increases with an increment of $M$.
	This trend is in agreement with Fig. \ref{M_SEE_1}.
	However, when $\kappa^h\!=\!0.1$,  $\kappa^G\!=\!\{0.2,0.3\}$, the SEE starts to decrease with the increase of $M$.
	This reveals that a serious channel estimation error will be introduced with a larger $M$ under high-level CSIU, which hinders the improvement of the SEE performance as $M$ grows.
	By contrast, when fixing $M$, the SEE still raises steadily as $N$ grows even at higher CSIU.
	The reason is that more DoFs to the active beamforming are offered by the increase of $N$, which is enough to compensate for the channel estimation error.
	
	Fig. \ref{Y-coordinate_SEE} illustrates the impact of STAR-RIS location on the SEE performance by varying the $X$-coordinate of STAR-RIS, $x_r$.
	It can be observed that the SEE	increases with respect to $x_r$ when $x_r\in[-6,0]$.
	This is due to the fact that the channel gain is a decreasing function of link distance.
	Specifically, with the decrease of the distances of BS-STAR-RIS and STAR-RIS-Bobs links, the channel gain increases subsequently, and the SEE performance improves  correspondingly.
	In particular, when the STAR-RIS is deployed in the vicinity of the BS and Bobs, i.e., $x_r=0$, it creates signal hot spot phenomenon for them.
	However, for STAR-RIS that continues to move in the positive direction of the horizontal axis, it will be far away from the BS and Bobs, but close to Eves. 
	Therefore, the legitimate reception is weakened while the risk of information leakage is greatly increased. 
	As a result, the SEE of all schemes shows a downward trend.
	In addition, another intuitive observation can be made that the performance of our proposed STAR-RIS assisted NOMA system is always significantly better than benchmark schemes.
	
	\begin{figure}[tbp]
		\centering
		\subfigure[$J=4:J_r=J_t=2$]
		{
			\begin{minipage}[t]{0.5\textwidth}
				\centering
				\includegraphics[width=2.5in,height=2in]{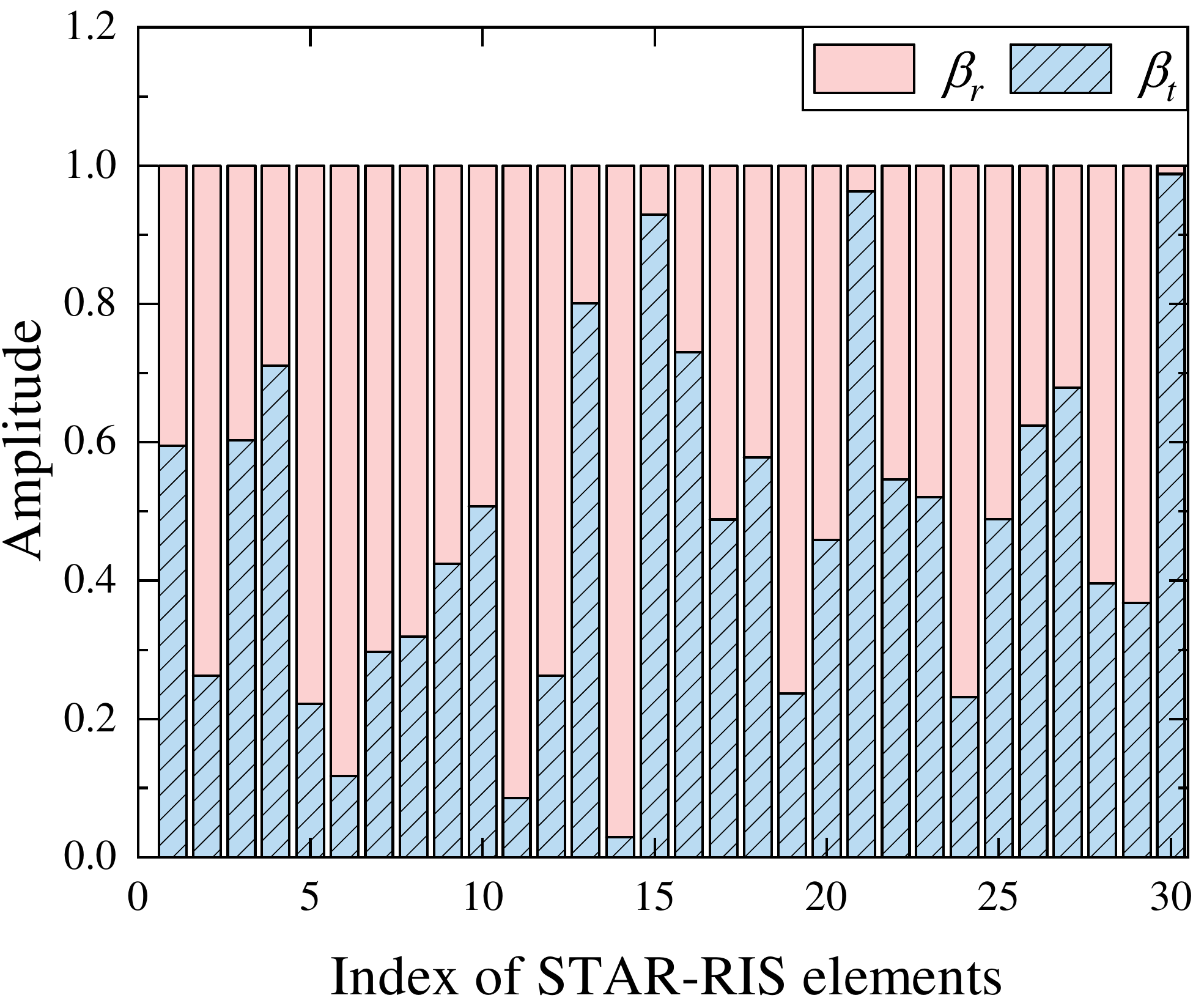}
				\label{amplitude_1}
			\end{minipage}
		}
		\subfigure[$J=6:J_r=2,J_t=4$]
		{
			\begin{minipage}[t]{0.5\textwidth}
				\centering
				\includegraphics[width=2.5in,height=2in]{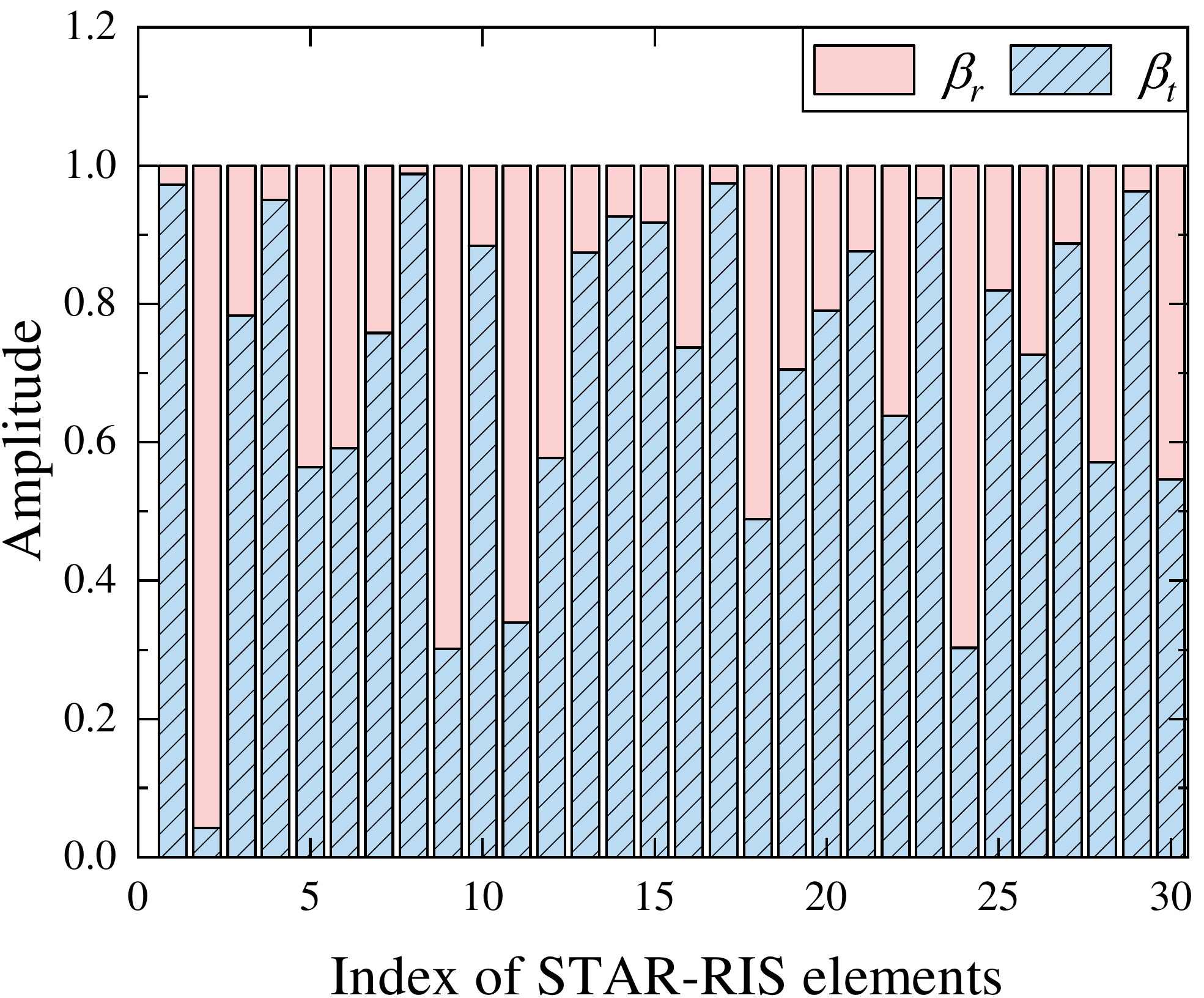}
				\label{amplitude_2}
			\end{minipage}
		}	
		\caption{The values of the transmission and reflection amplitudes for each STAR-RIS element with different number of Bobs.
			The other system parameters are set as $M\!=\!30$, $N\!=\!5$, $P_{\max}\!=\!40$ dBm, and $P_r(b)\!=\!10$ mW.
		} 
		\label{fig:amplitude}
	\end{figure}
	Lastly, the values of the transmission and reflection amplitudes for each STAR-RIS element with different number of Bobs are plotted in Fig. \ref{fig:amplitude}.
	Note that both Fig. \ref{amplitude_1} and Fig. \ref{amplitude_2} are based on the (NOMA, ES) scheme.
	Obviously, compared with Fig. \ref{amplitude_1}, the energy allocated to the transmission amplitudes $\{\beta_{m}^t\}$ in Fig. \ref{amplitude_2} are much more than that allocated to the reflection amplitudes $\{\beta_{m}^r\}$.
	This is expected because if there are more Bobs in the transmission space in Fig. \ref{amplitude_2}, STAR-RIS prefers to allocate more energy to this space, thereby enhancing the SEE performance of each Bob.
	Notably, these results in Fig. \ref{fig:amplitude} also reveal that STAR-RISs can provide new DoFs for system designs, which allows us to flexibly optimize the transmission and reflection coefficients according to the application scenarios.
	\section{Conclusion}\label{Conclusion}
	This paper investigated the full-space secure communication system using a novel reconfigurable dual-functional surface. 
	In particular, we took STAR-RIS as an example to study secure and energy-efficient communications in NOMA networks.
	Under three operating protocols of STAR-RISs, different non-convex resource allocation problems were formulated for the SEE maximization while considering imperfect CSI.
	Relying on the AO algorithm, the formulated problems were efficiently solved in the ES and MS cases.
	For the TS case, a two-layer algorithm combining bisection search and iterative algorithms was proposed.
	Simulation results verified the SEE advantages of the proposed system over conventional RIS assisted systems.
	Additionally, three remarks can be drawn from numerical results, providing helpful insights for practical system designs:
	1) SEE increases rapidly with the maximum transmit power and gradually flattens, indicating that all the available BS transmit power should be utilized in the low power domain, but once saturated, no further power is required;
	2) It is necessary to select an appropriate protocol based on the application scenario, such as employing ES and MS for sufficient power and TS for limited power;
	3) Deploying large-scale DF-RISs does not always improve the SEE, which also depends on the CSIU level and the static power consumption of DF-RISs.
	
	In addition to assisting secure communications, there are other potential applications for DF-RISs, such as coverage extension, coordinated multi-point communication, as well as localization and sensing\cite{STAR-Liu,IOS-Zhang}.
	However, the introducing of DF-RISs brings new research challenges, some of which are listed as follows:
		\begin{itemize}
			\item \emph{Channel acquisition:}  	
			Since both reflection and transmission coefficients need to be jointly designed with pilot sequences, channel acquisition methods for SF-RISs are difficult to be directly applied to DF-RISs\cite{CSI estimation-STAR-RIS}. 
			In particular, given the fact that the STAR-RIS phase shifts are intricately coupled in practice, the channel acquisition is more complicated and deserves further investigation.
			
			\item \emph{DF-RIS deployment:}
			The performance gains achieved by DF-RIS depend on its relative location to the BS and users, especially for multi-user scenarios.
			Identifying the optimal deployment of DF-RIS by balancing the number of users in reflecting and transmitting half-spaces is mandatory, and therefore a promising research direction.
		\end{itemize}  
	
		\section*{Appendix A: Proof of Proposition 1} \label{APPENDIX A}
		According to the FTS, the following inequality holds
		\begin{eqnarray}
			\label{Appendix_A_1}
			\left|x\right|^2\geq 2{\rm Re}\{(x^{(\ell)})^{\ast}x\}-(x^{(\ell)})^{\ast}x^{(\ell)},
		\end{eqnarray}
		where $x$ is a complex scalar variable and $x^{(\ell)}$ represents a feasible point in the $\ell$-th iteration.
		Then, by replacing  $x$ and $x^{(\ell)}$ in (\ref{Appendix_A_1}) with $(\mathbf{h}_{k,l}^{\mathrm H}+\mathbf{u}_{k}^{\mathrm H} \mathbf{G}_{k,l})\mathbf{w}_{k,j}$ and $(\mathbf{h}_{k,l}^{\mathrm H}+(\mathbf{u}_{k}^{(\ell)})^{\mathrm H} \mathbf{G}_{k,l})\mathbf{w}_{k,j}^{(\ell)}$, respectively, a lower bound of the convex term $|\bar{\mathbf{h}}_{k,l} \mathbf{w}_{k,j}|^2$ can be obtained as
		\begin{eqnarray}
			\label{P2_C_1.1_proof}
			\nonumber
			\!\!\!\!\!\!\!\!\!\!\!\!\!\!\!\!&{}&2 {\rm Re} \big\{\underbrace{(\mathbf{h}_{k,l}^{\mathrm H}+ (\mathbf{u}_{k}^{(\ell)})^{\mathrm H} \mathbf{G}_{k,l})\mathbf{w}_{k,j}^{(\ell)}\mathbf{w}_{k,j}^{\mathrm H} (\mathbf{h}_{k,l}+\mathbf{G}_{k,l}^{\mathrm H}\mathbf{u}_{k})}_{f_1}\big\}\\
			\!\!\!\!\!\!\!\!\!\!\!\!\!\!\!\!&{}& -\underbrace{(\mathbf{h}_{k,l}^{\mathrm H}+(\mathbf{u}_{k}^{(\ell)})^{\mathrm H} \mathbf{G}_{k,l})\mathbf{w}_{k,j}^{(\ell)}(\mathbf{w}_{k,j}^{(\ell)})^{\mathrm H}(\mathbf{h}_{k,l}\!+\!\mathbf{G}_{k,l}^{\mathrm H}\mathbf{u}_{k}^{(\ell)})}_{f_2}.
		\end{eqnarray}
		Moreover, by substituting $\mathbf{h}_{k,l}=\hat{\mathbf{h}}_{k,l}+\triangle\mathbf{h}_{k,l}$ and  $\mathbf{G}_{k,l}=\hat{\mathbf{G}}_{k,l}+\triangle\mathbf{G}_{k,l}$, $\forall k, \forall j\in\mathcal{J}_k,\forall l\in\mathcal{L}_k$, into (\ref{P2_C_1.1_proof}) and performing further mathematical transformations\cite{Appendix-derivation,TSP-A framework}, the expression $f_1$ in (\ref{P2_C_1.1_proof}) can be reformulated as (\ref{Appendix_A_2}) at the top of the next page.
		\begin{figure*}[tp]
				\begin{eqnarray}
					\nonumber
					f_1~~\!\!\!\!\!\!\!\!\!\!\!\!\!\!&{}&=[(\hat{\mathbf{h}}_{k,l}^{\mathrm{H}}+\triangle\mathbf{h}_{k,l}^{\mathrm{H}})+ (\mathbf{u}_{k}^{(\ell)})^{\mathrm H}(\hat{\mathbf{G}}_{k,l}+\triangle\mathbf{G}_{k,l} )]\mathbf{w}_{k,j}^{(\ell)}\mathbf{w}_{k,j}^{\mathrm H} [(\hat{\mathbf{h}}_{k,l}+\triangle\mathbf{h}_{k,l})+(\hat{\mathbf{G}}_{k,l}^{\mathrm H}+\triangle\mathbf{G}_{k,l}^{\mathrm H})\mathbf{u}_{k}]\\
					\nonumber
					\!\!\!\!\!\!\!\!\!\!\!\!\!\!&{}&=(\hat{\mathbf{h}}_{k,l}^{\mathrm{H}}+(\mathbf{u}_{k}^{(\ell)})^{\mathrm H}\hat{\mathbf{G}}_{k,l})\mathbf{w}_{k,j}^{(\ell)}\mathbf{w}_{k,j}^{\mathrm H}
					(\hat{\mathbf{h}}_{k,l}^{\mathrm{H}}+\hat{\mathbf{G}}_{k,l}^{\mathrm H}\mathbf{u}_{k})
					+(\hat{\mathbf{h}}_{k,l}^{\mathrm{H}}+(\mathbf{u}_{k}^{(\ell)})^{\mathrm H}\hat{\mathbf{G}}_{k,l})\mathbf{w}_{k,j}^{(\ell)}\mathbf{w}_{k,j}^{\mathrm H}
					(\triangle\mathbf{h}_{k,l}+\triangle\mathbf{G}_{k,l}^{\mathrm H}\mathbf{u}_{k})\\
					\nonumber
					\!\!\!\!\!\!\!\!\!\!\!\!\!\!&{}&~~+(\triangle\mathbf{h}_{k,l}^{\mathrm{H}}+(\mathbf{u}_{k}^{(\ell)})^{\mathrm H}\triangle\mathbf{G}_{k,l})\mathbf{w}_{k,j}^{(\ell)}\mathbf{w}_{k,j}^{\mathrm H}
					(\hat{\mathbf{h}}_{k,l}+\hat{\mathbf{G}}_{k,l}^{\mathrm H}\mathbf{u}_{k})
					+(\triangle\mathbf{h}_{k,l}^{\mathrm{H}}+(\mathbf{u}_{k}^{(\ell)})^{\mathrm H}\triangle\mathbf{G}_{k,l})\mathbf{w}_{k,j}^{(\ell)}\mathbf{w}_{k,j}^{\mathrm H}
					(\triangle\mathbf{h}_{k,l}+\triangle\mathbf{G}_{k,l}^{\mathrm H}\mathbf{u}_{k})\\
					\nonumber
					\!\!\!\!\!\!\!\!\!\!\!\!\!\!&{}&=(\hat{\mathbf{h}}_{k,l}^{\mathrm{H}}+(\mathbf{u}_{k}^{(\ell)})^{\mathrm H}\hat{\mathbf{G}}_{k,l})\mathbf{w}_{k,j}^{(\ell)}\mathbf{w}_{k,j}^{\mathrm H}
					(\hat{\mathbf{h}}_{k,l}^{\mathrm{H}}+\hat{\mathbf{G}}_{k,l}^{\mathrm H}\mathbf{u}_{k})
					+(\hat{\mathbf{h}}_{k,l}^{\mathrm{H}}+(\mathbf{u}_{k}^{(\ell)})^{\mathrm H}\hat{\mathbf{G}}_{k,l})\mathbf{w}_{k,j}^{(\ell)}\mathbf{w}_{k,j}^{\mathrm H}\triangle\mathbf{h}_{k,l}\\
					\nonumber
					\!\!\!\!\!\!\!\!\!\!\!\!\!\!&{}&~~+{\rm vec}^{\mathrm H}{(\triangle\mathbf{G}_{k,l})}{\rm vec}(\mathbf{u}_{k}(\hat{\mathbf{h}}_{k,l}^{\mathrm{H}}+(\mathbf{u}_{k}^{(\ell)})^{\mathrm H}\hat{\mathbf{G}}_{k,l})\mathbf{w}_{k,j}^{(\ell)}\mathbf{w}_{k,j}^{\mathrm H})
					+\triangle\mathbf{h}_{k,l}^{\mathrm{H}}\mathbf{w}_{k,j}^{(\ell)}\mathbf{w}_{k,j}^{\mathrm H}
					(\hat{\mathbf{h}}_{k,l}+\hat{\mathbf{G}}_{k,l}^{\mathrm{H}}\mathbf{u}_{k})
					+\triangle\mathbf{h}_{k,l}^{\mathrm{H}}\mathbf{w}_{k,j}^{(\ell)}\mathbf{w}_{k,j}^{\mathrm H}\triangle\mathbf{h}_{k,l}\\
					\nonumber
					\!\!\!\!\!\!\!\!\!\!\!\!\!\!&{}&~~+{\rm vec}^{\mathrm H}(\mathbf{u}_{k}^{(\ell)}(\hat{\mathbf{h}}_{k,l}+\mathbf{u}_{k}^{\mathrm{H}}\hat{\mathbf{G}}_{k,l}^{\mathrm{H}})\mathbf{w}_{k,j}(\mathbf{w}_{k,j}^{(\ell)})^{\mathrm H}
					){\rm vec}(\triangle\mathbf{G}_{k,l})
					+{\rm vec}^{\mathrm H}(\triangle\mathbf{G}_{k,l})(\mathbf{w}_{k,j}^{\ast}(\mathbf{w}_{k,j}^{(\ell)})^{\mathrm T}\otimes \mathbf{u}_{k})\triangle\mathbf{h}_{k,l}^{\ast}\\
					\nonumber
					\!\!\!\!\!\!\!\!\!\!\!\!\!\!&{}&~~+\triangle\mathbf{h}_{k,l}^{\mathrm T}(\mathbf{w}_{k,j}^{\ast}(\mathbf{w}_{k,j}^{(\ell)})^{\mathrm T}\otimes (\mathbf{u}_{k}^{(\ell)})^{\mathrm H}{\rm vec}(\triangle\mathbf{G}_{k,l}) 
					+{\rm vec}^{\mathrm H}(\triangle\mathbf{G}_{k,l})( \mathbf{w}_{k,j}^{\ast}(\mathbf{w}_{k,j}^{(\ell)})^{\mathrm T}\otimes \mathbf{u}_{k} (\mathbf{u}_{k}^{(\ell)})^{\mathrm H}){\rm vec}(\triangle\mathbf{G}_{k,l})\\
					\label{Appendix_A_2}
					\!\!\!\!\!\!\!\!\!\!\!\!\!\!&{}&=(\mathbf{x}_{k,j}^l)^{\mathrm H}\mathbf{\widetilde{A}}_{k,j}{\mathbf{x}_{k,j}^l}
					+(\mathbf{\widetilde{a}}_{k,j}^l)^{\mathrm H}{\mathbf{x}_{k,j}^l}+(\mathbf{x}_{k,j}^l)^{\mathrm H}\mathbf{\widehat{a}}_{k,j}^l+\widetilde{a}_{k,j}^l.
				\end{eqnarray}
			\hrulefill
		\end{figure*}
		Similarly, the term $f_2$ in (\ref{P2_C_1.1_proof}) is rewritten as
		\begin{eqnarray}
			\label{Appendix_A_3}
			(\mathbf{x}_{k,j}^l)^{\mathrm H}\mathbf{\widehat{A}}_{k,j}\mathbf{x}_{k,j}^l+(\mathbf{\bar{a}}_{k,j}^l)^{\mathrm H}\mathbf{x}_{k,j}^l+(\mathbf{x}_{k,j}^l)^{\mathrm H}\mathbf{\bar{a}}_{k,j}^l+\widehat{a}_{k,j}^l.
		\end{eqnarray}
		Finally, we can obtain (\ref{C_proposition1}), which completes the proof.
	
		\section*{Appendix B: Proof of Proposition 2} \label{APPENDIX B}
		Based on the Schur's complement Lemma\cite{schur}, constraint (\ref{P2_C_1.2}) can be reformulated as 
		\begin{eqnarray}
			\label{Appendix-B-1}
			\!\!\!\!\!&\mathbf{0} \preceq\left[\begin{array}{cc}\!\!\!\!
				\eta_{k,j}^l-\sigma^{2} & \!\!\!\!\bar{\mathbf{h}}_{k,l} \mathbf{w}_{k,-j}\\
				(\bar{\mathbf{h}}_{k,l} \mathbf{w}_{k,-j})^{\mathrm H}& \!\!\!\!\mathbf{I}
			\end{array}\!\!\right], \forall k, \forall j\!\in\!\mathcal{J}_k, \forall l\!\in\!\mathcal{L}_k.
		\end{eqnarray}
		Then, by using $\bar{\mathbf{h}}_{k,l} \mathbf{w}_{k,-j}=(\mathbf{h}_{k,l}^{\mathrm H}+\mathbf{u}_{k}^{\mathrm H} \mathbf{G}_{k,l})\mathbf{w}_{k,-j}$, and inserting $\mathbf{h}_{k,l}=\hat{\mathbf{h}}_{k,l}+\triangle\mathbf{h}_{k,l}$ and  $\mathbf{G}_{k,l}=\hat{\mathbf{G}}_{k,l}+\triangle\mathbf{G}_{k,l}$, $\forall k, \forall j\in\mathcal{J}_k, \forall l\in\mathcal{L}_k$, into (\ref{Appendix-B-1}), (\ref{Appendix-B-1}) can be rewritten as 
		\begin{eqnarray}
			\label{Appendix-B-2}
			\nonumber
			&{}&\!\!\!\!\!\!\!\!\!\!\mathbf{0} \preceq\left[\begin{array}{cc}
				\eta_{k,j}^l-\sigma^{2} & (\boldsymbol{\pi}_{k,j}^l)^{\mathrm{H}} \\
				\boldsymbol{\pi}_{k,j}^l & \mathbf{I}
			\end{array}\right]	\\
			\nonumber
			&{}&\!\!\!\!\!\!\!\!\!\!+\!\left[\!\!\begin{array}{cc}\!\!\!\!
				0 & \!\!\!\!\! (\triangle \mathbf{h}_{k,l}^{\mathrm{H}}\!+\!\mathbf{u}_{k}^{\mathrm{H}} \triangle \mathbf{G}_{k,l}) \mathbf{w}_{k,-j} \\
				\mathbf{w}_{k,-j}^{\mathrm{H}}(\triangle \mathbf{h}_{k,l}\!+\!\triangle \mathbf{G}_{k,l}^{\mathrm{H}} \mathbf{u}_{k}) & \!\!\!\!\! \mathbf{0}
			\end{array}\!\!\right],\\
			&{}&\!\!\!\!\!\!\!\!\forall k, \forall j\in\mathcal{J}_k, \forall l\in\mathcal{L}_k.
		\end{eqnarray}
		Consequently, (\ref{Appendix-B-2}) can be finally converted into (\ref{P2_C_1.3.1}) by performing simple matrix operations and the proof is completed.


\begin{thebibliography}{99}
		\bibliographystyle{IEEEtran}
		\bibitem{2022-6G}
			K. Nikitopoulos, ``Massively parallel, nonlinear processing for 6G: Potential gains and further research challenges," \emph{IEEE Commun, Mag.}, vol. 60, no. 1, pp. 81-87, Jan. 2022.
		\bibitem{6G_vision}
		W. Saad, M. Bennis, and M. Chen, ``A vision of 6G wireless systems: Applications, trends, technologies, and open research problems," \emph{IEEE Netw.}, vol. 34, no. 3, pp. 134-142, May 2020.
		
		\bibitem{RIS-review-1}
			M. A. {ElMossallamy}, H. {Zhang}, L. {Song}, K. G. {Seddik}, Z. {Han}, and G. Y. {Li}, ``Reconfigurable intelligent surfaces for wireless communications: Principles, challenges, and opportunities,'' \emph{IEEE Trans. Cogn. Commun. Netw.}, vol. 6, no. 3, pp. 990-1002, Sept. 2020.
			\bibitem{RIS-review-2}
			Q. {Wu} and R. {Zhang}, ``Towards smart and reconfigurable environment: Intelligent reflecting surface aided wireless network,'' \emph{IEEE Commun. Mag.}, vol. 58, no. 1, pp. 106-112, Jan. 2020.
		\bibitem{RIS-review-3}
		C. {Huang}, A. {Zappone}, G. C. {Alexandropoulos}, M. {Debbah}, and C. {Yuen}, ``Reconfigurable intelligent surfaces for energy efficiency in wireless communication," \emph{IEEE Trans. Wireless Commun.}, vol. 18, no. 8, pp. 4157-4170, Aug. 2019.
		
			\bibitem{JSAC-SRE}
			M. Di Renzo  \textit{et al.}, ``Smart radio environments empowered by reconfigurable intelligent surfaces: How it works, state of research, and the road ahead," \emph{IEEE J. Sel. Areas Commun.}, vol. 38, no. 11, pp. 2450-2525, Nov. 2020.
		\bibitem{Wen-WCL_SOP}
		W. Wang, H. Tian, and W. Ni, ``Secrecy performance analysis of IRS-aided UAV relay system," \emph{IEEE Wireless Commun. Lett.}, vol. 10, no. 12, pp. 2693-2697, Dec. 2021.
		
		\bibitem{STAR-CL}
		J. {Xu}, Y. {Liu}, X. {Mu}, and O. A. {Dobre.}, ``STAR-RISs: Simultaneous transmitting and reflecting reconfigurable intelligent surfaces," \emph{IEEE Commun. Lett.}, vol. 25, no. 9, pp. 3134-3138, Sept. 2021.
		\bibitem{STAR-IOS}
		J. Xu \textit{et al.}, ``Simultaneously transmitting and reflecting (STAR) intelligent omni-surfaces, their modeling and implementation," Sept. 2021. [Online]. Available: https://arxiv.org/abs/2108.06233v2.
		\bibitem{STAR-Liu}
		Y. {Liu} \textit{et al.}, ``STAR: Simultaneous transmission and reflection for $360^{\circ}$ coverage by intelligent surfaces," \emph{IEEE Wireless Commun.}, vol. 28, no. 6, pp. 102-109, Dec. 2021.
		\bibitem{STAR-RIS-TWC}
		X. Mu, Y. Liu, L. Guo, J. Lin, and R. Schober, "Simultaneously transmitting and reflecting (STAR) RIS aided wireless communications," \emph{IEEE Trans. Wireless Commun.}, 2021, early access, doi: 10.1109/TWC.2021.3118225.
		
		\bibitem{IOS-TVT}
		S. {Zhang}, H. {Zhang}, B. {Di}, Y. {Tan}, Z. {Han}, and L. {Song}, ``Beyond intelligent reflecting surfaces: Reflective-transmissive metasurface aided communications for full-dimensional coverage extension,'' \emph{IEEE Trans. Veh. Technol.}, vol. 69, no. 11, pp. 13905-13909, Nov. 2020.
		
			\bibitem{NOMA-Mag}
			A. S. d. Sena \emph{et al.}, ``What role do intelligent reflecting surfaces play in multi-antenna non-orthogonal multiple access?" \emph{IEEE Wireless Commun.}, vol. 27, no. 5, pp. 24-31, Oct. 2020.
			\bibitem{NOMA-WCNC}
			G. Yang, X. Xu, and Y.-C. Liang, ``Intelligent reflecting surface assisted non-orthogonal multiple access," in \emph{Proc. IEEE Wireless Commun. Netw. Conf. (WCNC)}, Seoul, South Korea, May 2020, pp. 1-6.
			\bibitem{Ding-NOMA}
			Z. Ding and H. V. Poor, ``A simple design of IRS-NOMA transmission," \emph{IEEE Commun. Lett.}, vol. 24, no. 5, pp. 1119-1123, May 2020.
		\bibitem{NOMA_RIS_OMA}
		B. Zheng, Q. Wu, and R. Zhang, ``Intelligent reflecting surface-assisted multiple access with user pairing: NOMA or OMA?" \emph{IEEE Commun. Lett.}, vol. 24, no. 4, pp. 753-757, Apr. 2020.
		
		\bibitem{NOMA_RIS_Liu}
		Y. Liu, X. Mu, X. Liu, M. Di Renzo, Z. Ding, and R. Schober, ``Reconfigurable intelligent surface (RIS) aided multi-user networks: Interplay between NOMA and RIS,” Oct. 2021. [Online]. Available: https://arxiv.org/abs/2011.13336v2.
		\bibitem{NOMA_RIS_UAV}
		X. Mu, Y. Liu, L. Guo, J. Lin, and H. V. Poor, ``Intelligent reflecting surface enhanced multi-UAV NOMA networks," \emph{IEEE J. Sel. Areas Commun.}, vol. 39, no. 10, pp. 3051-3066, Oct. 2021.
		
		\bibitem{Ni_NOMA_RIS_TWC}
		W. Ni, X. Liu, Y. Liu, H. Tian, and Y. Chen, ``Resource allocation for multi-cell IRS-aided NOMA networks," \emph{IEEE Trans. Wireless Commun.}, vol. 20, no. 7, pp. 4253-4268, Jul. 2021.
		\bibitem{Mu_NOMA_RIS_TWC}
		X. Mu, Y. Liu, L. Guo, J. Lin, and N. Al-Dhahir, ``Exploiting intelligent reflecting surfaces in NOMA networks: Joint beamforming optimization," \emph{IEEE Trans. Wireless Commun.}, vol. 19, no. 10, pp. 6884-6898, Oct. 2020.
		\bibitem{SEE_TVT}
		F. Fang, Y. Xu, Q. -V. Pham, and Z. Ding, ``Energy-efficient design of IRS-NOMA networks," \emph{IEEE Trans. Veh. Technol.}, vol. 69, no. 11, pp. 14088-14092, Nov. 2020.
		
		\bibitem{NOMA-EE-deployment}
		X. Liu, Y. Liu, Y. Chen, and H. V. Poor, ``RIS enhanced massive non-orthogonal multiple access networks: Deployment and passive beamforming design," \emph{IEEE J. Sel. Areas Commun.}, vol. 39, no. 4, pp. 1057-1071, Apr. 2021.
		
		\bibitem{RIS-secure-GC}
		X. Yu, D. Xu, and R. Schober, ``Enabling secure wireless communications via intelligent reflecting surfaces," in \emph{Proc. IEEE Global Commun. Conf. (GLOBECOM)}, Waikoloa, HI, USA, Dec. 2019, pp. 1-6.
		\bibitem{RIS-secure-CL}
		H. Shen, W. Xu, S. Gong, Z. He, and C. Zhao, ``Secrecy rate maximization for intelligent reflecting surface assisted multi-antenna communications,”\emph{IEEE Commun. Lett.}, vol. 23, no. 9, pp. 1488-1492,
		Sept. 2019.
		\bibitem{RIS-secure-WCL}
		M. Cui, G. Zhang, and R. Zhang, ``Secure wireless communication via intelligent reflecting surface," \emph{IEEE Wireless Commun. Lett.}, vol. 8, no. 5, pp. 1410-1414, Oct. 2019.
		
		\bibitem{NOMA_RIS_secure_EL}
		L. Yang and Y. Yuan, ``Secrecy outage probability analysis for RIS assisted NOMA systems," \emph{Electronics Letters}, vol. 56, no. 23, pp. 1254-1256, Nov. 2020.
		\bibitem{NOMA_RIS_secure_CL}
		Z. Tang, T. Hou, Y. Liu, J. Zhang, and C. Zhong, ``A novel design of RIS for enhancing the physical layer security for RIS-aided NOMA networks," \emph{IEEE Wireless Commun. Lett.}, vol. 10, no. 11, pp. 2398-2401, Nov. 2021.
		
			\bibitem{NOMA_RIS_secure_AN}
			Z. Zhang, C. Zhang, C. Jiang, F. Jia, J. Ge, and F. Gong, ``Improving physical layer security for reconfigurable intelligent surface aided NOMA 6G networks," \emph{IEEE Trans. Veh. Technol.}, vol. 70, no. 5, pp. 4451-4463, May 2021.
		
		\bibitem{NOMA_RIS_secure_robust}
		Z. Zhang, L. Lv, Q. Wu, H. Deng, and J. Chen, ``Robust and secure communications in intelligent reflecting surface assisted NOMA networks," \emph{IEEE Commun. Lett.}, vol. 25, no. 3, pp. 739-743, Mar. 2021.
			\bibitem{Eve-SIC}
			Z. Zhang, J. Chen, Q. Wu, Y. Liu, L. Lv, and X. Su, "Securing NOMA networks by exploiting intelligent reflecting surface," \emph{IEEE Trans. Commun.}, vol. 70, no. 2, pp. 1096-1111, Feb. 2022.
		
		\bibitem{CL_IOS_secure}
		H. Niu, Z. Chu, F. Zhou, and Z. Zhu, ``Simultaneous transmission and reflection reconfigurable intelligent surface assisted secrecy MISO networks," \emph{IEEE	Commun. Lett.}, vol. 25, no. 11, pp. 3498-3502, Nov. 2021.
		\bibitem{IOS-Zhang}
		H. Zhang \textit{et al.}, ``Intelligent omni-surfaces for full-dimensional wireless communications: Principle, technology, and implementation," \emph{IEEE Commun. Mag.}, vol. 60, no. 2, pp. 39-45, Feb. 2022.
		
		\bibitem{STAR-RIS-NOMA-CL}
		C. Wu, Y. Liu, X. Mu, X. Gu, and O. A. Dobre, ``Coverage characterization of STAR-RIS networks: NOMA and OMA," \emph{IEEE Commun. Lett.}, vol. 25, no. 9, pp. 3036-3040, Sept. 2021.
		\bibitem{STAR-RIS-NOMA-Zuo}
		J. Zuo, Y. Liu, Z. Ding, L. Song, and H. V. Poor, ``Joint design for simultaneously transmitting and reflecting (STAR) RIS assisted NOMA systems," Jun. 2021. [Online]. Available: https://arxiv.org/abs/2106.03001v1.
		\bibitem{STAR-RIS-NOMA-Hou}
		T. Hou, J. Wang, Y. Liu, X. Sun, A. Li, and B. Ai, ``A joint design for STAR-RIS enhanced NOMA-CoMP networks: A simultaneous-signal-enhancement-and-cancellation-based (SSECB) design," \emph{IEEE Trans. Veh. Technol.}, vol. 71, no. 1, pp. 1043-1048, Jan. 2022.
		\bibitem{STAR-RIS-NOMA-uplink}
		J. Zuo, Y. Liu, Z. Ding, and X. Wang, ``Uplink NOMA for STAR-RIS networks," Oct. 2021. [Online]. Available: https://arxiv.org/abs/2110.05686.
			\bibitem{STAR-Coupled}
			Y. Liu, X. Mu, R. Schober, and H. V. Poor, ``Simultaneously transmitting and reflecting (STAR)-RISs: A coupled phase-shift model," Oct. 2021. [Online]. Available: https://arxiv.org/abs/2110.02374v1.
		\bibitem{Wen-IOS-TVT}
		W. Wang, W. Ni, H. Tian, and L. Song, ``Intelligent omni-surface enhanced aerial secure offloading," \emph{IEEE Trans. Veh. Technol.}, 2022, early access, doi: 10.1109/TVT.2022.3150769.
		\bibitem{TVT-NOMA-decoding}
		P. Liu, Y. Li, W. Cheng, X. Gao, and X. Huang, ``Intelligent reflecting surface aided NOMA for millimeter-wave massive MIMO with lens antenna array," \emph{IEEE Trans. Veh. Technol.}, vol. 70, no. 5, pp. 4419-4434, May 2021.
		\bibitem{EE_SCA}
		H. M. {Al-Obiedollah}, K. {Cumanan}, J. Thiyagalingam, A. G. Burr, Z. Ding, and O. A. Dobre, ``Energy efficient beamforming design for MISO non-orthogonal multiple access systems," \emph{IEEE Trans. Commun.}, vol. 67, no. 6, pp. 4117-4131, Jun. 2019.
		\bibitem{RIS-TWC-Rate}
		W. Wang \textit{et al.}, ``Beamforming and jamming optimization for IRS-aided secure NOMA networks," \emph{IEEE Trans. Wireless Commun.}, 2021, early access, doi: 10.1109/TWC.2021.3104856.
		
		\bibitem{TSP-A framework}
		G. Zhou, C. Pan, H. Ren, K. Wang, and A. Nallanathan, ``A framework	of robust transmission design for IRS-aided MISO communications with imperfect cascaded channels," \emph{IEEE Trans. Signal Process.}, vol. 68, pp. 5092-5106, Jan. 2020.
		
			\bibitem{CSI-1}
			Z.-Q. He and X. Yuan, ``Cascaded channel estimation for large intelligent metasurface assisted massive MIMO," \emph{IEEE Wireless Commun. Lett.}, vol. 9, no. 2, pp. 210-214, Feb. 2020.
			\bibitem{CSI-2}
			H. Liu, X. Yuan, and Y.-J. A. Zhang, ``Matrix-calibration-based cascaded channel estimation for reconfigurable intelligent surface assisted multiuser MIMO," \emph{IEEE J. Sel. Areas Commun.}, vol. 38, no. 11, pp. 2621-2636, Nov. 2020.
			\bibitem{CSI-3}
			Z. Wang, L. Liu, and S. Cui, ``Channel estimation for intelligent reflecting surface assisted multiuser communications: Framework, algorithms, and analysis," \emph{IEEE Trans. Wireless Commun.}, vol. 19, no. 10, pp. 6607-6620, Oct. 2020.
		
			\bibitem{Appendix-derivation}
			X.-D. Zhang, \emph{Matrix Analysis and Applications}. Cambridge, U.K.: Cambridge Univ. Press, 2017.
		\bibitem{s_procedure}
		Z.-Q. {Luo}, J. F. {Sturm}, and S. {Zhang}, ``Multivariate nonnegative quadratic mappings," \emph{SIAM J. Optim.}, vol. 14, no. 4, pp. 1140-1162, Jan. 2004.
		\bibitem{schur}
		S. {Boyd} and L. Vandenberghe, \emph{Convex Optimization. Cambridge}, U.K.: Cambridge Univ. Press, 2004.
		\bibitem{general sign_definiteness}
		I. R. Petersen, ``A stabilization algorithm for a class of uncertain linear	systems,"  \emph{Syst. Contr. Lett.}, no. 8, pp. 351-357, Jan. 1987.
		\bibitem{general sign_definiteness_proof}
		E. A. {Gharavol} and E. G. {Larsson}, ``The sign-definiteness lemma and its applications to robust transceiver optimization for multiuser MIMO systems," \emph{IEEE Trans. Signal Process.}, vol. 61, no. 2, pp. 238-252, Jan. 2013.
		\bibitem{PCCP}
		Y. {Chen}, Y. {Wang}, and L. Jiao, "Robust transmission for reconfigurable intelligent surface aided millimeter wave vehicular communications with statistical CSI," \emph{IEEE Trans. Wireless Commun.}, vol. 21, no. 2, pp. 928-944, Feb. 2022.
		\bibitem{interior point}
		A. Ben-Tal and A. Nemirovski, \emph{Lectures on Modern Convex Optimization:	Analysis, Algorithms, and Engineering Applications}. Philadelphia, PA, USA: SIAM, 2001.	
			\bibitem{Prb_reference}
			T. Shafique, H. Tabassum, and E. Hossain, ``Optimization of wireless relaying with flexible UAV-borne reflecting surfaces," \emph{IEEE Trans. Commun.}, vol. 69, no. 1, pp. 309-325, Jan. 2021.
		\bibitem{convergence}
		X. {Yu}, D. {Xu}, Y. {Sun}, D. W. K. {Ng}, and R. {Schober}, ``Robust and secure wireless communications via intelligent reflecting surfaces," \emph{IEEE J. Sel. Areas Commun.}, vol. 38, no. 11, pp. 2637-2652, Nov. 2020.
			\bibitem{CSI estimation-STAR-RIS}
			C. Wu, C. You, Y. Liu, X. Gu, and Y. Cai, ``Channel estimation for STAR-RIS-aided wireless communication," \emph{IEEE Commun. Lett.}, 2021, early access, doi: 10.1109/LCOMM.2021.3139198.
	\end{thebibliography}
\end{document}